\documentclass[a4paper,11pt]{article}

\usepackage{jheppubmod}
\usepackage{amsmath}
\usepackage{amssymb}
\usepackage{bbm}
\usepackage{tipa}
\usepackage{mathrsfs}
\usepackage{mathtools}
\usepackage{grffile}
\usepackage{graphicx}
\usepackage{hyperref}

\title{Reformulations of Yang-Mills Theories with Space-time Tensor Fields}

\author{Zhi-Qiang Guo}
\emailAdd{zhiqiang.guo@usm.cl}
\affiliation{Departamento de F\'{i}sica y Centro Cient\'{i}fico
Tecnol\'{o}gico de Valpara\'{i}so,\\ Universidad T\'{e}cnica Federico
Santa Mar\'{i}a,\\ Casilla 110-V, Valpara\'{i}so, Chile}

\abstract{We provide the reformulations of Yang-Mills theories in terms of gauge invariant metric-like variables in three and four dimensions. The reformulations are used to analyze the dimension two gluon condensate and give gauge invariant descriptions of gluon polarization. In three dimensions, we obtain a non-zero dimension two gluon condensate by one loop computation, whose value is similar to the square of photon mass in the Schwinger model. In four dimensions, we obtain a Lagrangian with the dual property, which shares the similar but different property with the dual superconductor scenario. We also make discussions on the effectiveness of one loop approximation.}

\keywords{Higher Derivative Theory, Metric-like Formulation, Gauge Invariant Condensate}

\notoc

\begin{document}

\maketitle

\vspace{3mm}

\section{Introduction}\label{sec:1}

Regarding quantum Yang-Mills theories as highly nonlinear theories, it is difficult to achieve an understanding of its infrared region by the perturbative method. There are several approaches which suggest that the infrared region could be described through reformulating the Yang-Mills fields in terms of new variables. In~\cite{Faddeev:1998eq}, it was proposed that the infrared limit of the $SU(2)$ Yang-Mills theory in 4 dimensions could be given by a nonlinear sigma model by using partially dual variables\footnote{For its application to the confinement problem, see~\cite{Shabanov:1999xy,Gies:2001hk,Cho:2002iv} and the review article~\cite{Kondo:2014sta}.}, which are the decomposition of $SU(2)$ gauge field in terms of the coset variables of its $U(1)$ subgroup~\cite{Duan:1979,Cho:1981ww,Evslin:2011ti}. In~\cite{Faddeev:2006sw}, it was further proposed that a complete off-shell decomposition of $SU(2)$ field can be implemented through the view of spin-charge separation inspired by the strong correlated electron system.

There are also proposals to reformulate Yang-Mills theories by making use of field strength variables or gauge invariant metric-like variables~\cite{Lunev:1992zt}. Similar to the $U(1)$ Maxwell theory, which can be expressed by the field strength variables, the $SU(2)$ Yang-Mills theory can also be expressed by the field strength variables albeit in terms of an infinite series~\cite{Halpern:1977fw}. In~\cite{Haagensen:1994sy}, the frame-like fields are used as the pre-potential, and the $SU(2)$ Yang-Mills theory is recast into a $R^2$ gravity theory in 3 dimensions. In~\cite{Diakonov:2000pq}, the authors employed the metric-like fields and proposed that Yang-Mills theories could be regarded as the diffeomorphism invariant gravity theory broken by the background dependent ether term~\cite{Diakonov:2001xg}.

From another different angle, the reformulation or decomposition of gauge field is useful to address physical issues which are closely related to the gauge invariance. In~\cite{Verschelde:2001ia}, the transverse part of gauge field is used to analyze the gauge invariant dimension two condensate~\cite{Boucaud:2000nd,Gubarev:2000eu,Kondo:2001nq}, which is also been discussed by using the ``remaining'' part after subtracting an ``Abelian'' part from the original gauge field~\cite{Kondo:2006ih}. In~\cite{Chen:2008ag}, it was proposed that the gauge invariant contribution of gluon polarization to the nucleon spin can be described by decomposing the gauge field into its physical part and its pure gauge part. In~\cite{Sijacki:1990xp}, many properties of quantum chromodynamics are discussed by dressed gluon fields, which are gluon fields with its pure gauge parts subtracted as background fields.

 Inspired by the above investigations, we attempt to provide analysis on the infrared region of Yang-Mills theory through a novel reformulation and decomposition of the gauge field. At first, we decompose the gauge field into two parts as $A^{a}_{\mu}=B^{a}_{\mu}+e^{a}_{\mu}$. Here $e^{a}_{\mu}$ can be regarded as the frame-like fields in gravity theory, which have the same transformation properties as the gauge field strength. $B^{a}_{\mu}$ can be regarded as the gauge connection in gravity theory, which can be solved in terms of $e^{a}_{\mu}$ through imposing compatibility conditions, then a reformulation of $A^{a}_{\mu}$ in terms of $e^{a}_{\mu}$ is obtained. The Yang-Mills Lagrangian can then be reformulated by using the metric variable $g_{\mu\nu}$. We show that $g_{\mu\nu}$ provides facilities to analyze the gauge invariant dimension two condensate and give gauge invariant descriptions of gluon polarization.

This paper is organized as follows. In section~\ref{sec:2}, we discuss the $SU(2)$ Yang-Mills theory in three dimensions~(3D), where a dimension two condensate with the value $\frac{\lambda^2}{\pi^4}\kappa^4$ is derived by considering the one loop quantum correction. Here $\kappa^2$ is the coupling constant in 3D, which has the mass dimension; And $\lambda$ is a numerical constant produced by the one loop correction. We also provide arguments to demonstrate that the higher loop expansions are actually strong coupling expansions in subsection~\ref{sec:2.2}. In section~\ref{sec:3}, we discuss the $SU(2)$ Yang-Mills theory in four dimensions~(4D). We first derive the scalar-vector sector of the full Lagrangian in subsection~\ref{sec:3.2a}. By means of this scalar-vector sector, we propose a duality property between a gauge invariant order parameter and a nonzero dimension two condensate. This phenomenon is similar to but different from the dual superconductor scenario in~\cite{Faddeev:1998eq,Faddeev:2006sw}. In subsection~\ref{sec:3.2}, we give the one-loop result of the dimension two condensate in 4D, and the effectiveness of the one-loop computation is also discussed. Two subsections~\ref{sec:2.2-pf} and~\ref{sec:3.2-pf} are used to discuss the relations about the partition function between reformulated theories and Yang-Mills theories. We provide conclusions in section~\ref{sec:4}. Several appendices are used to provide more details of the paper.

\section{\texorpdfstring{$SU(2)$}{SU(2)} Yang-Mills Theory in Three Dimensions}\label{sec:2}

\subsection{Formulation with space-time tensor field}\label{sec:2.1}

In 3 dimensional space-time, the Lagrangian of $SU(2)$ Yang-Mills theory is
\begin{eqnarray}
\label{sec2-ym-lag}
\mathscr{L}&=&-\frac{1}{4\kappa^2}\eta^{\alpha\mu}\eta^{\beta\nu}F^{a}_{\alpha\beta}F^{a}_{\mu\nu},\\
F^{a}_{\mu\nu}&=&\partial_{\mu}A^{a}_{\nu}-\partial_{\nu}A^{a}_{\mu}+\epsilon^{abc}A^{b}_{\mu}A^{c}_{\nu},\nonumber
\end{eqnarray}
where $\eta^{\mu\nu}=\mathrm{diag}(1,-1,-1)$ is the Lorentz metric in 3D. We first decompose the gauge field as\footnote{For a different approach based on matrix parametrization of $SU(N)$ Yang-Mills theory in 3D, see~\cite{Karabali:1995ps}.}
\begin{eqnarray}
\label{sec2-gf-dec}
A^{a}_{\mu}=B^{a}_{\mu}+e^{a}_{\mu}.
\end{eqnarray}
The decomposition~(\ref{sec2-gf-dec}) is familiar with the setting in the 3 dimensional gravity, where this decomposition is used to recast the Chern-Simons theory into a geometrical formulation~\cite{Achucarro:1987vz,Witten:1988hc}. Under the gauge transformation, $e_{\mu}=e^{a}_{\mu}\tau^{a}$ and $B_{\mu}=B^{a}_{\mu}\tau^{a}$, where $\tau^{a}$ takes values in the $SU(2)$ Lie algebra, have the gauge transformation
\begin{eqnarray}
\label{sec2-gf-dec-gt-e}
{e}_{\mu}&\rightarrow&{U}{e}_{\mu}U^{-1},\\
\label{sec2-gf-dec-gt-b}
{B}_{\mu}&\rightarrow&{U}{B}_{\mu}U^{-1}-i\partial_{\mu}UU^{-1}.
\end{eqnarray}
These are similar to that we decompose the gauge field into its background part and its quantum part, then the quantum part transforms as Eq.~(\ref{sec2-gf-dec-gt-e}) and the background part transforms as Eq.~(\ref{sec2-gf-dec-gt-b}). Under this decomposition, the field strength has the decomposition
\begin{eqnarray}
\label{sec2-gf-dec-fstr-0}
F^{a}_{\mu\nu}&=&\mathcal{B}^{a}_{\mu\nu}+\mathcal{T}^{a}_{\mu\nu}
+\epsilon^{abc}e^{b}_{\mu}e^{c}_{\nu},\\
\label{sec2-gf-dec-fstr-1}
\mathcal{B}^{a}_{\mu\nu}&=&\partial_{\mu}B^{a}_{\nu}-\partial_{\nu}B^{a}_{\mu}+\epsilon^{abc}B^{b}_{\mu}B^{c}_{\nu},\\
\label{sec2-gf-dec-fstr-2}
\mathcal{T}^{a}_{\mu\nu}&=&(\partial_{\mu}e^{a}_{\nu}+\epsilon^{abc}B^{b}_{\mu}e^{c}_{\nu})-(\partial_{\nu}e^{a}_{\mu}
+\epsilon^{abc}B^{b}_{\nu}e^{c}_{\mu}).
\end{eqnarray}
Here $\mathcal{B}^{a}_{\mu\nu}$ is the curvature of $B^{a}_{\mu}$, and $\mathcal{T}^{a}_{\mu\nu}$ has the similar structure to the torsion in gravity theory. In the conventional background approach, $B^{a}_{\mu}$ is a fixed classical background, which is independent of $e^{a}_{\mu}$. In this paper, we require that $B^{a}_{\mu}$ and $e^{a}_{\mu}$ satisfy the equation
\begin{eqnarray}
\label{sec2-gf-dec-res}
\partial_{\mu}e^{a}_{\nu}+\epsilon^{abc}B^{b}_{\mu}e^{c}_{\nu}=\Gamma^{\rho}_{\mu\nu}e^{a}_{\rho}.
\end{eqnarray}
This constraint is consistent with the transformations~(\ref{sec2-gf-dec-gt-e}) and~(\ref{sec2-gf-dec-gt-b}). This equation is also similar to the compatibility equation in gravity theory. From Eq.~(\ref{sec2-gf-dec-res}), we can solve $B^{a}_{\mu}$ as
\begin{eqnarray}
\label{sec2-gf-dec-res-sol}
B^{a}_{\mu}=-\frac{1}{2}\epsilon^{abc}E^{\rho}_{b}(\partial_{\mu}e_{\rho}^{c}-\Gamma^{\sigma}_{\mu\rho}e^{c}_{\sigma}).
\end{eqnarray}
Here $E^{\rho}_{a}$ is the inverse of $e_{\rho}^{a}$, that is, $E^{\rho}_{a}e_{\rho}^{b}=\delta^{a}_{b}$ and $E^{\alpha}_{a}e_{\beta}^{a}=\delta^{\alpha}_{\beta}$. As in gravity theory, from Eq.~(\ref{sec2-gf-dec-res}), we can obtain the metric compatibility condition
\begin{eqnarray}
\label{sec2-gf-res-g}
\partial_{\mu}g_{\alpha\beta}=\Gamma^{\rho}_{\mu\alpha}g_{\rho\beta}+\Gamma^{\rho}_{\mu\beta}g_{\rho\alpha}.
\end{eqnarray}
Here $g_{\alpha\beta}=e^{a}_{\alpha}e^{a}_{\beta}$ is the metric tensor, and $g^{\alpha\beta}=E_{a}^{\alpha}E_{a}^{\beta}$ is its inverse. A general solution of Eq.~(\ref{sec2-gf-res-g}) is
\begin{eqnarray}
\label{sec2-gf-res-H-g-sol}
\Gamma^{\rho}_{\mu\nu}=\frac{1}{2}g^{\rho\sigma}(\partial_{\mu}g_{\sigma\nu}
+\partial_{\nu}g_{\sigma\mu}-\partial_{\sigma}g_{\mu\nu})
-g^{\rho\sigma}(T^{\tau}_{\mu\sigma}g_{\tau\nu}+T^{\tau}_{\nu\sigma}g_{\tau\mu})+T^{\rho}_{\mu\nu}.
\end{eqnarray}
Here $T^{\rho}_{\mu\nu}$ is the torsion tensor. In the following discussions, we shall use the torsion-free condition
\begin{eqnarray}
\label{sec2-gf-res-H-g-tor}
T^{\rho}_{\mu\nu}=0,
\end{eqnarray}
so the connection~(\ref{sec2-gf-res-H-g-sol}) is reduced to the Levi-Civita connection. Under the torsion-free condition, $B^{a}_{\mu}$ is completely determined by $e^{a}_{\mu}$. Because $e^{a}_{\mu}$ has the same superficial degrees of freedoms with $A^{a}_{\mu}$, the torsion-free condition ensures that no extra field variables are introduced. From Eq.~(\ref{sec2-gf-dec-res}) and using the connections~(\ref{sec2-gf-dec-res-sol}), (\ref{sec2-gf-res-H-g-sol}) and~(\ref{sec2-gf-res-H-g-tor}), we can obtain
\begin{eqnarray}
\label{sec2-gf-fsol-1}
\mathcal{B}^{a}_{\mu\nu}=\frac{1}{2}R^{\sigma}_{\hspace{1mm}\rho\mu\nu}e^{c}_{\sigma}E^{\rho}_{b}\epsilon^{abc},
~~\mathcal{T}^{a}_{\mu\nu}=0,
\end{eqnarray}
where
\begin{eqnarray}
\label{sec2-gf-fsol-2}
R^{\sigma}_{\hspace{1mm}\rho\mu\nu}=\partial_{\mu}\Gamma^{\sigma}_{\nu\rho}-\partial_{\nu}\Gamma^{\sigma}_{\mu\rho}+
\Gamma^{\tau}_{\nu\rho}\Gamma^{\sigma}_{\mu\tau}-\Gamma^{\tau}_{\mu\rho}\Gamma^{\sigma}_{\nu\tau}
\end{eqnarray}
is the Riemann curvature. Using these results, the Lagrangian~(\ref{sec2-ym-lag}) can be rewritten as\footnote{Here we have two metrics $\eta_{\mu\nu}$ and $g_{\mu\nu}$. The conventions of usage about these metrics are given in appendix~\ref{convention}.}
\begin{eqnarray}
\label{sec2-ym-lag-re}
\mathscr{L}&=&\mathscr{L}^{(4)}+\mathscr{L}^{(2)}+\mathscr{L}^{(0)},\\
\label{sec2-ym-lag-re-0}
-4\kappa^2\mathscr{L}^{(4)}&=&\frac{1}{4}\eta^{\mu\alpha}\eta^{\nu\beta}
(g^{\rho\theta}g_{\sigma\tau}-\delta^{\rho}_{\tau}\delta^{\theta}_{\sigma})
{R}^{\sigma}_{\hspace{1mm}\rho\mu\nu}{R}^{\tau}_{\hspace{1mm}\theta\alpha\beta},\\
\label{sec2-ym-lag-re-2}
-4\kappa^2\mathscr{L}^{(2)}&=&\eta^{\mu\alpha}\eta^{\nu\beta}
(\delta^{\rho}_{\alpha}g_{\sigma\beta}-\delta^{\rho}_{\beta}g_{\sigma\alpha}){R}^{\sigma}_{\hspace{1mm}\rho\mu\nu},\\
\label{sec2-ym-lag-re-4}
-4\kappa^2\mathscr{L}^{(0)}&=&\eta^{\mu\alpha}\eta^{\nu\beta}(g_{\mu\alpha}g_{\nu\beta}-g_{\mu\beta}g_{\nu\alpha}).
\end{eqnarray}
The Lagrangian $\mathscr{L}$ is divided into three parts in the above. The $\mathscr{L}^{(4)}$ part is not new, as it has been given in~\cite{Haagensen:1994sy} in a Hamiltonian analysis of the $SU(2)$ Yang-Mills theory and recently in~\cite{Nojiri:2002ad,Mitra:2013wma} through the similar approach as we just presented in the above. Because we have decomposed $A^{a}_{\mu}$ into two parts in Eq.~(\ref{sec2-gf-dec}), compared to the results in~\cite{Haagensen:1994sy,Nojiri:2002ad,Mitra:2013wma}, we obtain two additional parts $\mathscr{L}^{(2)}$ and $\mathscr{L}^{(0)}$. From the view of effective field theory, the terms $\mathscr{L}^{(2)}$ and $\mathscr{L}^{(0)}$ are both relevant operators. The Lagrangian $\mathscr{L}$ is also similar to the higher derivative gravity of Stelle~\cite{Stelle:1976gc}. In~\cite{Sijacki:1990xp}, it has been suggested that the infrared region of quantum chromodynamics could be described by the Stelle action, in which the higher derivative terms are responsible for the confinement.

\subsection{Relations between Partition Functions}\label{sec:2.2-pf}

In the previous section, we have expressed the Yang-Mills fields $A^{a}_{\mu}$ in terms of $e^{a}_{\mu}$. A question is that if the reformulated theory is equivalent to the original Yang-Mills theory from the quantum perspective. Obviously, there is a Jacobian matrix corresponding to the variable transformation~(\ref{sec2-gf-dec-res-sol}). This Jacobian matrix has been discussed in~\cite{Mitra:2013wma}. In this subsection, we address this question from a different approach. We consider the partition function
\begin{eqnarray}
\label{sec2-ym-pf}
Z[B^{a}_{\mu},e^{a}_{\mu}]=\int [\mathscr{D}B^{a}_{\mu}][\mathscr{D}e^{a}_{\mu}][\mathscr{D}j^{a}_{\mu}]\mathrm{exp}
\bigl(i\int{d^3x}(\mathscr{L}+\varepsilon^{\mu\alpha\beta}{j}^{a}_{\mu}\mathcal{T}^{a}_{\alpha\beta})\bigr).
\end{eqnarray}
Here $\mathscr{L}$ is the Yang-Mills action~(\ref{sec2-ym-lag}) with the decomposition $A^{a}_{\mu}=B^{a}_{\mu}+ e^{a}_{\mu}$. $\mathcal{T}^{a}_{\alpha\beta}$ is the torsion constraint condition~(\ref{sec2-gf-dec-fstr-2}). By integrating out ${j}^{a}_{\mu}$, we can obtain
\begin{eqnarray}
\label{sec2-ym-pf-j}
Z[B^{a}_{\mu},e^{a}_{\mu}]=\int [\mathscr{D}B^{a}_{\mu}][\mathscr{D}e^{a}_{\mu}]\delta\bigl(\varepsilon^{\mu\alpha\beta}\mathcal{T}^{a}_{\alpha\beta}\bigr)
\mathrm{exp}\bigl(i\int{d^3x}\mathscr{L}\bigr).
\end{eqnarray}
So $Z[B^{a}_{\mu},e^{a}_{\mu}]$ can be understood as the path integral of $B^{a}_{\mu}$ and $e^{a}_{\mu}$ with the torsion free condition $\mathcal{T}^{a}_{\alpha\beta}=0$. From Eq.~(\ref{sec2-ym-pf-j}), we plan to integrate out $B^{a}_{\mu}$, which requires the solution of the torsion free condition $\mathcal{T}^{a}_{\alpha\beta}=0$. The solution has been given in Eq.~(\ref{sec2-gf-dec-res-sol})
\begin{eqnarray}
\label{sec2-ym-pf-zb-b}
B^{a}_{\mu}(e)=-\frac{1}{2}\epsilon^{abc}E^{\rho}_{b}
(\partial_{\mu}e_{\rho}^{c}-\Gamma^{\sigma}_{\mu\rho}e^{c}_{\sigma}),
\end{eqnarray}
From Eq.~(\ref{sec2-ym-pf-j}), we obtain
\begin{eqnarray}
\label{sec2-ym-pf-zb}
Z[B^{a}_{\mu},e^{a}_{\mu}]=\int [\mathscr{D}e^{a}_{\mu}] [\mathscr{D}B^{a}_{\mu}]\frac{1}{J_{e}}
\delta\left(B^{a}_{\mu}-B^{a}_{\mu}(e)\right)\mathrm{exp}\bigl(i\int{d^3x}
\mathscr{L}(B^{a}_{\mu},e^{a}_{\mu})\bigr).
\end{eqnarray}
In the above, $J_{e}$ is the Jacobian factor associated with the variable transformation of the delta function, which is given by
\begin{eqnarray}
\label{sec2-ym-pf-norm-jb}
J_{e}&=&\mathrm{det}\bigl(\mathcal {M}^{\alpha\beta}_{ab}\bigr),\\
\label{sec2-ym-pf-trans-mdef}
\mathcal {M}^{\mu\nu}_{ab}&=&\frac{\partial\left(\varepsilon^{\mu\alpha\beta}\mathcal{T}^{a}_{\alpha\beta}\right)}{\partial{B}^{b}_{\nu}}
=2\varepsilon^{\mu\nu\beta}\varepsilon^{abc}e^{c}_{\beta},
\end{eqnarray}
where $\mathcal {M}^{\mu\nu}_{ab}$ is a $9\times{9}$ matrix. Integrating out $B^{a}_{\mu}$ in Eq.~(\ref{sec2-ym-pf-zb}), we obtain
\begin{eqnarray}
\label{sec2-ym-pf-zb-pi}
Z[e^{a}_{\mu}]=\int [\mathscr{D}e^{a}_{\mu}]\frac{1}{J_{e}}\mathrm{exp}\bigl(i\int{d^3x}
\mathscr{L}(e^{a}_{\mu})\bigr),
\end{eqnarray}
where $\mathscr{L}(e^{a}_{\mu})$ solely depends on $e^{a}_{\mu}$, which is equivalent to the reformulated Lagrangian in Eq.~(\ref{sec2-ym-lag-re}). From Eq.~(\ref{sec2-ym-pf-norm-jb}), we see that additional Jacobian factor $J_{e}$ depends only on $e^{a}_{\mu}$ but not its derivative, which does not influence the path integral~\cite{Stelle:1976gc,Hamber:2009} and can be abandoned.

In the above, we firstly integrated out $B^{a}_{\mu}$, then we obtain the partition function of $e^{a}_{\mu}$. Alternatively, we can integrate out $e^{a}_{\mu}$ at first, then we obtain the partition function of $B^{a}_{\mu}$. In this case, it is difficult to solve the constraint $\mathcal{T}^{a}_{\alpha\beta}=0$ regarding $e^{a}_{\mu}$ as the arguments , because $\mathcal{T}^{a}_{\alpha\beta}$ are partial differential equations of $e^{a}_{\mu}$, compared to that they are algebraic equations of $B^{a}_{\mu}$. However, the path integral about $e^{a}_{\mu}$ in Eq.~(\ref{sec2-ym-pf}) is similar to the calculations in the background field method. We add the gauge fixing term and the Faddeev-Popov ghost term to Eq.~(\ref{sec2-ym-pf})
\begin{eqnarray}
\label{sec2-ym-pf-ze}
Z[B^{a}_{\mu},e^{a}_{\mu}]&=&\int [\mathscr{D}B^{a}_{\mu}][\mathscr{D}j^{a}_{\mu}][\mathscr{D}\bar{\eta}][\mathscr{D}\eta][\mathscr{D}e^{a}_{\mu}]
\nonumber\\
&\times&\mathrm{exp}\biggl(i\int{d^3x}
(\mathscr{L}(B^{a}_{\mu},e^{a}_{\mu})+\varepsilon^{\mu\alpha\beta}{j}^{a}_{\mu}\mathcal{T}^{a}_{\alpha\beta}
+\mathscr{L}_{\mathrm{GF}}+\mathscr{L}_{\mathrm{FP}})\biggr),\\
\label{sec2-ym-pf-ze-gf}
-2\kappa^2\mathscr{L}_{\mathrm{GF}}&=&(D^{\mu}e^{a}_{\mu})(D^{\nu}e^{a}_{\nu}),\\
\label{sec2-ym-pf-ze-fp}
\mathscr{L}_{\mathrm{FP}}&=&
\bar{\eta}^{a}\left(-(D^{\mu}D_{\mu})^{ac}-(D^{\mu})^{ad}\varepsilon^{dbc}e^{b}_{\mu}\right)\eta^{c},
\end{eqnarray}
where
\begin{eqnarray}
\label{sec2-ym-pf-ze-der}
D^{ac}_{\mu}=\partial_{\mu}\delta^{ac}+\epsilon^{abc}{B}^{b}_{\mu}.
\end{eqnarray}
Integrating out $e^{a}_{\mu}$ to one loop order, we have the determinant
\begin{eqnarray}
\label{sec2-ym-pf-norm-trans-mdef-e}
\left[\mathrm{det}\left(-\eta_{\alpha\beta}(\eta^{\mu\nu}D_{\mu}D_{\nu})^{ab}
+2\varepsilon^{abc}\mathcal{B}^{c}_{\alpha\beta}\right)\right]^{-\frac{1}{2}}.
\end{eqnarray}
Integrating out the ghost fields to one loop order, we have the determinant
\begin{eqnarray}
\label{sec2-ym-pf-norm-trans-mdef-e-gh}
\mathrm{det}\left(-(\eta^{\mu\nu}D_{\mu}D_{\nu})^{ab}\right).
\end{eqnarray}
The functional determinant~(\ref{sec2-ym-pf-norm-trans-mdef-e}) and~(\ref{sec2-ym-pf-norm-trans-mdef-e-gh}) can be computed in a formal series of $B^{a}_{\mu}$. The one-loop result for $SU(N)$ Yang-Mills theories can be found in~\cite{Peskin:1995ev}. The total effects of these determinants are to renormalize the coupling constant.

In the above discussions, we begin with the path integral in Eq.~(\ref{sec2-ym-pf}). On one hand, we integrate out $B^{a}_{\mu}$ at first, then we obtain the reformulated Lagrangian~(\ref{sec2-ym-lag-re}). Alternatively, we integrate out $e^{a}_{\mu}$, then we obtain the effective action of $B^{a}_{\mu}$ in the background field method. The above discussions provide the relations of the reformulated Lagrangian and the effective action of $B^{a}_{\mu}$.

\subsection{Scalar Sector}\label{sec:2.3-sr}

In Eq.~(\ref{sec2-ym-lag-re}), we have obtained a pure space-time tensor formulation of the Yang-Mills Lagrangian. The dynamical variable of this Lagrangian is a spin-2 tensor $g_{\mu\nu}$. In this subsection, we discuss a simple case of this Lagrangian. Using the Lorentz metric,  we decompose $g_{\mu\nu}$ as
\begin{eqnarray}
\label{sec2-a-g-def-sc}
g_{\mu\nu}=\frac{1}{3}(\eta^{\rho\sigma}g_{\rho\sigma})\eta_{\mu\nu}
+\bigl(g_{\mu\nu}-\frac{1}{3}(\eta^{\rho\sigma}g_{\rho\sigma})\eta_{\mu\nu}\bigr),
\end{eqnarray}
where
\begin{eqnarray}
\label{sec2-a-g-def-sc-def}
\varphi^2(x)=\frac{1}{3}(\eta^{\rho\sigma}g_{\rho\sigma})
\end{eqnarray}
is the trace part, which is the scalar degree of freedom of $g_{\mu\nu}$. Because the $SU(2)$ group is compact, the diagonal elements of $g_{\mu\nu}$ defined in Eq.~(\ref{sec2-gf-res-g}) are all non-negative. Hence we cannot suppose the tensor part on the right hand of Eq.~(\ref{sec2-a-g-def-sc}) is zero. However, for the $SU(2)$ group in the Euclidean space-time or the non-compact $SO(1,2)$ group in the Minkowski space-time, the tensor part can be supposed to be zero. In the following, we shall only give the explicit expression of the scalar sector. The sector including the tensor part can be obtained as a formal series. The scalar sector of the Lagrangian $\mathscr{L}$ in Eq.~(\ref{sec2-ym-lag-re}) can be written as
\begin{eqnarray}
\label{sec2-ym-lag-re-dec-sca}
\mathscr{L}&=&\frac{1}{2}\eta^{\mu\nu}\partial_{\mu}\phi\partial_{\nu}\phi-\frac{\kappa^2}{24}\phi^4+\mathscr{L}_{\mathrm{bound}}\\
&-&\frac{1}{\kappa^2}\frac{1}{\phi^2}(\square\phi)^2
+\frac{1}{2\kappa^2}\frac{1}{\phi^3}(\partial^{\nu}\phi\partial_{\nu}\phi)\square\phi,\nonumber
\end{eqnarray}
where we have defined $\phi=\frac{\sqrt{6}}{\kappa}\varphi$, $\square=\eta^{\mu\nu}\partial_{\mu}\partial_{\nu}$, and the boundary term is
\begin{eqnarray}
\label{sec2-ym-lag-re-dec-sca-bound}
\mathscr{L}_{\mathrm{bound}}&=&-\frac{1}{3}\partial^{\mu}(\phi\partial_{\mu}\phi)
-\frac{1}{2\kappa^2}\partial^{\mu}\bigl(\frac{1}{\phi^2}\partial^{\nu}\phi\partial_{\mu}\partial_{\nu}\phi\bigr)\\
&+&\frac{1}{2\kappa^2}\partial^{\mu}\bigl(\frac{1}{\phi^2}\partial_{\mu}\phi\square\phi\bigr)
+\frac{1}{2\kappa^2}\partial^{\mu}\bigl(\frac{1}{\phi^3}\partial_{\mu}\phi\partial^{\nu}\phi\partial_{\nu}\phi\bigr).\nonumber
\end{eqnarray}
From Eq.~(\ref{sec2-ym-lag-re-dec-sca}), we see that the Lagrangian has higher derivative terms and also has dilation-like couplings. We can introduce two auxiliary fields $\theta(x)$ and $\sigma(x)$ to rewrite the Lagrangian~(\ref{sec2-ym-lag-re-dec-sca}) in the polynomial formulation
\begin{eqnarray}
\label{sec2-ym-lag-re-dec-sca-aux}
\mathscr{L}&=&\frac{1}{2}\eta^{\mu\nu}\partial_{\mu}\phi\partial_{\nu}\phi-\frac{\kappa^2}{24}\phi^4
-\partial^{\mu}\theta\partial_{\mu}\phi\\
&-&\frac{1}{2}\theta\sigma\phi^4-\frac{1}{4\kappa^2}\sigma^2\phi^6
+\frac{1}{4\kappa^2}\sigma\phi(\partial^{\mu}\phi\partial_{\mu}\phi).\nonumber
\end{eqnarray}
This Lagrangian is similar to the Lee-Wick Lagrangian~\cite{Lee:1969fy,Lee:1970iw}. The Lee-Wick Lagrangian includes higher derivative terms, but it can be recast into lower derivative formulation by introducing auxiliary ghost fields~\cite{Grinstein:2007mp}. In Eq.~(\ref{sec2-ym-lag-re-dec-sca-aux}), $\theta(x)$ can obtain a ghost-like kinetic term by its mixing with $\phi(x)$.

\subsection{Dimension Two Condensate}\label{sec:2.2}

The nonzero value\footnote{However, for a null result from the operator product expansion analysis of the lattice result, see~\cite{Bali:2014sja}.} of dimension two gluon condensate has been suggested~\cite{Verschelde:2001ia,Boucaud:2000nd,Gubarev:2000eu,Kondo:2001nq} through several different approaches. From our results in the last subsection, we learned that the metric tensor $g_{\mu\nu}$ is gauge invariant, and it is constructed from the part $e^{a}_{\mu}$ of $A^{a}_{\mu}$. So it provides a natural candidate for the dimension two gluon condensate. In this subsection, we provide analysis to show that a nonzero vacuum expectation of $g_{\mu\nu}$ can be generated by quantum corrections.

We expand $g_{\mu\nu}$ around a constant background $v\cdot\eta_{\mu\nu}$
\begin{eqnarray}
\label{sec2-a-g-def}
g_{\mu\nu}=v\cdot\eta_{\mu\nu}+h_{\mu\nu}.
\end{eqnarray}
where $h_{\mu\nu}$ is the fluctuation around the constant background. Assuming that $h_{\mu\nu}$ is small, we can express $g^{\mu\nu}$ as
\begin{eqnarray}
\label{sec2-a-g-def-inv}
g^{\mu\nu}=\frac{1}{v}\eta^{\mu\nu}-\frac{1}{v^2}\eta^{\mu\rho}\eta^{\nu\sigma}h_{\rho\sigma}+\cdots.
\end{eqnarray}
The quadratic terms of the Lagrangian $\mathscr{L}$ can then be written as
\begin{eqnarray}
\label{sec2-ym-lag-re-dec}
\mathscr{L}&=&\mathscr{L}^{(4)}+\mathscr{L}^{(2)}+\mathscr{L}^{(1)}+\mathscr{L}^{(0)},\\
\label{sec2-ym-lag-re-dec-0}
-4\kappa^2\mathscr{L}^{(4)}&=&\frac{1}{2v^2}\left(\partial^{\mu}\partial_{\nu}h_{\alpha\beta}
\partial_{\mu}\partial^{\nu}h^{\alpha\beta}-2\partial_{\mu}\partial_{\nu}h_{\alpha\beta}
\partial^{\mu}\partial^{\alpha}h^{\beta\nu}+\partial_{\mu}\partial_{\nu}h_{\alpha\beta}
\partial^{\alpha}\partial^{\beta}h^{\mu\nu}\right),\\
\label{sec2-ym-lag-re-dec-2}
-4\kappa^2\mathscr{L}^{(2)}&=&\frac{1}{2v}\left(-3\partial^{\mu}h_{\alpha\beta}\partial_{\mu}h^{\alpha\beta}
+\partial^{\mu}h^{\alpha}_{\alpha}\partial_{\mu}h^{\beta}_{\beta}
-4\partial_{\mu}h^{\alpha}_{\alpha}\partial_{\nu}h^{\mu\nu}\right)\\
&+&\frac{1}{v}\left(\partial_{\mu}h_{\nu\alpha}\partial^{\nu}h^{\mu\alpha}
+2\partial_{\mu}h^{\mu\alpha}\partial^{\nu}h_{\nu\alpha}\right),\nonumber\\
\label{sec2-ym-lag-re-dec-1}
-4\kappa^2\mathscr{L}^{(1)}&=&-2\partial_{\mu}\partial_{\nu}h^{\mu\nu}+2\partial^{\mu}\partial_{\mu}h^{\alpha}_{\alpha},\\
\label{sec2-ym-lag-re-dec-4}
-4\kappa^2\mathscr{L}^{(0)}&=&6v^2+4v\cdot{h}^{\alpha}_{\alpha}
+\eta^{\mu\alpha}\eta^{\nu\beta}(h_{\mu\alpha}h_{\nu\beta}-h_{\mu\beta}h_{\nu\alpha}).
\end{eqnarray}
In the above, the linear term $\mathscr{L}^{(1)}$ is a total divergence term. From Eqs.~(\ref{sec2-ym-lag-re-dec-0})-(\ref{sec2-ym-lag-re-dec-4}), using the projection operators in appendix~\ref{propagator}, we obtain the kinetic operator
\begin{eqnarray}
\label{sec2-ym-lag-re-prop}
D_{\mu\nu,\alpha\beta}&=&a_{1}P^{(2)}_{\mu\nu,\alpha\beta}+a_{2}P^{(1)}_{\mu\nu,\alpha\beta}
+a_{3}P^{(0,s)}_{\mu\nu,\alpha\beta}\\
&+&a_{4}P^{(0,\omega)}_{\mu\nu,\alpha\beta}+a_{5}P^{(0,s\omega)}_{\mu\nu,\alpha\beta}
+a_{6}P^{(0,\omega{s})}_{\mu\nu,\alpha\beta},\nonumber
\end{eqnarray}
where the coefficients are
\begin{eqnarray}
\label{sec2-ym-lag-re-prop-co-1}
a_{1}&=&-\frac{1}{2v^2}\square^2-\frac{3}{2v}\square+1,~~
a_{2}=1,\\
\label{sec2-ym-lag-re-prop-co-2}
a_{3}&=&-\frac{1}{2v^2}\square^2-\frac{1}{2v}\square-1,~~
a_{4}=0,\\
\label{sec2-ym-lag-re-prop-co-3}
a_{5}&=&a_{6}=-\frac{1}{\sqrt{2}}\bigl(\frac{1}{v}\square+2\bigr).
\end{eqnarray}
The inverse of this operator is
\begin{eqnarray}
\label{sec2-ym-lag-re-prop-inv}
\left(D^{-1}\right)_{\mu\nu,\alpha\beta}&=&\frac{1}{a_{1}}P^{(2)}_{\mu\nu,\alpha\beta}
+\frac{1}{a_{2}}P^{(1)}_{\mu\nu,\alpha\beta}
+\frac{a_{4}}{a_{3}a_{4}-a_{5}a_{6}}P^{(0,s)}_{\mu\nu,\alpha\beta}\\
&+&\frac{a_{3}}{a_{3}a_{4}-a_{5}a_{6}}P^{(0,\omega)}_{\mu\nu,\alpha\beta}
-\frac{a_{5}}{a_{3}a_{4}-a_{5}a_{6}}P^{(0,s\omega)}_{\mu\nu,\alpha\beta}
-\frac{a_{6}}{a_{3}a_{4}-a_{5}a_{6}}P^{(0,\omega{s})}_{\mu\nu,\alpha\beta}.\nonumber
\end{eqnarray}
$a_{1}$ can be factorized into
\begin{eqnarray}
\label{sec2-ym-lag-re-prop-co-1-fac}
a_{1}&=&-\frac{1}{2}\left(\frac{1}{v}\square+b_{+}\right)
\left(\frac{1}{v}\square-b_{-}\right),\\
\label{sec2-ym-lag-re-prop-co-1-fac-root}
b_{+}&=&\frac{1}{2}(\sqrt{17}+3),~~b_{-}=\frac{1}{2}(\sqrt{17}-3).
\end{eqnarray}
For $v>0$, we see that the $b_{-}$ term contributes the tachyonic mode in the propagator~(\ref{sec2-ym-lag-re-prop-inv}); While the $b_{+}$ term and the $a_{5}$ term contribute two massive modes.

In Eq.~(\ref{sec2-ym-lag-re-dec}), we only give the quadratic part of the Lagrangian. The interactive part can also be derived, but we shall not give them here due to the lengthy expressions. Using $V_{2}$ and $V_{4}$ as the number of the interactive vertex of two derivatives and four derivatives respectively, and $N_{I}$ and $L$ as the numbers of internal lines and loops, we have the superficial degree of divergence
\begin{eqnarray}
\label{sec2-ym-lag-3d-dod}
\omega=3L-4N_{I}+2V_{2}+4V_{4}.
\end{eqnarray}
Note that the coefficient of $N_{I}$ is $4$ due to the higher power of the propagator. We do not need the gauge fixing term and ghost fields because the propagator~(\ref{sec2-ym-lag-re-prop-inv}) is invertible. And we also have the relation from the topology of the diagram
\begin{eqnarray}
\label{sec2-ym-lag-3d-dod-l}
L=N_{I}-V_{2}-V_{4}+1.
\end{eqnarray}
Using Eq.~(\ref{sec2-ym-lag-3d-dod-l}) to eliminate $N_{I}$ from Eq.~(\ref{sec2-ym-lag-3d-dod}), finally we have
\begin{eqnarray}
\label{sec2-ym-lag-3d-dod-l-1}
\omega=3-L-2V_{2}.
\end{eqnarray}
Thanks to the higher derivative behavior of the propagator, the superficial degree of divergence $\omega$ is suppressed by the numbers of loop and two derivative vertex. So the higher derivative Lagrangian~(\ref{sec2-ym-lag-re-dec}) is actually super-renormalizable by power counting.

To compute the quantum correction to the effective potential to one loop order, we have the determinant
\begin{eqnarray}
\label{sec2-ym-lag-3d-eff-det}
\left[\mathrm{det}\left(D_{\mu\nu,\alpha\beta}\right)\right]^{-\frac{1}{2}}.
\end{eqnarray}
Regarding the tensor sub-sector as $6\times{6}$ matrix, and computing the determinant of the tensor sub-sector, we have
\begin{eqnarray}
\label{sec2-ym-lag-3d-eff-det-t}
\mathrm{det}\left(D_{\mu\nu,\alpha\beta}\right)=
\mathrm{det}\biggl(-\frac{1}{8}\bigl(\frac{1}{v}\square+b_{+}\bigr)^2
\bigl(\frac{1}{v}\square-b_{-}\bigr)^2
\bigl(\frac{1}{v}\square+2\bigr)^2\biggr).
\end{eqnarray}
In Eq.~(\ref{sec2-ym-lag-3d-eff-det-t}), it is clear that the contributions come from one tachyonic mode and two massive modes assuming $v>0$. Using this determinant, we obtain the one-loop effective potential
\begin{eqnarray}
\label{sec2-ym-lag-3d-eff}
V_{\mathrm{eff}}&=&\frac{3}{2\kappa^2}v^2
+\int\frac{d^3k_{\mathrm{E}}}{(2\pi)^3}\mathrm{log}\left(\frac{1}{v}k^2_{\mathrm{E}}+2\right)\nonumber\\
&+&\int\frac{d^3k_{\mathrm{E}}}{(2\pi)^3}\left[\mathrm{log}\left(\frac{1}{v}k^2_{\mathrm{E}}+b_{+}\right)
+\mathrm{log}\bigl\lvert\frac{1}{v}k^2_{\mathrm{E}}-b_{-}\bigr\rvert\right],
\end{eqnarray}
where $k_{\mathrm{E}}$ is the Euclidean momentum in 3D. We see that the contributions to the effective potential are divided into two massive modes and one tachyonic mode. Because the tachyonic mode is square in the determinant~(\ref{sec2-ym-lag-3d-eff-det-t}), the abstract sign is required in Eq.~(\ref{sec2-ym-lag-3d-eff}). We use the dimensional regularization to compute the effective potential. The details of calculation are given in appendix~\ref{dimreg}. The calculations with the dimensional regularization show that the integrals in Eq.~(\ref{sec2-ym-lag-3d-eff}) are finite in 3D. And finally we obtain the effective potential
\begin{eqnarray}
\label{sec2-ym-lag-3d-eff-dg}
V_{\mathrm{eff}}&=&\frac{3}{2\kappa^2}v^2
-\frac{1}{6\pi^2}v^{\frac{3}{2}}\bigl(c_1+c_2+c_3\bigr),\\
c_{1}&=&2^{\frac{3}{2}},~~c_{2}=(b_{+})^{\frac{3}{2}},
~~c_{3}=0.\nonumber
\end{eqnarray}
$c_{1}$ and $c_{2}$ are contributed by two massive modes, and $c_{3}$ is contributed by the tachyonic mode, which is zero in 3D by dimensional regularization. $V_{\mathrm{eff}}$ has the minimum at
\begin{eqnarray}
\label{sec2-ym-lag-3d-eff-dg-min}
\langle{v}\rangle=\frac{\lambda^2}{\pi^4}\kappa^4,~~\lambda=\frac{1}{12}\bigl(c_1+c_2+c_3\bigr),
\end{eqnarray}
where $\lambda$ can be regarded as the effective charge. In the Schwinger model~\cite{Schwinger:1962tp} in two dimensional space-time, a nonzero photon mass can be generated from the fermion loop contribution. We see that the dimension two condensate~(\ref{sec2-ym-lag-3d-eff-dg-min}) have the same structure with the square of the photon mass in two dimensions.

In the above, we have expanded $g_{\mu\nu}$ around a constant background $v\cdot\eta_{\mu\nu}$. Because the Lagrangian~(\ref{sec2-ym-lag-re}) includes the inverse metric $g^{\mu\nu}$, the Lagrangian~(\ref{sec2-ym-lag-re}) is an infinite series under the expansion of small $h_{\mu\nu}$. However, because of the higher derivative terms, we show that the quantum theory is renormalizable by power counting. This situation is similar to the higher derivative gravity of Stelle~\cite{Stelle:1976gc}, where the higher derivative terms are the main cause for the renormalizability.

In this subsection, we have computed the effective potential at one-loop. One question is that on what kind of condition this one-loop result can be trusted. In quantum Yang-Mills theory, the coupling $\kappa$ is the unique parameter, and the perturbative computations at lower loops are viable when $\kappa$ is small. However, a different point appears in the above reformulation of Yang-Mills theory. The difference is that all interactive vertices of the reformulated theory actually are suppressed by the dimension-2 condensate $v$. By the redefinition $\tilde{h}_{\mu\nu}=\frac{1}{\kappa\sqrt{v}}{h}_{\mu\nu}$,  $\tilde{h}_{\mu\nu}$ has the canonical dimension in 3D. The interactive vertex of the lowest dimension has the formulation $\frac{\kappa}{\sqrt{v}}\tilde{h}\partial\tilde{h}\partial\tilde{h}$, which is contributed by Eq.~(\ref{sec2-ym-lag-re-2}). The vertices of higher dimensions are suppressed by higher powers of $v$. Therefore, in the reformulated theory, the higher loop computations shall be suppressed by $v$ generally. The one-loop result in the above remains as a good approximation if $v$ is large. If we use the result of Eq.~(\ref{sec2-ym-lag-3d-eff-dg-min}) as an approximation of $v$, then the expansion around the large $v$ corresponds to the large $\kappa$. Hence the higher loop expansions basically correspond to the expansions of strong coupling.

\subsection{Gauge Invariant Gluon Polarization}\label{sec:2.3}

The gauge invariant description~\cite{Ji:1996ek} of gluon angular momentum is an important ingredient to address the nucleon spin problem. For two recent reviews on this problem, see~\cite{Leader:2013jra,Wakamatsu:2014zza}. In~\cite{Chen:2008ag}, a gauge invariant expression is provided through decomposing the gauge field into its physical parts and its pure gauge part
\begin{eqnarray}
\label{sec2.3-gp-def-phys}
A^{a\mu}=A^{a\mu}_{\mathrm{phys}}+A^{a\mu}_{\mathrm{pure}},
\end{eqnarray}
in which $A^{a\mu}_{\mathrm{phys}}$ transforms as Eq.~(\ref{sec2-gf-dec-gt-e}), and $A^{a\mu}_{\mathrm{pure}}$ transforms as Eq.~(\ref{sec2-gf-dec-gt-b}). The descriptions of gluon polarization and orbital angular momentum are given by~\cite{Wakamatsu:2010cb,Zhang:2011rn,Guo:2012wv}
\begin{eqnarray}
\label{sec2.3-gp-def-gs}
\kappa^2M_{gs}^{\mu\alpha\beta}&=&-F^{a\mu\alpha}A^{a\beta}_{\mathrm{phys}}+F^{a\mu\beta}A^{a\alpha}_{\mathrm{phys}},\\
\label{sec2.3-gp-def-go}
\kappa^2M_{go}^{\mu\alpha\beta}&=&(x^{\beta}\mathscr{T}^{\mu\alpha}-x^{\alpha}\mathscr{T}^{\mu\beta})\\
&+&x^{\beta}\eta_{\rho\sigma}F^{a\mu\rho}
D^{\sigma}A^{a\alpha}_{\mathrm{phys}}-x^{\alpha}\eta_{\rho\sigma}F^{a\mu\rho}
D^{\sigma}A^{a\beta}_{\mathrm{phys}},\nonumber
\end{eqnarray}
where
\begin{eqnarray}
\label{sec2.3-gp-def-go-cd}
D^{\sigma}A^{a\alpha}_{\mathrm{phys}}=\partial^{\sigma}A^{a\alpha}_{\mathrm{phys}}
+\epsilon^{abc}A^{b\sigma}A^{c\alpha}_{\mathrm{phys}}
\end{eqnarray}
is the gauge covariant derivative, and $\mathscr{T}^{\mu\nu}$ is proportional to the energy-momentum tensor
\begin{eqnarray}
\label{sec2.3-gp-def-gs-em}
\mathscr{T}^{\mu\nu}=\eta_{\rho\sigma}F^{\mu\rho}_{a}F^{a\sigma\nu}+\kappa^2\eta_{\mu\nu}\mathscr{L}.
\end{eqnarray}
From Eq.~(\ref{sec2-gf-dec}), we know that $A^{a}_{\mu}$ has been decomposed into two parts $B^{a}_{\mu}$ and $e^{a}_{\mu}$. $B^{a}_{\mu}$ has the same transform as $A^{a\mu}_{\mathrm{pure}}$, though $B^{a}_{\mu}$ is not a pure gauge field. Now it is appropriate to make the designation
\begin{eqnarray}
\label{sec2.3-gp-phys}
A^{a\mu}_{\mathrm{phys}}=\eta^{\mu\sigma}e^{a}_{\sigma},
\end{eqnarray}
as they have the same transformation. Through this designation, $M_{gs}^{\mu\alpha\beta}$ can be expressed completely with the tensor fields as
 \begin{eqnarray}
\label{sec2.3-gp-def-gs-t}
\kappa^2M_{gs}^{\mu\alpha\beta}=-\eta^{\mu\tau}
(\eta^{\alpha\theta}\eta^{\beta\gamma}-\eta^{\beta\theta}\eta^{\alpha\gamma})\sqrt{\lvert{g}\rvert}
(\frac{1}{2}R^{\sigma}_{\hspace{1mm}\rho\tau\theta}
g^{\rho\lambda}\epsilon_{\gamma\lambda\sigma}+\epsilon_{\gamma\tau\theta}),
\end{eqnarray}
where $\lvert{g}\rvert$ is the absolute value of the determinant of $g_{\mu\nu}$. In the low momentum limit, the derivative term is relatively small, then $M_{gs}^{\mu\alpha\beta}$ shall be dominated by
 \begin{eqnarray}
\label{sec2.3-gp-def-gs-t-l}
\kappa^2M_{gs}^{\mu\alpha\beta}\approx -2\sqrt{\lvert{g}\rvert}\epsilon^{\mu\alpha\beta}.
\end{eqnarray}
Similarly, $M_{go}^{\mu\alpha\beta}$ can also be expressed by space-time tensor fields and has the decomposition
 \begin{eqnarray}
\label{sec2.3-gp-def-g0-t}
\kappa^2M_{go}^{\mu\alpha\beta}=M_{gs(0)}^{\mu\alpha\beta}+M_{go(1)}^{\mu\alpha\beta}+M_{go(2)}^{\mu\alpha\beta}
+M_{go(3)}^{\mu\alpha\beta}+\frac{1}{\ell^4}M_{go(4)}^{\mu\alpha\beta}.
\end{eqnarray}
Here $M_{go(0)}^{\mu\alpha\beta}$, $M_{go(2)}^{\mu\alpha\beta}$ and $M_{go(4)}^{\mu\alpha\beta}$ are
\begin{eqnarray}
\label{sec2.3-gp-def-go-t-0}
M_{go(0)}^{\mu\alpha\beta}&=&x^{\beta}\mathscr{T}_{(0)}^{\mu\alpha}-x^{\alpha}\mathscr{T}_{(0)}^{\mu\beta}\\
&+&x^{\beta}\eta^{\mu\rho}\eta^{\alpha\sigma}\eta^{\tau\theta}(g_{\rho\theta}g_{\tau\sigma}-g_{\rho\sigma}g_{\tau\theta})
-(\alpha\leftrightarrow\beta ),\\
\label{sec2.3-gp-def-go-t-2}
M_{go(2)}^{\mu\alpha\beta}&=&x^{\beta}\mathscr{T}_{(2)}^{\mu\alpha}-x^{\alpha}\mathscr{T}_{(2)}^{\mu\beta}\\
&+&\frac{1}{2}x^{\beta}\eta^{\mu\theta}\eta^{\alpha\gamma}\eta^{\tau\sigma}R^{\lambda}_{\hspace{1mm}\rho\theta\tau}
(\delta^{\rho}_{\sigma}g_{\lambda\gamma}-\delta^{\rho}_{\gamma}g_{\lambda\sigma})-(\alpha\leftrightarrow\beta ),\nonumber\\
\label{sec2.3-gp-def-go-t-4}
M_{go(4)}^{\mu\alpha\beta}&=&x^{\beta}\mathscr{T}_{(4)}^{\mu\alpha}-x^{\alpha}\mathscr{T}_{(4)}^{\mu\beta}.\nonumber
\end{eqnarray}
In the above, $\mathscr{T}_{(0)}^{\mu\nu}$, $\mathscr{T}_{(2)}^{\mu\nu}$ and $\mathscr{T}_{(4)}^{\mu\nu}$ are the components of $\mathscr{T}^{\mu\nu}$
 \begin{eqnarray}
\label{sec2.3-gp-def-em-t}
\mathscr{T}^{\mu\nu}&=&\mathscr{T}_{(0)}^{\mu\nu}
+\mathscr{T}_{(2)}^{\mu\nu}+\mathscr{T}_{(4)}^{\mu\nu},
\end{eqnarray}
and they have the expressions
\begin{eqnarray}
\label{sec2.3-gp-def-em-t-0}
\mathscr{T}_{(0)}^{\mu\nu}&=&\eta^{\mu\theta}\eta^{\nu\lambda}\eta^{\tau\gamma}
(g_{\theta\gamma}g_{\tau\lambda}-g_{\theta\lambda}g_{\tau\gamma})-\frac{1}{4}\eta^{\mu\nu}\mathscr{L}^{(0)},\\
\label{sec2.3-gp-def-em-t-2}
\mathscr{T}_{(2)}^{\mu\nu}&=&\eta^{\mu\theta}\eta^{\nu\tau}\eta^{\gamma\lambda}
(\delta^{\rho}_{\theta}g_{\sigma\gamma}-\delta^{\rho}_{\gamma}g_{\sigma\theta}){R}^{\sigma}_{\hspace{1mm}\rho\lambda\tau}
-\frac{1}{4}\eta^{\mu\nu}\mathscr{L}^{(2)},\\
\label{sec2.3-gp-def-em-t-4}
\mathscr{T}_{(4)}^{\mu\nu}&=&\frac{1}{4}
\eta^{\mu\alpha}\eta^{\nu\beta}\eta^{\gamma\theta}(g^{\rho\lambda}g_{\sigma\tau}-\delta^{\rho}_{\tau}\delta^{\lambda}_{\sigma})
{R}^{\sigma}_{\hspace{1mm}\rho\alpha\gamma}{R}^{\tau}_{\hspace{1mm}\lambda\beta\theta}
-\frac{1}{4}\eta^{\mu\nu}\mathscr{L}^{(4)}.
\end{eqnarray}
The parts $M_{go(1)}^{\mu\alpha\beta}$ and $M_{go(3)}^{\mu\alpha\beta}$ are
\begin{eqnarray}
\label{sec2.3-gp-def-go-t-1}
M_{go(1)}^{\mu\alpha\beta}&=&\sqrt{\lvert{g}\rvert}x^{\beta}\eta^{\mu\theta}\eta^{\alpha\tau}\eta^{\lambda\gamma}
\Gamma^{\sigma}_{\theta\lambda}\epsilon_{\sigma\gamma\tau}-(\alpha\leftrightarrow\beta),\\
\label{sec2.3-gp-def-go-t-3}
M_{go(3)}^{\mu\alpha\beta}&=&\frac{1}{2}x^{\beta}\eta^{\mu\gamma}\eta^{\alpha\lambda}
\sqrt{\lvert{g}\rvert}{R}^{\sigma}_{\hspace{1mm}\rho\gamma\theta}\eta^{\theta\tau}g^{\rho\nu}
\Gamma^{\delta}_{\tau\lambda}\epsilon_{\delta\nu\sigma}-(\alpha\leftrightarrow\beta).
\end{eqnarray}
In the low momentum limit, $M_{go}^{\mu\alpha\beta}$ shall be mainly dominated by $M_{go(0)}^{\mu\alpha\beta}$.

In the above, the gauge invariant descriptions of gluon angular momentum have been expressed by the geometrical objects.
$M_{gs}^{\mu\alpha\beta}$ in Eq.~(\ref{sec2.3-gp-def-gs-t}) and $M_{go(2)}^{\mu\alpha\beta}$ in Eq.~(\ref{sec2.3-gp-def-go-t-2}) both include the contribution from the Riemann curvature, though they have the different tensor structure. $M_{go(1)}^{\mu\alpha\beta}$ and $M_{go(3)}^{\mu\alpha\beta}$ in Eqs.~(\ref{sec2.3-gp-def-go-t-1}) and (\ref{sec2.3-gp-def-go-t-3}) also includes the contribution from the Levi-Civita connection. We shall give more discussions on the gluon angular momentum in subsection~\ref{sec:3.3}.

In the above, we have used the condition that $A^{a}_{\mu}$ has the decomposition~(\ref{sec2-gf-dec}), in which $B^{a}_{\mu}$ is determined by the torsion-free condition~(\ref{sec2-gf-res-H-g-tor}). A question is if $e^{a}_{\mu}$ exists for a given $A^{a}_{\mu}$. In subsection~\ref{sec:2.1}, we solve $B^{a}_{\nu}$ using the torsion-free equation
\begin{eqnarray}
\label{sec23-b-e-tor-eq}
(\partial_{\mu}B^{a}_{\nu}+\epsilon^{abc}B^{b}_{\mu}e^{c}_{\nu})-(\mu\leftrightarrow\nu)=0,
\end{eqnarray}
which is equivalent to
\begin{eqnarray}
\label{sec23-a-e-tor-eq}
(\partial_{\mu}A^{a}_{\nu}+\epsilon^{abc}A^{b}_{\mu}e^{c}_{\nu})-(\mu\leftrightarrow\nu)=2\epsilon^{abc}e^{b}_{\mu}e^{c}_{\nu}.
\end{eqnarray}
 In order to obtain $e^{a}_{\mu}$ in terms of $A^{a}_{\mu}$, we need to solve Eq.~(\ref{sec23-a-e-tor-eq}).  Eq.~(\ref{sec23-a-e-tor-eq}) is a partial differential equation about $e^{a}_{\mu}$, which is not viable to derive explicit solution generally. Here we provide a formal solution based on the Taylor expansion. Supposing that $A^{a}_{\mu}(x)$ have the expansion around $x^{\mu}_0$
\begin{eqnarray}
\label{sec2-a-exp}
A^{a}_{\mu}(x)=A^{a}_{\mu}(x_0)+(x-x_0)^{\alpha}A^{a}_{\mu,\alpha}(x_0)
+\frac{1}{2}(x-x_0)^{\alpha}(x-x_0)^{\beta}A^{a}_{\mu,\alpha\beta}(x_0)+\cdots.
\end{eqnarray}
We can solve $e^{a}_{\mu}(x)$ using the ansatz
\begin{eqnarray}
\label{sec2-e-exp}
e^{a}_{\mu}(x)=e^{a}_{\mu}(x_0)+(x-x_0)^{\alpha}e^{a}_{\mu,\alpha}(x_0)
+\frac{1}{2}(x-x_0)^{\alpha}(x-x_0)^{\beta}e^{a}_{\mu,\alpha\beta}(x_0)+\cdots.
\end{eqnarray}
From the zero order of Eq.~(\ref{sec23-a-e-tor-eq}), we obtain
\begin{eqnarray}
\label{sec2-e-eq-sol}
e^{a}_{[\mu,\nu]}(x_0)
=-\frac{1}{2}\epsilon^{abc}(A^{b}_{\mu}e^{c}_{\nu}-A^{b}_{\nu}e^{c}_{\mu})(x_0)+\epsilon^{abc}e^{b}_{\mu}(x_0)e^{c}_{\nu}(x_0),
\end{eqnarray}
where $e^{a}_{[\mu,\nu]}=\frac{1}{2}(e^{a}_{\mu,\nu}-e^{a}_{\nu,\mu})$. The antisymmetrical part of $e^{a}_{\mu,\nu}(x_0)$ can be determined by $A^{a}_{\mu}(x_0)$ and $e^{a}_{\mu}(x_0)$. From the first order of Eq.~(\ref{sec23-a-e-tor-eq}), we obtain
\begin{eqnarray}
\label{sec2-e-eq-sol-1}
e^{a}_{[\mu,\nu]\alpha}(x_0)
=-\frac{1}{2}\epsilon^{abc}(A^{b}_{\mu,\alpha}e^{c}_{\nu}+A^{b}_{\mu}e^{c}_{\nu,\alpha})(x_0)
+\epsilon^{abc}(e^{b}_{\mu,\alpha}e^{c}_{\nu}+e^{b}_{\mu}e^{c}_{\nu,\alpha})(x_0)-(\mu\leftrightarrow\nu).
\end{eqnarray}
This equation expresses $e^{a}_{[\mu,\nu]\alpha}(x_0)$ with $A^{a}_{\mu,\nu}(x_0)$, $e^{a}_{\mu,\nu}(x_0)$ and $e^{a}_{\mu}(x_0)$. This procedure can be performed repeatedly, and the coefficients of  $e^{a}_{\mu}(x)$ can be determined in order. Therefore, for a given $A^{a}_{\mu}(x)$, $e^{a}_{\mu}(x)$ can be found by the above Taylor expansion, but $e^{a}_{\mu}(x)$ is not uniquely determined.

\section{\texorpdfstring{$SU(2)$}{SU(2)} Gauge Theory in Four Dimensions}\label{sec:3}

\subsection{Formulation with space-time tensor fields}\label{sec:3.1}

After the discussion on 3 dimensions, we begin to discuss the $SU(2)$ Yang-Mills theory in 4 dimensions in this section. Although the $SU(2)$ Yang-Mills theories in 3D and 4D have the same Lie Algebra, the 4D case is more complicated than the 3D case because of the number of space-time dimension. In 4 dimensional space-time, the Lagrangian of $SU(2)$ Yang-Mills theory is
\begin{eqnarray}
\label{sec3-ym-lag}
\mathscr{L}&=&-\frac{1}{4\kappa^2}\eta^{\alpha\mu}\eta^{\beta\nu}F^{a}_{\alpha\beta}F^{a}_{\mu\nu},\\
F^{a}_{\mu\nu}&=&\partial_{\mu}A^{a}_{\nu}-\partial_{\nu}A^{a}_{\mu}+\epsilon^{abc}A^{b}_{\mu}A^{c}_{\nu},\nonumber
\end{eqnarray}
where $\eta^{\mu\nu}=\mathrm{diag}(1,-1,-1,-1)$ is the Lorentz metric in 4D. As in section~\ref{sec:2}, we decompose the gauge field as
\begin{eqnarray}
\label{sec3-gf-dec}
A^{a}_{\mu}=B^{a}_{\mu}+e^{a}_{\mu}.
\end{eqnarray}
Their transformations are similar to that in Eqs.~(\ref{sec2-gf-dec-gt-e}) and (\ref{sec2-gf-dec-gt-b}). Accordingly the field strength has the decomposition
\begin{eqnarray}
\label{sec3-gf-dec-fstr-0}
F^{a}_{\mu\nu}&=&\mathcal{B}^{a}_{\mu\nu}+\mathcal{T}^{a}_{\mu\nu}
+\epsilon^{abc}e^{b}_{\mu}e^{c}_{\nu},\\
\label{sec3-gf-dec-fstr-1}
\mathcal{B}^{a}_{\mu\nu}&=&\partial_{\mu}B^{a}_{\nu}-\partial_{\nu}B^{a}_{\mu}+\epsilon^{abc}B^{b}_{\mu}B^{c}_{\nu},\\
\label{sec3-gf-dec-fstr-2}
\mathcal{T}^{a}_{\mu\nu}&=&(\partial_{\mu}e^{a}_{\nu}+\epsilon^{abc}B^{b}_{\mu}e^{c}_{\nu})-(\partial_{\nu}e^{a}_{\mu}
+\varepsilon^{abc}B^{b}_{\nu}e^{c}_{\mu}).
\end{eqnarray}
We also require that $B^{a}_{\mu}$ and $e^{a}_{\mu}$ satisfy
\begin{eqnarray}
\label{sec3-gf-dec-res}
\partial_{\mu}e^{a}_{\nu}+\epsilon^{abc}B^{b}_{\mu}e^{c}_{\nu}=\Gamma^{\rho}_{\mu\nu}e^{a}_{\rho} .
\end{eqnarray}
Before solving the above equation, we provide some additional information about  $e^{a}_{\alpha}$. For $SU(2)$ gauge theory in 4D, $e^{a}_{\alpha}$ can be regarded as a $3\times{4}$ matrix, which does not have an inverse in the conventional meaning. However, we can define a right-inverse matrix of $e^{a}_{\alpha}$ as
\begin{eqnarray}
\label{sec3-gf-res-e-inv}
e^{a}_{\alpha}E^{\alpha}_{b}=\delta^{a}_{b}.
\end{eqnarray}
$E^{\alpha}_{a}$ can be regarded as a $4\times{3}$ matrix, and Eq.~(\ref{sec3-gf-res-e-inv}) imposes $3\times{3}$ constraints, so $E^{\alpha}_{a}$ exists but are not determined uniquely. Notice that $E^{\alpha}_{b}$ is only a right-inverse matrix of $e^{a}_{\alpha}$. The quantity
\begin{eqnarray}
\label{sec3-gf-res-e-inv-n}
n_{\alpha}^{\beta}=e^{a}_{\alpha}E^{\beta}_{a}
\end{eqnarray}
is not equal to the identity matrix but is an independent tensor. Now we define the metric-like fields
\begin{eqnarray}
\label{sec3-gf-dec-def}
g_{\alpha\beta}=e^{a}_{\alpha}e^{a}_{\beta},~~g^{\alpha\beta}=E^{\alpha}_{a}E^{\alpha}_{a}.
\end{eqnarray}
They are the gauge invariant space-time tensors. From Eq.~(\ref{sec3-gf-dec-def}), we know that $g_{\alpha\beta}$ is the product of two $4\times{3}$ matrices. So $g_{\alpha\beta}$ is not an invertible matrix, and $g^{\alpha\beta}$ is also not invertible. From Eq.~(\ref{sec3-gf-res-e-inv}), we obtain the relation between $g_{\alpha\beta}$ and $g^{\alpha\beta}$
\begin{eqnarray}
\label{sec3-gf-res-e-inv-rel-1}
g_{\alpha\rho}g^{\rho\sigma}g_{\sigma\beta}=g_{\alpha\beta},~~
g^{\alpha\rho}g_{\rho\sigma}g^{\sigma\beta}=g^{\alpha\beta},~~
g_{\alpha\rho}g^{\rho\beta}=n_{\alpha}^{\beta}.
\end{eqnarray}
The relation~(\ref{sec3-gf-res-e-inv-rel-1}) can be identified with the definition of the Moore-Penrose generalized inverse matrix. So $g^{\alpha\beta}$ can be regarded as a generalized inverse matrix of $g_{\alpha\beta}$. From Eq.~(\ref{sec3-gf-res-e-inv}), we also have the identities
\begin{eqnarray}
\label{sec3-gf-res-e-inv-rel-2}
g_{\alpha\rho}n_{\beta}^{\rho}=g_{\alpha\beta},~~g^{\alpha\rho}n_{\rho}^{\beta}=g^{\alpha\beta},
~~n^{\rho}_{\alpha}n_{\rho}^{\beta}=n_{\alpha}^{\beta}.
\end{eqnarray}
So $n_{\alpha}^{\beta}$ behaves like a projection tensor, and more discussions about this projection tensor are given in appendix~\ref{append-0}. From $e_{\alpha}^{a}$ and $E^{\alpha}_{a}$, we can define the gauge invariant variables
\begin{eqnarray}
\label{sec3-gf-dec-def-asym}
\mathcal{E}_{\alpha\beta\gamma}=\epsilon_{abc}e_{\alpha}^{a}e_{\beta}^{b}e^{\gamma}_{c},~~
\mathcal{E}^{\alpha\beta\gamma}=\epsilon^{abc}E^{\alpha}_{a}E^{\beta}_{b}E^{\gamma}_{c}.
\end{eqnarray}
They are totally antisymmetrical space-time tensors. They satisfy the identity
\begin{eqnarray}
\label{sec3-gf-dec-def-asym-id-n}
\mathcal{E}^{\alpha\beta\rho}\mathcal{E}_{\mu\nu\sigma}&=&
n^{\alpha}_{\mu}n^{\beta}_{\nu}n^{\rho}_{\sigma}-n^{\alpha}_{\mu}n^{\beta}_{\sigma}n^{\rho}_{\nu}
+n^{\alpha}_{\sigma}n^{\beta}_{\mu}n^{\rho}_{\nu}\nonumber\\
&-&n^{\alpha}_{\sigma}n^{\beta}_{\nu}n^{\rho}_{\mu}
+n^{\alpha}_{\nu}n^{\beta}_{\sigma}n^{\rho}_{\mu}-n^{\alpha}_{\nu}n^{\beta}_{\mu}n^{\rho}_{\sigma},
\end{eqnarray}
or in an equivalent formulation
\begin{eqnarray}
\label{sec3-gf-dec-def-asym-id-n-g}
\mathcal{E}_{\alpha\beta\rho}\mathcal{E}_{\mu\nu\sigma}&=&
g_{\alpha\mu}g_{\beta\nu}g_{\rho\sigma}-g_{\alpha\mu}g_{\beta\sigma}g_{\rho\nu}
+g_{\alpha\sigma}g_{\beta\mu}g_{\rho\nu}\nonumber\\
&-&g_{\alpha\sigma}g_{\beta\nu}g_{\rho\mu}
+g_{\alpha\nu}g_{\beta\sigma}g_{\rho\mu}-g_{\alpha\nu}g_{\beta\mu}g_{\rho\sigma}.
\end{eqnarray}
There is also a similar identity for $g^{\alpha\beta}$. From Eq.~(\ref{sec3-gf-res-e-inv}), we have the projection relations similar to Eq.~(\ref{sec3-gf-res-e-inv-rel-2})
\begin{eqnarray}
\label{sec3-gf-dec-def-asym-pr}
\mathcal{E}_{\alpha\beta\rho}n^{\rho}_{\gamma}=\mathcal{E}_{\alpha\beta\gamma},~~
\mathcal{E}^{\alpha\beta\rho}n_{\rho}^{\gamma}=\mathcal{E}^{\alpha\beta\gamma}.
\end{eqnarray}
$\mathcal{E}_{\alpha\beta\gamma}$ and $\mathcal{E}^{\alpha\beta\gamma}$ are totally antisymmetrical tensors in 4D, so they have 4 independent components respectively. We can define
\begin{eqnarray}
\label{sec3-gf-res-e-inv-n-v}
V_{\mu}=\frac{1}{6}\epsilon_{\mu\alpha\beta\gamma}\mathcal{E}^{\alpha\beta\gamma},~~
U^{\mu}=\frac{1}{6}\epsilon^{\mu\alpha\beta\gamma}\mathcal{E}_{\alpha\beta\gamma},
\end{eqnarray}
or equivalently
\begin{eqnarray}
\label{sec3-gf-res-e-inv-n-v-eq}
\mathcal{E}_{\alpha\beta\gamma}=-U^{\mu}\epsilon_{\mu\alpha\beta\gamma},~~
\mathcal{E}^{\alpha\beta\gamma}=-V_{\mu}\epsilon^{\mu\alpha\beta\gamma},
\end{eqnarray}
where $\epsilon^{\mu\alpha\beta\gamma}$ is totally antisymmetric with the convention $\epsilon^{0123}=1$, and $\epsilon_{\mu\alpha\beta\gamma}$ is obtained from $\epsilon^{\mu\alpha\beta\gamma}$ by raising down the indices with $\eta_{\mu\nu}$
\begin{eqnarray}
\label{sec31-anti}
\epsilon_{\mu\alpha\beta\gamma}=\eta_{\mu\nu}\eta_{\alpha\rho}\eta_{\beta\sigma}\eta_{\gamma\tau}
\epsilon^{\nu\rho\sigma\tau}.
\end{eqnarray}
We have the identities for the vectors $U^{\mu}$ and $V_{\mu}$
\begin{eqnarray}
\label{sec3-gf-res-e-inv-n-v-iden}
g_{\alpha\beta}U^{\beta}=0,~~g^{\alpha\beta}V_{\beta}=0.
\end{eqnarray}
Eq.~(\ref{sec3-gf-res-e-inv-n-v-iden}) shows $U^{\mu}$ is an eigenvector of $g_{\alpha\beta}$ with the eigenvalue $0$, which is an equal statement that $g_{\alpha\beta}$ is not invertible. From Eq.~(\ref{sec3-gf-dec-def-asym-id-n-g}), we have
\begin{eqnarray}
\label{sec3-gf-res-e-iden-uu}
U^{\alpha}U^{\beta}&=&
\frac{1}{6}\varepsilon^{\alpha\rho\mu\tau}\varepsilon^{\beta\sigma\nu\theta}g_{\rho\sigma}g_{\mu\nu}g_{\tau\theta},\\
\label{sec3-gf-res-e-iden-vv}
V_{\alpha}V_{\beta}&=&
\frac{1}{6}\varepsilon_{\alpha\rho\mu\tau}\varepsilon_{\beta\sigma\nu\theta}g^{\rho\sigma}g^{\mu\nu}g^{\tau\theta},
\end{eqnarray}
which provide another relations between $U^{\mu}$ and $g_{\alpha\beta}$. Using Eqs.~(\ref{sec3-gf-res-e-inv}), (\ref{sec3-gf-res-e-inv-n}) and~(\ref{sec3-gf-res-e-inv-n-v-eq}), we have the identities
\begin{eqnarray}
\label{sec3-gf-res-e-inv-n-v-u}
n_{\beta}^{\alpha}&=&\delta_{\beta}^{\alpha}+U^{\alpha}V_{\beta},\\
\label{sec3-gf-res-e-inv-n-v-n}
n_{\rho}^{\rho}&=&3,~~U^{\rho}V_{\rho}=-1.
\end{eqnarray}

With the above identities, now we can start to solve $B^{a}_{\mu}$. From Eq.~(\ref{sec3-gf-dec-res}), $B^{a}_{\mu}$ is solved as
\begin{eqnarray}
\label{sec3-gf-dec-res-sol}
B^{a}_{\mu}=-\frac{1}{2}\epsilon^{abc}E^{\rho}_{b}(\partial_{\mu}e_{\rho}^{c}-\Gamma^{\sigma}_{\mu\rho}e^{c}_{\sigma}),
\end{eqnarray}
and Eq.~(\ref{sec3-gf-dec-res}) also yields the condition
\begin{eqnarray}
\label{sec3-gf-res-g}
\partial_{\mu}g_{\alpha\beta}=\Gamma^{\rho}_{\mu\alpha}g_{\rho\beta}+\Gamma^{\rho}_{\mu\beta}g_{\rho\alpha}.
\end{eqnarray}
Substituting Eq.~(\ref{sec3-gf-dec-res-sol}) into Eq.~(\ref{sec3-gf-dec-res}), we obtain
\begin{eqnarray}
\label{sec3-gf-dec-res-con}
\partial_{\mu}e^{a}_{\nu}+\epsilon^{abc}B^{b}_{\mu}e^{c}_{\nu}=
\frac{1}{2}e^{a}_{\tau}(\partial_{\mu}n^{\tau}_{\nu}+\Gamma^{\tau}_{\mu\rho}n^{\rho}_{\nu})
+\frac{1}{2}E_{a}^{\tau}(\partial_{\mu}g_{\tau\nu}-\Gamma^{\rho}_{\mu\tau}g_{\rho\nu}) .
\end{eqnarray}
If $n^{\tau}_{\nu}$ is the identity matrix, then Eq.~(\ref{sec3-gf-res-g}) would imply that Eq.~(\ref{sec3-gf-dec-res-con}) is consistent with Eq.~(\ref{sec3-gf-dec-res}) automatically. However, here $n^{\tau}_{\nu}$ is an independent tensor, so the consistency among Eqs.~(\ref{sec3-gf-dec-res}), (\ref{sec3-gf-res-g}) and (\ref{sec3-gf-dec-res-con}) requires the condition
\begin{eqnarray}
\label{sec3-gf-dec-res-con-h}
n^{\sigma}_{\tau}(\partial_{\mu}n^{\tau}_{\nu}+\Gamma^{\tau}_{\mu\rho}n^{\rho}_{\nu}-\Gamma^{\rho}_{\mu\nu}n^{\tau}_{\rho})=0.
\end{eqnarray}
Using Eqs.~(\ref{sec3-gf-res-e-inv-rel-1}), (\ref{sec3-gf-res-e-inv-rel-2}) and~(\ref{sec3-gf-res-g}), we have
\begin{eqnarray}
\label{sec3-gf-dec-res-con-h-inv-1}
\partial_{\mu}g^{\alpha\beta}&=&\partial_{\mu}(g^{\alpha\rho}g_{\rho\sigma}g^{\sigma\beta})
=g^{\alpha\rho}g^{\sigma\beta}\partial_{\mu}g_{\rho\sigma}+n^{\alpha}_{\sigma}\partial_{\mu}g^{\sigma\beta}
+n^{\beta}_{\sigma}\partial_{\mu}g^{\sigma\alpha}\nonumber\\
&=&g^{\alpha\rho}g^{\sigma\beta}\partial_{\mu}g_{\rho\sigma}+2\partial_{\mu}g^{\alpha\beta}
-g^{\sigma\beta}\partial_{\mu}n^{\alpha}_{\sigma}-g^{\sigma\alpha}\partial_{\mu}n^{\beta}_{\sigma},
\end{eqnarray}
which is equivalent to
\begin{eqnarray}
\label{sec3-gf-dec-res-con-h-inv}
\partial_{\mu}g^{\alpha\beta}&=&g^{\sigma\beta}\partial_{\mu}n^{\alpha}_{\sigma}+g^{\sigma\alpha}\partial_{\mu}n^{\beta}_{\sigma}
-g^{\alpha\rho}g^{\sigma\beta}\partial_{\mu}g_{\rho\sigma}\nonumber\\
&=&g^{\alpha\rho}(\partial_{\mu}n^{\beta}_{\rho}-\Gamma^{\tau}_{\mu\rho}n^{\beta}_{\tau})
+g^{\beta\rho}(\partial_{\mu}n^{\alpha}_{\rho}-\Gamma^{\tau}_{\mu\rho}n^{\alpha}_{\tau}).
\end{eqnarray}
This is a equation about the generalized inverse metric $g^{\alpha\beta}$. We see that because $g_{\alpha\beta}$ is not invertible, the connection $\Gamma^{\tau}_{\mu\rho}$ is not compatible with $g^{\alpha\beta}$ any more\footnote{However, the connection, which is compatible with the metric and the generalized inverse metric, exists and we give the derivation in appendix~\ref{append-1}.}. From Eqs.~(\ref{sec3-gf-res-g}) and (\ref{sec3-gf-dec-res-con-h-inv}), we further have
\begin{eqnarray}
\label{sec3-gf-dec-res-con-h-inv-h}
\partial_{\mu}n^{\alpha}_{\beta}&=&
g^{\alpha\sigma}\partial_{\mu}g_{\sigma\beta}+g_{\sigma\beta}\partial_{\mu}g^{\alpha\sigma}\nonumber\\
&=&\Gamma^{\tau}_{\mu\beta}n^{\alpha}_{\tau}
+n^{\sigma}_{\beta}(\partial_{\mu}n^{\alpha}_{\sigma}-\Gamma^{\tau}_{\mu\sigma}n^{\alpha}_{\tau}),
\end{eqnarray}
which is equivalent to Eq.~(\ref{sec3-gf-dec-res-con-h}). So Eq.~(\ref{sec3-gf-dec-res-con-h}) is not an independent equation to be satisfied, but it can de derived from Eqs.~(\ref{sec3-gf-res-g}). Therefore, to find the connection, we only need to solve Eq.~(\ref{sec3-gf-res-g}). A connection which satisfies Eq.~(\ref{sec3-gf-res-g}) is given as
\begin{eqnarray}
\label{sec3-gf-res-g-sol-0}
\Gamma^{\rho}_{\alpha\beta}=g^{\rho\sigma}\Gamma_{\sigma,\alpha\beta}
-g^{\rho\sigma}(\Sigma_{\alpha\sigma}^{\tau}g_{\tau\beta}+\Sigma_{\beta\sigma}^{\tau}g_{\tau\alpha})
+\Sigma_{\alpha\beta}^{\rho},
\end{eqnarray}
in which
\begin{eqnarray}
\label{sec3-gf-res-g-sol-1}
\Sigma^{\rho}_{\alpha\beta}=\frac{1}{2}(\delta^{\tau}_{\beta}-n^{\tau}_{\beta})g^{\rho\sigma}\Gamma_{\tau,\alpha\sigma}
-\frac{1}{2}(\delta^{\tau}_{\alpha}-n^{\tau}_{\alpha})g^{\rho\sigma}\Gamma_{\tau,\beta\sigma},
\end{eqnarray}
and
\begin{eqnarray}
\label{sec3-gf-res-g-sol-2}
\Gamma_{\sigma,\alpha\beta}=\frac{1}{2}(\partial_{\alpha}g_{\beta\sigma}+\partial_{\beta}g_{\alpha\sigma}-\partial_{\sigma}g_{\alpha\beta})
\end{eqnarray}
is the Christoffel symbols of the first kind. The detail to derive the above solution is given in appendix~\ref{append-1}. The above connection~(\ref{sec3-gf-res-g-sol-0}) is much more complicated than the Levi-Civita connection, which is mainly because $g_{\alpha\beta}$ is not invertible. If we suppose $g_{\alpha\beta}$ is invertible, then $n^{\beta}_{\alpha}=\delta^{\beta}_{\alpha}$ is the identity matrix. So the corrections from $n^{\beta}_{\alpha}$ vanish, and the connection~(\ref{sec3-gf-res-g-sol-0}) is reduced to the Levi-Civita connection.

Using the above results, we can obtain
\begin{eqnarray}
\label{sec3-gf-fsol-1}
\mathcal{B}^{a}_{\mu\nu}=\frac{1}{2}R^{\sigma}_{\hspace{1mm}\rho\mu\nu}e^{c}_{\sigma}E^{\rho}_{b}\varepsilon^{abc},
~~\mathcal{T}^{a}_{\mu\nu}=2\Sigma^{\rho}_{\mu\nu}e^{a}_{\rho},
\end{eqnarray}
where
\begin{eqnarray}
\label{sec3-gf-fsol-2}
R^{\sigma}_{\hspace{1mm}\rho\mu\nu}=\partial_{\mu}\Gamma^{\sigma}_{\nu\rho}-\partial_{\nu}\Gamma^{\sigma}_{\mu\rho}+
\Gamma^{\tau}_{\nu\rho}\Gamma^{\sigma}_{\mu\tau}-\Gamma^{\tau}_{\mu\rho}\Gamma^{\sigma}_{\nu\tau}
\end{eqnarray}
is the Riemann curvature with the connection defined by~(\ref{sec3-gf-res-g-sol-0}). The Lagrangian~(\ref{sec2-ym-lag}) has a natural decomposition according to the power counting of derivative
\begin{eqnarray}
\label{sec3-ym-lag-re}
\mathscr{L}&=&\mathscr{L}^{(0)}+\mathscr{L}^{(1)}+\mathscr{L}^{(2)}
+\mathscr{L}^{(3)}+\mathscr{L}^{(4)}.
\end{eqnarray}
Here the parts $\mathscr{L}^{(0)}$, $\mathscr{L}^{(2)}$ and $\mathscr{L}^{(4)}$ are
\begin{eqnarray}
\label{sec3-ym-lag-re-4}
-4\kappa^2\mathscr{L}^{(4)}&=&\frac{1}{4}\eta^{\mu\alpha}\eta^{\nu\beta}
(g^{\rho\theta}g_{\sigma\tau}-n^{\rho}_{\tau}n^{\theta}_{\sigma})
{R}^{\sigma}_{\hspace{1mm}\rho\mu\nu}{R}^{\tau}_{\hspace{1mm}\theta\alpha\beta},\\
\label{sec3-ym-lag-re-2}
-4\kappa^2\mathscr{L}^{(2)}&=&\eta^{\mu\alpha}\eta^{\nu\beta}
(n^{\rho}_{\alpha}g_{\sigma\beta}-n^{\rho}_{\beta}g_{\sigma\alpha}){R}^{\sigma}_{\hspace{1mm}\rho\mu\nu}
+4\eta^{\mu\alpha}\eta^{\nu\beta}g_{\rho\sigma}\Sigma^{\rho}_{\alpha\beta}\Sigma^{\sigma}_{\mu\nu},\\
\label{sec3-ym-lag-re-0}
-4\kappa^2\mathscr{L}^{(0)}&=&\eta^{\mu\alpha}\eta^{\nu\beta}(g_{\mu\alpha}g_{\nu\beta}-g_{\mu\beta}g_{\nu\alpha}).
\end{eqnarray}
They are similar to the situation in 3D, but with the identity matrix replaced by $n^{\rho}_{\tau}$. In 4D, the torsion tensor is not zero any more, and we also have contribution from the torsion in Eq.~(\ref{sec3-ym-lag-re-2}). $\mathscr{L}^{(1)}$ and $\mathscr{L}^{(3)}$ are relevant to the torsion and the totally antisymmetric tensor
\begin{eqnarray}
\label{sec3-ym-lag-re-3}
-4\kappa^2\mathscr{L}^{(3)}&=&2\eta^{\mu\alpha}\eta^{\nu\beta}
{R}^{\sigma}_{\hspace{1mm}\rho\mu\nu}g^{\rho\tau}\Sigma^{\theta}_{\alpha\beta}\mathcal{E}_{\theta\tau\sigma},\\
\label{sec3-ym-lag-re-1}
-4\kappa^2\mathscr{L}^{(1)}&=&4\eta^{\mu\alpha}\eta^{\nu\beta}\Sigma^{\theta}_{\mu\nu}\mathcal{E}_{\theta\alpha\beta}.
\end{eqnarray}
From Eq.~(\ref{sec3-gf-res-e-inv-n-v-eq}), we know that the totally anti-symmetric tensor $\mathcal{E}_{\theta\alpha\beta}$ is equivalent to the vector $U^{\mu}$. So in the expression of $\mathscr{L}^{(1)}$ and $\mathscr{L}^{(3)}$, $\mathcal{E}_{\theta\alpha\beta}$ can also be replaced by the vector $U^{\mu}$.

\subsection{Relations between Partition Functions}\label{sec:3.2-pf}

In subsection~\ref{sec:2.2-pf}, we have discussed the relations between partition functions in 3D case. In this subsection, we discuss the similar relations in 4D. We consider the partition function
\begin{eqnarray}
\label{sec3-ym-pf}
Z[B^{a}_{\mu},e^{a}_{\mu}]&=&\int [\mathscr{D}B^{a}_{\mu}][\mathscr{D}e^{a}_{\mu}][\mathscr{D}V_{\mu}][\mathscr{D}j^{a}_{\mu}]
[\mathscr{D}N_{\mu}]\\
&\times&\mathrm{exp}\bigl(i\int{d^4x}
(\mathscr{L}+\varepsilon^{\mu\nu\alpha\beta}{j}^{a}_{\mu}V_{\nu}\mathcal{T}^{a}_{\alpha\beta}
+\varepsilon^{\mu\nu\alpha\beta}N_{\mu}{e}^{a}_{\nu}\mathcal{T}^{a}_{\alpha\beta})\bigr).\nonumber
\end{eqnarray}
Here $\mathscr{L}$ is the Yang-Mills Lagrangian~(\ref{sec2-ym-lag}) with the decomposition $A^{a}_{\mu}=B^{a}_{\mu}+ e^{a}_{\mu}$. $\mathcal{T}^{a}_{\alpha\beta}$ is the torsion constraint condition~(\ref{sec3-gf-dec-fstr-2}). Different from Eq.~(\ref{sec2-ym-pf}), the vector $V_{\mu}$ is introduced to the path integral, and we have two constraints associated to ${j}^{a}_{\mu}$ and $N_{\mu}$. The first reason to introduce $V_{\mu}$ is that $\varepsilon^{\mu\nu\alpha\beta}$ has four indices in 4D; So an extra vector is needed to form a scalar expression in Eq.~(\ref{sec3-ym-pf}). The second reason is that $V_{\mu}$ is required in the reformulation of the Yang-Mills Lagrangian, as $V_{\mu}$ is needed in Eqs.~(\ref{sec3-ym-lag-re-3}), (\ref{sec3-ym-lag-re-1}) and~(\ref{sec3-gf-res-e-inv-n-v-u}). Using $V_{\mu}$, $E_{a}^{\mu}$ can be expressed as
\begin{eqnarray}
\label{sec32-e-v}
E_{a}^{\mu}=\frac{1}{2}\varepsilon^{\mu\nu\alpha\beta}\epsilon^{abc}V_{\nu}e_{\alpha}^{b}e_{\beta}^{c}.
\end{eqnarray}
By integrating out ${j}^{a}_{\mu}$ and $N^{a}_{\mu}$, we can obtain
\begin{eqnarray}
\label{sec3-ym-pf-j}
Z[B^{a}_{\mu},e^{a}_{\mu}]&=&\int [\mathscr{D}B^{a}_{\mu}][\mathscr{D}e^{a}_{\mu}][\mathscr{D}V_{\mu}]
\mathrm{exp}\bigl(i\int{d^4x}\mathscr{L}\bigr)\\
&{\times}&\delta\left(\varepsilon^{\mu\nu\alpha\beta}V_{\nu}\mathcal{T}^{a}_{\alpha\beta}\right)
\delta\left(\varepsilon^{\mu\nu\alpha\beta}{e}^{a}_{\nu}\mathcal{T}^{a}_{\alpha\beta}\right).\nonumber
\end{eqnarray}
These integrals yield two constraint conditions in the delta functions. In order to integrate out $B^{a}_{\mu}$, we have to solve the constraints
\begin{eqnarray}
\label{sec3-ym-pf-tor-j}
\varepsilon^{\mu\nu\alpha\beta}V_{\nu}\mathcal{T}^{a}_{\alpha\beta}&=&0,\\
\label{sec3-ym-pf-tor-n}
\varepsilon^{\mu\nu\alpha\beta}{e}^{a}_{\nu}\mathcal{T}^{a}_{\alpha\beta}&=&0.
\end{eqnarray}
Using Eq.~(\ref{sec3-gf-dec-res}), $B^{a}_{\mu}$ is given by
\begin{eqnarray}
\label{sec3-ym-pf-zb-b}
B^{a}_{\mu}(e)=-\frac{1}{2}\epsilon^{abc}E^{\rho}_{b}
(\partial_{\mu}e_{\rho}^{c}-\Gamma^{\sigma}_{\mu\rho}e^{c}_{\sigma}).
\end{eqnarray}
So the constraints~(\ref{sec3-ym-pf-tor-j}) and~(\ref{sec3-ym-pf-tor-n}) are equivalent to
\begin{eqnarray}
\label{sec3-ym-pf-tor-j-re}
\varepsilon^{\mu\nu\alpha\beta}V_{\nu}\mathcal{T}^{a}_{\alpha\beta}
&=&\varepsilon^{\mu\nu\alpha\beta}V_{\nu}\Sigma^{\rho}_{\alpha\beta}e^{a}_{\rho}=0,\\
\label{sec3-ym-pf-tor-n-re}
\varepsilon^{\mu\nu\alpha\beta}{e}^{a}_{\nu}\mathcal{T}^{a}_{\alpha\beta}&=&
\varepsilon^{\mu\nu\alpha\beta}{e}^{a}_{\nu}e^{a}_{\rho}\Sigma^{\rho}_{\alpha\beta}=0.
\end{eqnarray}
or in tensor formulations
\begin{eqnarray}
\label{sec3-ym-pf-tor-ten-j}
\varepsilon^{\mu\nu\alpha\beta}V_{\nu}\Sigma^{\rho}_{\alpha\beta}n^{\sigma}_{\rho}
&=&\mathcal{E}^{\mu\alpha\beta}\Sigma^{\rho}_{\alpha\beta}n^{\sigma}_{\rho}=0,\\
\label{sec3-ym-pf-tor-ten-n}
\varepsilon^{\mu\nu\alpha\beta}g_{\rho\nu}\Sigma^{\rho}_{\alpha\beta}&=&0.
\end{eqnarray}
By the identity~(\ref{sec3-gf-dec-def-asym-id-n}),  Eq.~(\ref{sec3-ym-pf-tor-ten-j}) is equivalent to
\begin{eqnarray}
\label{sec3-ym-pf-tor-ten-sol-j}
\frac{1}{2}\mathcal{E}_{\mu\alpha\beta}\mathcal{E}^{\mu\theta\tau}\Sigma^{\rho}_{\theta\tau}n^{\sigma}_{\rho}=
n^{\theta}_{\alpha}n^{\tau}_{\beta}\Sigma^{\rho}_{\theta\tau}n^{\sigma}_{\rho}=0.
\end{eqnarray}
So the constraint~(\ref{sec3-ym-pf-tor-ten-j}) determines the $n^{\theta}_{\alpha}n^{\tau}_{\beta}\Sigma^{\rho}_{\theta\tau}n^{\sigma}_{\rho}$ part of $\Sigma^{\rho}_{\alpha\beta}$ in Eq.~(\ref{ap1-gf-tor-dec-c-re}). After contracting the Levi-Civita symbols, Eq.~(\ref{sec3-ym-pf-tor-ten-n}) can be rewritten as
\begin{eqnarray}
\label{sec3-ym-pf-tor-ten-sol-n}
\Sigma^{\rho}_{\alpha\beta}g_{\rho\sigma}+\Sigma^{\rho}_{\sigma\alpha}g_{\rho\beta}+
\Sigma^{\rho}_{\beta\sigma}g_{\rho\alpha}=0.
\end{eqnarray}
From Eq.~(\ref{sec3-ym-pf-tor-ten-sol-n}), we obtain
\begin{eqnarray}
\label{sec3-ym-pf-tor-ten-sol-n-con}
(\delta^{\sigma}_{\mu}-n^{\sigma}_{\mu})(\Sigma^{\rho}_{\alpha\sigma}g_{\rho\beta}-
\Sigma^{\rho}_{\beta\sigma}g_{\rho\alpha})=0,
\end{eqnarray}
which implies
\begin{eqnarray}
\label{sec3-ym-pf-tor-ten-sol-n-con-1}
(\delta^{\sigma}_{\mu}-n^{\sigma}_{\mu})(n^{\theta}_{\alpha}\Sigma^{\rho}_{\theta\sigma}n^{\beta}_{\rho}-
g^{\beta\theta}\Sigma^{\rho}_{\theta\sigma}g_{\rho\alpha})=0.
\end{eqnarray}
By this equation, the part of $\Sigma^{\rho}_{\alpha\beta}$
\begin{eqnarray}
\label{se3-gf-tor-dec-c-re-1}
(\delta^{\sigma}_{\mu}-n^{\sigma}_{\mu})(n^{\theta}_{\alpha}\Sigma^{\rho}_{\theta\sigma}n^{\beta}_{\rho}-
g^{\beta\theta}\Sigma^{\rho}_{\theta\sigma}g_{\rho\alpha})
\end{eqnarray}
is determined to be zero. With Eqs.~(\ref{sec3-ym-pf-tor-ten-sol-j}) and~(\ref{sec3-ym-pf-tor-ten-sol-n-con-1}), there are four undetermined parts of $\Sigma^{\rho}_{\alpha\beta}$ in Eq.~(\ref{ap1-gf-tor-dec-c-re}) left
\begin{eqnarray}
\label{se3-gf-tor-dec-c-re-3}
(\delta^{\tau}_{\mu}-n^{\tau}_{\mu})n^{\theta}_{\alpha}\Sigma^{\rho}_{\theta\tau}(\delta^{\sigma}_{\rho}-n^{\sigma}_{\rho})
,~~(\delta^{\tau}_{\alpha}-n^{\tau}_{\alpha})n^{\theta}_{\mu}\Sigma^{\rho}_{\theta\tau}(\delta^{\sigma}_{\rho}-n^{\sigma}_{\rho}),\\
\label{se3-gf-tor-dec-c-re-4}
(\delta^{\tau}_{\mu}-n^{\tau}_{\mu})(\delta^{\theta}_{\alpha}-n^{\theta}_{\alpha})
\Sigma^{\rho}_{\theta\tau}(\delta^{\sigma}_{\rho}-n^{\sigma}_{\rho})
,~~n^{\tau}_{\mu}n^{\theta}_{\alpha}\Sigma^{\rho}_{\theta\tau}(\delta^{\sigma}_{\rho}-n^{\sigma}_{\rho}),
\end{eqnarray}
and one undetermined part of $\Gamma^{\rho}_{\alpha\beta}$ in Eq.~(\ref{ap1-gf-con})
\begin{eqnarray}
\label{sec3-gf-con}
\Gamma^{\rho}_{(\alpha\beta)}(\delta_{\rho}^{\sigma}-n_{\rho}^{\sigma}).
\end{eqnarray}
However, using the properties of the projection tensor in appendix~\ref{append-0}, we know that the above five undetermined parts actually do not contribute to $B^{a}_{\mu}$ in Eq.~(\ref{sec3-ym-pf-zb-b}) and the Lagrangian~(\ref{sec3-ym-lag-re}). That is, the solution given by Eqs.~(\ref{ap1-gf-con}) and~(\ref{ap1-gf-tor-dec-c-re}) with the constraints~(\ref{sec3-ym-pf-tor-j}) and~(\ref{sec3-ym-pf-tor-n}) gives the same $B^{a}_{\mu}$ and the Lagrangian~(\ref{sec3-ym-lag-re}) as the solution in~(\ref{sec3-gf-res-g-sol-0}). The above discussions show that the constraints~(\ref{sec3-ym-pf-tor-j}) and~(\ref{sec3-ym-pf-tor-n}) are enough to give the results in the subsection~\ref{sec:3.1}.

After analyzing constraints in the above, the next step is to integrate out $B^{a}_{\mu}$. We have
\begin{eqnarray}
\label{sec3-ym-pf-zb}
Z[B^{a}_{\mu},e^{a}_{\mu}]&=&\int [\mathscr{D}e^{a}_{\mu}][\mathscr{D}V_{\mu}] [\mathscr{D}B^{a}_{\mu}]
\mathrm{exp}\bigl(i\int{d^3x}\mathscr{L}(B^{a}_{\mu},e^{a}_{\mu})\bigr)\nonumber\\
&\times&\frac{1}{J_{e}}\delta\bigl(B^{a}_{\mu}-\hat{B}^{a}_{\mu}(e)\bigr)
\frac{1}{N_{e}}\delta\bigl(\varepsilon^{\mu\nu\alpha\beta}
\epsilon^{abc}e^{a}_{\nu}(B^{b}_{\alpha}-\tilde{B}^{b}_{\alpha}(e))e^{c}_{\beta}\bigr).
\end{eqnarray}
In the above, $\hat{B}^{a}_{\mu}(e)$ is the solution given by Eqs.~(\ref{sec3-ym-pf-zb-b}) and (\ref{ap1-gf-tor-dec-c-re}) with the constraint~(\ref{sec3-ym-pf-tor-ten-sol-j}). $\tilde{B}^{b}_{\alpha}(e)$ is the solution given by Eqs.~(\ref{sec3-ym-pf-zb-b})
and (\ref{ap1-gf-tor-dec-c-re}) with the constraint~(\ref{sec3-ym-pf-tor-ten-sol-n-con-1}). $J_{e}$ and $N_{e}$ are the Jacobian factor associated with the variable transformation of the delta function. $J_{e}$ is given by
\begin{eqnarray}
\label{sec3-ym-pf-norm-jb-j}
J_{e}&=&e_{12-n}\bigl(\mathcal {M}^{\mu\nu}_{ab}\bigr),\\
\label{sec3-ym-pf-trans-mdef-j}
\mathcal{M}^{\mu\nu}_{ab}&=&\frac{\partial
\left(\varepsilon^{\mu\tau\alpha\beta}V_{\tau}\mathcal{T}^{a}_{\alpha\beta}\right)}{\partial{B}^{b}_{\nu}}
=2\varepsilon^{\mu\tau\nu\beta}V_{\tau}\varepsilon^{abc}e^{c}_{\beta},
\end{eqnarray}
and $N_{e}$ is given by
\begin{eqnarray}
\label{sec3-ym-pf-norm-jb-n}
N_{e}&=&\mathrm{Det}\bigl(\mathcal {N}^{\mu\nu}\bigr),\\
\label{sec3-ym-pf-trans-mdef-n}
\mathcal{N}^{\mu\nu}&=&\frac{\partial
\left(\varepsilon^{\mu\tau\alpha\beta}{e}^{a}_{\tau}\mathcal{T}^{a}_{\alpha\beta}\right)}
{\partial
\left(\varepsilon^{\nu\theta\rho\sigma}e^{a}_{\theta}B^{b}_{\rho}e^{c}_{\sigma}\right)}
=2\eta^{\mu\nu},
\end{eqnarray}
where $\mathcal {M}^{\mu\nu}_{ab}$ is a $12\times{12}$ matrix. Note that we have used the elementary symmetric polynomials $e_{12-n}$ in Eq.~(\ref{sec3-ym-pf-norm-jb-j}), where $n$ is the number of zero eigenvalue of matrix, and $e_{12-n}$ can be computed by Eq.~(\ref{ap3-ch-th-co}) in appendix~\ref{append-3}. This is because $\mathcal {M}^{\mu\nu}_{ab}$ is not invertible, and it has one zero eigenvalue, so $n=1$ in Eq.~(\ref{sec3-ym-pf-norm-jb-j}). For an invertible $k\times{k}$ matrix $M$, we have $e_{k}(M)=\mathrm{det}(M)$, so Eq.~(\ref{sec3-ym-pf-norm-jb-j}) shall give the conventional Jacobian factor in the invertible situation. Integrating out $B^{a}_{\mu}$ in Eq.~(\ref{sec2-ym-pf-zb}), we obtain
\begin{eqnarray}
\label{sec3-ym-pf-zb-pi}
Z[e^{a}_{\mu}]=\int [\mathscr{D}e^{a}_{\mu}][\mathscr{D}V_{\mu}]\frac{1}{J_{e}}\frac{1}{N_{e}}\mathrm{exp}\bigl(i\int{d^3x}
\mathscr{L}(e^{a}_{\mu})\bigr),
\end{eqnarray}
where $\mathscr{L}(e^{a}_{\mu})$ solely depends on $e^{a}_{\mu}$, which is equivalent to the expression in Eq.~(\ref{sec2-ym-lag-re}). From Eqs.~(\ref{sec3-ym-pf-norm-jb-j}) and~(\ref{sec3-ym-pf-norm-jb-n}), we see that $J_{e}$ and $N_{e}$ depend only on $e^{a}_{\mu}$ but not its derivative, which shall not influence the path integral~\cite{Stelle:1976gc,Hamber:2009}.

In the above, we obtain the reformulated theory in subsection~\ref{sec:3.1} by integrate out $B^{a}_{\mu}$ at first. Following the discussions in subsection~\ref{sec:2.2-pf}, we plan to integrate out $e^{a}_{\mu}$. Similar to the situation in subsection~\ref{sec:2.2-pf}, solving the constraints in terms of $B^{a}_{\mu}$ is difficult because they are partial differential equations of $e^{a}_{\mu}$. However, because the constraints~(\ref{sec3-ym-pf-tor-j}) and~(\ref{sec3-ym-pf-tor-n}) are polynomials of $e^{a}_{\mu}$, the path integral about $e^{a}_{\mu}$ in Eq.~(\ref{sec3-ym-pf}) can be performed using the similar method in the background field method. We have the path integral
\begin{eqnarray}
\label{sec3-ym-pf-ze}
Z[B^{a}_{\mu},e^{a}_{\mu}]&=&\int [\mathscr{D}B^{a}_{\mu}][\mathscr{D}N_{\mu}][\mathscr{D}j^{a}_{\mu}]
[\mathscr{D}V_{\mu}][\mathscr{D}\bar{\eta}][\mathscr{D}\eta][\mathscr{D}e^{a}_{\mu}]\nonumber\\
&\times&\mathrm{exp}\biggl(i\int{d^3x}
\bigl(\mathscr{L}(B^{a}_{\mu},e^{a}_{\mu})+\mathscr{L}_{\mathrm{GF}}+\mathscr{L}_{\mathrm{FP}}\bigr)\biggr)\\
&\times&\mathrm{exp}\bigl(i\int{d^3x}(\varepsilon^{\mu\nu\alpha\beta}{j}^{a}_{\mu}V_{\nu}\mathcal{T}^{a}_{\alpha\beta}
+\varepsilon^{\mu\nu\alpha\beta}N_{\mu}{e}^{a}_{\nu}\mathcal{T}^{a}_{\alpha\beta})\bigr),\nonumber\\
\end{eqnarray}
where
\begin{eqnarray}
\label{sec3-ym-pf-ze-gf}
-2\kappa^2\mathscr{L}_{\mathrm{GF}}&=&(D^{\mu}e^{a}_{\mu})(D^{\nu}e^{a}_{\nu}),\\
\label{sec3-ym-pf-ze-fp}
\mathscr{L}_{\mathrm{FP}}&=&
\bar{\eta}^{a}\left(-(D^{\mu}D_{\mu})^{ac}-(D^{\mu})^{ad}\varepsilon^{dbc}e^{b}_{\mu}\right)\eta^{c},
\end{eqnarray}
are the gauge fixing term and the corresponding Faddeev-Popov ghost term, and the gauge covariant derivative is
\begin{eqnarray}
\label{sec3-ym-pf-ze-der}
D^{ac}_{\mu}=\partial_{\mu}\delta^{ac}+\epsilon^{abc}{B}^{b}_{\mu}.
\end{eqnarray}
To one loop order, we have the determinant from the integral of $e^{a}_{\mu}$
\begin{eqnarray}
\label{sec3-ym-pf-norm-trans-mdef-e-ja}
\bigl[\mathrm{det}\bigl(\mathcal{X}^{\alpha\beta}_{ab}\bigr)\bigr]^{-\frac{1}{2}},
\end{eqnarray}
where
\begin{eqnarray}
\label{sec3-ym-pf-norm-trans-mdef-e}
\mathcal{X}^{\alpha\beta}_{ab}=-\eta^{\alpha\beta}(\eta^{\mu\nu}D_{\mu}D_{\nu})^{ab}
+2\varepsilon^{abc}\mathcal{B}^{c\alpha\beta}
+2\varepsilon^{\alpha\beta\tau\sigma}N_{\tau}D_{\sigma}^{ab}.
\end{eqnarray}
We also have the determinant from the ghost term
\begin{eqnarray}
\label{sec3-ym-pf-norm-trans-mdef-e-gh}
\mathrm{det}\left(-(\eta^{\mu\nu}D_{\mu}D_{\nu})^{ab}\right).
\end{eqnarray}
Note that the constraint~(\ref{sec3-ym-pf-tor-n}) is quadratic polynomial of $e^{a}_{\mu}$, so we have the contribution from $N_{\mu}$ in Eq.~(\ref{sec3-ym-pf-norm-trans-mdef-e}), then the path integral of $N_{\mu}$ is also further required.  Because the gauge invariance remains intact in the background method, the total effects of these determinants can be accounted by renormalizing the coupling constant. The contribution from the $N_{\mu}$ term could modify the beta function, which makes the beta function different from that of the conventional Yang-Mills theory. However, if the beta function is still negative, we can think that the modified theory is still close to the Yang-Mills theory but not significantly different. The above discussions provide relations between partition functions in 4 dimensional case, which are more complicated compared to the situation in 3D.

\subsection{Scalar-Vector Sector}\label{sec:3.2a}

In the subsection~\ref{sec:3.1}, we have obtained a metric-like reformulation of the $SU(2)$ Yang-Mills theory in 4D. In this subsection, we shall discuss a simplified version of the Lagrangian~(\ref{sec3-ym-lag-re}). Because $g_{\mu\nu}$ is not invertible, we propose a parametrization of $g_{\mu\nu}$ as
\begin{eqnarray}
\label{sec32-ym-g-dec}
g_{\mu\nu}=\left(\frac{\gamma_{\rho\sigma}U^{\rho}U^{\sigma}}{\mathrm{det}(\gamma_{\rho\sigma})}\right)^{\frac{1}{3}}
\left(\gamma_{\mu\nu}-\frac{\gamma_{\mu\rho}U^{\rho}\gamma_{\nu\sigma}U^{\sigma}}{\gamma_{\rho\sigma}U^{\rho}U^{\sigma}}\right),
\end{eqnarray}
where $U^{\mu}$ has been defined in Eq.~(\ref{sec3-gf-res-e-inv-n-v}), which is the eigenvector of $g_{\mu\nu}$ with zero eigenvalue~(\ref{sec3-gf-res-e-inv-n-v-iden}). We also suppose that $\gamma_{\mu\nu}$ is invertible and its inverse is $\gamma^{\mu\nu}$. The parametrization~(\ref{sec32-ym-g-dec}) has three properties: Its determinant is zero; It has $U^{\nu}$ as the eigenvector with zero eigenvalue; It also makes Eq.~(\ref{sec3-gf-res-e-iden-uu}) satisfied. Using $\gamma_{\mu\nu}$ and $U^{\nu}$, we can also obtain the generalized inverse metric and the projection tensor
\begin{eqnarray}
\label{sec32-ym-g-inv-dec}
g^{\mu\nu}&=&\left(\frac{\mathrm{det}(\gamma_{\rho\sigma})}{\gamma_{\rho\sigma}U^{\rho}U^{\sigma}}\right)^{\frac{1}{3}}
\left(\gamma^{\mu\nu}-\frac{U^{\mu}U^{\nu}}{\gamma_{\rho\sigma}U^{\rho}U^{\sigma}}\right),\\
\label{sec32-ym-n-dec}
n^{\alpha}_{\beta}&=&\delta^{\alpha}_{\beta}-\frac{\gamma_{\beta\tau}U^{\tau}U^{\alpha}}{\gamma_{\rho\sigma}U^{\rho}U^{\sigma}}.
\end{eqnarray}
The expressions (\ref{sec32-ym-g-dec}), (\ref{sec32-ym-g-inv-dec}) and~(\ref{sec32-ym-n-dec}) are consistent with Eq.~(\ref{sec3-gf-res-e-inv-rel-1}). From Eqs.~(\ref{sec3-gf-res-e-inv-n-v}) and (\ref{sec3-gf-res-e-inv-n-v-n}), we also have
\begin{eqnarray}
\label{sec32-ym-g-dec-con}
\gamma^{\mu\tau}V_{\tau}+\frac{U^{\mu}}{\gamma_{\rho\sigma}U^{\rho}U^{\sigma}}=0,
\end{eqnarray}
which means that
\begin{eqnarray}
\label{sec32-ym-g-dec-con-re}
V_{\mu}=-\frac{\gamma_{\mu\tau}U^{\tau}}{\gamma_{\rho\sigma}U^{\rho}U^{\sigma}},
\end{eqnarray}
so $V_{\mu}$ is completely determined by $U^{\mu}$, and it is consistent with the constraint~(\ref{sec3-gf-res-e-inv-n-v-n}). As in Eq.~(\ref{sec2-a-g-def-sc}), we decompose $\gamma_{\mu\nu}$ into its trace part and its traceless part as
\begin{eqnarray}
\label{sec32-ym-g-dec-ga}
\gamma_{\mu\nu}=\varphi^2(x)\eta_{\mu\nu}+(\gamma_{\mu\nu}-\varphi^2(x)\eta_{\mu\nu}),
\varphi^2(x)=\frac{1}{4}\eta^{\rho\sigma}\gamma_{\rho\sigma}.
\end{eqnarray}
We also consider that $U^{\mu}$ is expressed as
\begin{eqnarray}
\label{sec32-ym-g-dec-u}
U^{\mu}=\varphi^3(x)\tilde{a}^{\mu}(x),
\end{eqnarray}
where the factor $\varphi^3(x)$ ensures that $\tilde{a}^{\mu}(x)$ is dimensionless. From $\tilde{a}^{\mu}(x)$, we can define a space-like vector field $a^{\mu}(x)$
\begin{eqnarray}
\label{sec32-ym-g-dec-u-a}
\eta_{\mu\nu}a^{\mu}a^{\nu}=-1,~~a^{\mu}=\frac{\tilde{a}^{\mu}(x)}{\sqrt{-\eta_{\mu\nu}\tilde{a}^{\mu}\tilde{a}^{\nu}}}.
\end{eqnarray}
Because the $SU(2)$ group is compact, we cannot suppose the tensor part of Eq.~(\ref{sec32-ym-g-dec-ga}) to be zero. In the following, we will only consider the sector of Eq.~(\ref{sec3-ym-lag-re}) including $a^{\mu}$ and $\varphi(x)$. The explicit expression including the tensor sector can be obtained as a formal series. $g_{\mu\nu}$ can be formally expanded as
\begin{eqnarray}
\label{sec32-ym-g-dec-ap}
g_{\mu\nu}\approx \varphi^2(x)(\eta_{\mu\nu}+a_{\mu}a_{\nu}),
\end{eqnarray}
where the tensor sector is not shown. Correspondingly, $g^{\mu\nu}$ and $n_{\alpha}^{\beta}$ can be given by
\begin{eqnarray}
\label{sec32-ym-g-dec-re-n}
g^{\mu\nu}\approx\varphi^{-2}(x)\left(\eta^{\mu\nu}+a^{\mu}a^{\nu}\right),~~
n_{\alpha}^{\beta}\approx\delta_{\alpha}^{\beta}+a_{\alpha}a^{\beta},
\end{eqnarray}
where $a_{\mu}=\eta_{\mu\sigma}a^{\sigma}$. Using the above results, the torsion~(\ref{sec3-gf-res-g-sol-1}) is
\begin{eqnarray}
\label{sec32-tor}
\Sigma^{\rho}_{\alpha\beta}&=&-\frac{1}{4} (\eta^{\rho\sigma}+a^{\rho}a^{\sigma})
(a_{\alpha}\partial_{\sigma}a_{\beta}-a_{\beta}\partial_{\sigma}a_{\alpha})\nonumber\\
&-&\frac{1}{4}(a_{\alpha}\partial_{\beta}a^{\rho}-a_{\beta}\partial_{\alpha}a^{\rho})
+\frac{1}{2\phi}(\delta^{\rho}_{\alpha}a_{\beta}-\delta^{\rho}_{\beta}a_{\alpha})a^{\sigma}\partial_{\sigma}\phi.
\end{eqnarray}
The connection~(\ref{sec3-gf-res-g-sol-0}) have the expression
\begin{eqnarray}
\label{sec32-con}
\Gamma^{\rho}_{\alpha\beta}&=&\bar{\Gamma}^{\rho}_{\alpha\beta}+\Psi^{\rho}_{\alpha\beta},
\end{eqnarray}
where
\begin{eqnarray}
\label{sec32-con-def-gam}
\bar{\Gamma}^{\rho}_{\alpha\beta}&=&-\frac{1}{2}(a_{\alpha}\partial^{\rho}a_{\beta}-a_{\beta}\partial_{\alpha}a^{\rho})
-\frac{1}{2}a^{\rho}(\partial_{\alpha}a_{\beta}+\partial_{\beta}a_{\alpha})\\
&-&\frac{1}{2}a_{\alpha}a^{\sigma}(a_{\beta}\partial_{\sigma}a^{\rho}-a^{\rho}\partial_{\sigma}a_{\beta})
+\frac{1}{2}(a_{\alpha}\partial_{\beta}a^{\rho}+a_{\beta}\partial_{\alpha}a^{\rho}),\nonumber
\end{eqnarray}
and
\begin{eqnarray}
\label{sec32-con-def-psi}
\Psi^{\rho}_{\alpha\beta}&=&\frac{1}{\phi}(\delta^{\rho}_{\alpha}\partial_{\beta}\phi+\delta^{\rho}_{\beta}\partial_{\alpha}\phi)
+\frac{1}{\phi}a^{\rho}(a_{\alpha}\partial_{\beta}\phi+a_{\beta}\partial_{\alpha}\phi)\\
&-&\frac{1}{\phi}(\eta_{\alpha\beta}a^{\rho}-\delta^{\rho}_{\alpha}\partial_{\beta})a^{\sigma}\partial_{\sigma}\phi
-\frac{1}{\phi}(\eta_{\alpha\beta}+a_{\alpha}a_{\beta})\partial^{\rho}\phi.\nonumber
\end{eqnarray}
In the above, we have defined $\phi=\frac{\sqrt{6}}{\kappa}\varphi$. Note that $\bar{\Gamma}^{\rho}_{\alpha\beta}$ is independent of $\phi$. Using $\bar{\Gamma}^{\rho}_{\alpha\beta}$ and $\Psi^{\rho}_{\alpha\beta}$, the Riemann tensor has the decomposition
\begin{eqnarray}
\label{sec32-rie}
R^{\sigma}_{\hspace{1mm}\rho\mu\nu}=\bar{R}^{\sigma}_{\hspace{1mm}\rho\mu\nu}
+K^{\sigma}_{\hspace{1mm}\rho\mu\nu},
\end{eqnarray}
where
\begin{eqnarray}
\label{sec32-rie-bar}
\bar{R}^{\sigma}_{\hspace{1mm}\rho\mu\nu}=\partial_{\mu}\bar{\Gamma}^{\sigma}_{\nu\rho}
-\partial_{\nu}\bar{\Gamma}^{\sigma}_{\mu\rho}+\bar{\Gamma}^{\tau}_{\nu\rho}\bar{\Gamma}^{\sigma}_{\mu\tau}
-\bar{\Gamma}^{\tau}_{\mu\rho}\bar{\Gamma}^{\sigma}_{\nu\tau}
\end{eqnarray}
is the Riemann tensor independent of $\phi$, and
\begin{eqnarray}
\label{sec32-rie-k}
K^{\sigma}_{\hspace{1mm}\rho\mu\nu}=\bar{\nabla}_{\mu}\Psi^{\sigma}_{\nu\rho}-\bar{\nabla}_{\nu}\Psi^{\sigma}_{\mu\rho}
+\Psi^{\sigma}_{\mu\tau}\Psi^{\tau}_{\nu\rho}-\Psi^{\sigma}_{\nu\tau}\Psi^{\tau}_{\mu\rho}.
\end{eqnarray}
Here we have defined the derivative
\begin{eqnarray}
\label{sec32-rie-nab}
\bar{\nabla}_{\mu}\Psi^{\sigma}_{\nu\rho}=\partial_{\mu}\Psi^{\sigma}_{\nu\rho}
+\bar{\Gamma}^{\sigma}_{\mu\tau}\Psi^{\tau}_{\nu\rho}-\bar{\Gamma}^{\tau}_{\mu\rho}\Psi^{\sigma}_{\nu\tau}.
\end{eqnarray}
Using the above results, we can obtain the expression for the Lagrangian~(\ref{sec3-ym-lag-re}). The complete expressions of this Lagrangian are lengthy. The lower derivative part of  $\mathscr{L}$ is given by
\begin{eqnarray}
\label{sec32-ym-lag-re-app}
\mathscr{L}&\approx&\frac{1}{2}\partial_{\alpha}\phi\partial^{\alpha}\phi-\frac{\kappa^2}{24}\phi^4\\
&+&\frac{3}{4}a^{\alpha}a^{\beta}\partial_{\alpha}\phi\partial_{\beta}\phi
+\frac{1}{6}(2a^{\beta}\partial_{\alpha}a^{\alpha}+a^{\alpha}\partial_{\alpha}a^{\beta})\phi\partial_{\beta}\phi\nonumber\\
&+&\frac{1}{24}\phi^2(3\partial_{\alpha}a^{\beta}\partial_{\beta}a^{\alpha}-\partial_{\alpha}a^{\beta}\partial^{\alpha}a_{\beta}
-a^{\alpha}a^{\beta}\partial_{\alpha}a^{\rho}\partial_{\beta}a_{\rho}),\nonumber
\end{eqnarray}
where a total divergence term
\begin{eqnarray}
\label{sec32-ym-lag-re-app-div}
\mathscr{L}_{\mathrm{bound}}=\partial_{\alpha}\bigl(\frac{1}{12}\phi^2
(a^{\beta}\partial_{\beta}a^{\alpha}-a^{\alpha}\partial_{\beta}a^{\beta})
-\frac{1}{3}\phi(a^{\alpha}a^{\beta}\partial_{\beta}\phi+\partial^{\alpha}\phi)\bigr)
\end{eqnarray}
has been subtracted. Eq.~(\ref{sec32-ym-lag-re-app}) is different from Eq.~(\ref{sec2-ym-lag-re-dec-sca}), which is a pure scalar theory. In Eq.~(\ref{sec32-ym-lag-re-app}), a nonzero value of the quantity
\begin{eqnarray}
\label{sec32-ym-lag-re-app-vev}
\langle3\partial_{\alpha}a^{\beta}\partial_{\beta}a^{\alpha}-\partial_{\alpha}a^{\beta}\partial^{\alpha}a_{\beta}
-a^{\alpha}a^{\beta}\partial_{\alpha}a^{\rho}\partial_{\beta}a_{\rho}\rangle=m^2
\end{eqnarray}
shall yield a nonzero virtual mass for $\phi$. This virtual massive term together with $\frac{\kappa^2}{24}\phi^4$ support a minimum for a nonzero value of $\phi$. Inversely, when $\phi$ takes a nonzero constant value, that is
\begin{eqnarray}
\label{sec32-ym-lag-re-app-vev-phi}
\langle\phi\rangle=v.
\end{eqnarray}
In this case, Eq.~(\ref{sec3-ym-lag-re}) is the Lagrangian of the unit vector $a_{\mu}$. It is possible that the Lagrangian~(\ref{sec3-ym-lag-re}) supports nontrivial classical solution of finite energy.

Now we give comparisons of the above result to the dual superconductor scenario~\cite{Mandelstam:1974pi,Polyakov:1976fu,tHooft:1979uj}. In~\cite{Faddeev:1998eq}, the $SU(2)$ gauge field is decomposed in terms of a unit 3-vector $\bf{n}$, an Abelian field $C_{\mu}$ and two scalar fields $\rho(x)$ and $\sigma(x)$. The condensate of the order parameter $\bf{n}$ yields the Abelian-Higgs model, which supports vortex solutions~\cite{Nielsen:1973cs}. The vortex solutions are responsible for the string-like force in the large distance. In the dual view, the condensate of $\rho(x)$ and $\sigma(x)$ yields a Lagrangian of the order parameter $\bf{n}$. This Lagrangian has the structure of non-linear $\sigma$ model, which supports knot-like solutions of finite energy~\cite{Faddeev:2006sw}. The picture of our above result of this subsection has the similar interpretation to this dual scenario. Here $\phi^2(x)$ as the scalar freedom of $\gamma_{\mu\nu}$, has the natural interpretation as the candidate of the dimension two condensate. While the unit vector $a_{\mu}$ serves as the order parameter. In the above discussions, the scalar-vector sector of Eq.~(\ref{sec32-ym-lag-re-app}) is a Lagrangian about $\phi$ and $a_{\mu}$. From Eq.~(\ref{sec32-ym-lag-re-app-vev}), we knew that the nonzero condensate of $a_{\mu}$ yields a nonzero condensate of $\phi$. If $\phi$ obtains a nonzero condensate as Eq.~(\ref{sec32-ym-lag-re-app-vev-phi}), then Eq.~(\ref{sec3-ym-lag-re}) produces a Lagrangian about $a_{\mu}$. We conjecture that it could support classical solutions of finite energy.

There is an additional question required to be discussed for the scalar-vector sector in the above. For the $SU(2)$ group in Euclidean space-time or the $SO(1,2)$ group in Minkowski space-time, the tensor part can be supposed to be zero, then the Lagrangian $\mathscr{L}$~(\ref{sec3-ym-lag-re}) can be reduced to the scalar-vector sector similar to the expression in Eq.~(\ref{sec32-ym-lag-re-app}). A question is that equations of motion~(EOMs) of the scalar-vector Lagrangian may produce solutions which does not solve the original Yang-Mills equations. For the Faddeev-Niemi decomposition in~\cite{Faddeev:1998eq}, this question has been analyzed in~\cite{Evslin:2010sb,Niemi:2010ms}. For the $SU(2)$ Yang-Mills theory in 4D, there are $12$ EOMs for $A^{a}_{\mu}$, $3$ of which are the Gaussian constraints. The Faddeev-Niemi decomposition in~\cite{Faddeev:1998eq} reduces the numbers of field variables. As a consequence,  there are $8$ EOMs, while the Gaussian constraints are missed. Therefore, EOMs of the Faddeev-Niemi decomposition produce solutions with an external source term, which are not solutions of the Yang-Mills equations. For the scalar-vector Lagrangian in the above, the numbers of the fields variables are reduced to be $4$, hence EOMs of this Lagrangian may also produce solutions which does not solve the Yang-Mills equations.

\subsection{Dimension Two Condensate}\label{sec:3.2}

Similar to that in the subsection~\ref{sec:2.2}, $g_{\mu\nu}$ is gauge invariant and has the natural interpretation as the dimension two condensate. However, an important point different from section~\ref{sec:2} is that $g_{\mu\nu}$ is not invertible as we discussed in the subsection~\ref{sec:3.1}. We have provided a parametrization of $g_{\mu\nu}$ in Eq.~(\ref{sec32-ym-g-dec}). For the convenience of analyzing the dimension two condensate, here we propose an alternative parametrization of $g_{\mu\nu}$ as
\begin{eqnarray}
\label{sec32-g-para-a}
g_{\mu\nu}=C^{\frac{1}{3}}p_{\mu}^{\rho}p_{\nu}^{\sigma}\gamma_{\rho\sigma},
\end{eqnarray}
where
\begin{eqnarray}
\label{sec32-g-para-a-p}
p_{\beta}^{\alpha}=\delta_{\beta}^{\alpha}-\frac{\eta_{\beta\rho}U^{\rho}U^{\alpha}}{\eta_{\rho\sigma}U^{\rho}U^{\sigma}}
\end{eqnarray}
is the project tensor, and
\begin{eqnarray}
\label{sec32-g-para-a-c}
C=\frac{1}{\mathrm{det}(\gamma_{\rho\sigma})}
\frac{(\eta_{\rho\sigma}U^{\rho}U^{\sigma})^2}{\gamma^{\rho\sigma}\eta_{\rho\tau}\eta_{\sigma\theta}U^{\tau}U^{\theta}}.
\end{eqnarray}
Similar to that in Eq.~(\ref{sec32-ym-g-dec}), the parametrization in Eq.~(\ref{sec32-g-para-a}) also satisfies three constraints: The determinant of $g_{\mu\nu}$ is zero; $g_{\mu\nu}$ has the eigenvector $U^{\mu}$ with zero eigenvalue; And Eq.~(\ref{sec3-gf-res-e-iden-uu}) is satisfied. With this parametrization, the expression for $g^{\mu\nu}$ cannot be derived straightforwardly as in Eq.~(\ref{sec32-ym-g-inv-dec}), but it can be derived using the Cayley-Hamilton theorem. We give the derivation of $g^{\mu\nu}$ in appendix~\ref{append-3}. In order to discuss the dimension two condensate, we expand $\gamma_{\mu\nu}$ and $U^{\mu}$ around constant background as
\begin{eqnarray}
\label{sec32-g-f}
\gamma_{\mu\nu}&=&v\cdot\eta_{\mu\nu}+f_{\mu\nu},\\
\label{sec32-u-n}
U^{\mu}&=&v^{\frac{3}{2}}n^{\mu}+v^{\frac{1}{2}}u^{\mu},
\end{eqnarray}
where $v$ is constant and positive, and $n^{\mu}$ is a constant space-like vector
\begin{eqnarray}
\label{sec32-u-nn}
\eta_{\mu\nu}n^{\mu}n^{\nu}=-1.
\end{eqnarray}
In Eqs.~(\ref{sec32-g-f}) and~(\ref{sec32-u-n}), the power of $v$ is chosen by accounting the dimension of $\gamma_{\mu\nu}$ and $U^{\mu}$. One question is that the constant vector $n^{\mu}$ may break the Lorentz invariance. However, in the subsequent computation of the effective potential, $n^{\mu}$ can be eliminated using the constraint~(\ref{sec32-u-nn}). So the effective potential shall only depend on $v$. With these expansions, $g_{\mu\nu}$ can be approximated as
\begin{eqnarray}
\label{sec32-g}
g_{\mu\nu}&=&(\eta_{\mu\nu}+n_{\mu}n_{\nu})
\bigl(v-\frac{1}{3}(2u_{\sigma}n^{\sigma}+f_{\sigma}^{\sigma}+f_{\rho\sigma}n^{\rho}n^{\sigma})\bigr)\\
&+&\bar{p}_{\mu}^{\rho}\bar{p}_{\nu}^{\sigma}f_{\rho\sigma}
+\bigl(n_{\mu}u_{\nu}+u_{\mu}n_{\nu}+2n_{\mu}n_{\nu}(u_{\sigma}n^{\sigma})\bigr)+\cdots,\nonumber
\end{eqnarray}
where $\bar{p}_{\beta}^{\alpha}$ is the constant project tensor
\begin{eqnarray}
\label{sec32-p-c}
\bar{p}_{\beta}^{\alpha}=\delta_{\beta}^{\alpha}+n_{\beta}n^{\alpha}.
\end{eqnarray}
The generalized inverse metric $g^{\mu\nu}$ in Eq.~(\ref{ap3-ch-gi-s-g}) can be expanded as
\begin{eqnarray}
\label{sec32-g-inv}
g^{\mu\nu}&=&(\eta^{\mu\nu}+n^{\mu}n^{\nu})
\bigl(\frac{1}{v}+\frac{1}{3v^2}(2u_{\sigma}n^{\sigma}+f_{\sigma}^{\sigma}+f_{\rho\sigma}n^{\rho}n^{\sigma})\bigr)\\
&-&\frac{1}{v^2}\bar{p}^{\mu\rho}\bar{p}^{\nu\sigma}f_{\rho\sigma}
+\frac{1}{v^2}\bigl(n^{\mu}u^{\nu}+u^{\mu}n^{\nu}+2n^{\mu}n^{\nu}(u_{\sigma}n^{\sigma})\bigr)+\cdots,\nonumber
\end{eqnarray}
and the projector vector $n_{\beta}^{\alpha}$ in Eq.~(\ref{sec3-gf-res-e-inv-rel-1}) is  expanded as
\begin{eqnarray}
\label{sec32-n}
n_{\beta}^{\alpha}=(\delta^{\alpha}_{\beta}+n_{\beta}n^{\alpha})
+\frac{1}{v}\bigl(n_{\beta}u^{\alpha}+u_{\beta}n^{\alpha}+2n_{\beta}n^{\alpha}(u_{\sigma}n^{\sigma})\bigr)+\cdots.
\end{eqnarray}
Although $f_{\mu\nu}$ in Eq.~(\ref{sec32-g-f}) has ten components, only its projected part $\bar{f}_{\mu\nu}=\bar{p}_{\mu}^{\rho}\bar{p}_{\nu}^{\sigma}f_{\rho\sigma}$ contributes to $g_{\mu\nu}$ and $g^{\mu\nu}$. $\bar{f}_{\mu\nu}$ satisfies the transverse and traceless conditions
\begin{eqnarray}
\label{sec32-n-f}
n^{\mu}\bar{f}_{\mu\nu}=0,~~n^{\mu}n^{\nu}\bar{f}_{\mu\nu}=0,
\end{eqnarray}
so $\bar{f}_{\mu\nu}$ has 5 effective freedoms, which are equivalent to the components of the massive spin-2 field. $u_{\mu}$ has four freedoms. So $g_{\mu\nu}$ has 9 freedoms totally, which is consistent with the fact that $g_{\mu\nu}$ is a degenerate and symmetric tensor. We can define a new tensor $h_{\mu\nu}$
\begin{eqnarray}
\label{sec32-h}
h_{\mu\nu}=\bar{p}_{\mu}^{\rho}\bar{p}_{\nu}^{\sigma}f_{\rho\sigma}
+\bigl(n_{\mu}u_{\nu}+u_{\mu}n_{\nu}+n_{\mu}n_{\nu}(u_{\sigma}n^{\sigma})\bigr),
\end{eqnarray}
then we have
\begin{eqnarray}
\label{sec32-h-u}
u_{\mu}=-h_{\mu\nu}n^{\nu},
\end{eqnarray}
that is, $u_{\mu}$ can be projected out from $h_{\mu\nu}$. So the tensor $\bar{f}_{\mu\nu}$ and the vector $u_{\mu}$ can be packed into a single tensor $h_{\mu\nu}$.  In the following discussions, we shall use $h_{\mu\nu}$ as the dynamical variables. In terms of $h_{\mu\nu}$, $g_{\mu\nu}$ and $g^{\mu\nu}$ can be simplified as
\begin{eqnarray}
\label{sec32-g-h}
g_{\mu\nu}&=&(\eta_{\mu\nu}+n_{\mu}n_{\nu})
\bigl(v-\frac{1}{3}(h+\varphi)\bigr)\\
&+&h_{\mu\nu}+\varphi{n}_{\mu}n_{\nu}+\cdots,\nonumber\\
\label{sec32-g-inv-h}
g^{\mu\nu}&=&(\eta^{\mu\nu}+n^{\mu}n^{\nu})
\bigl(\frac{1}{v}+\frac{1}{3v^2}(h+\varphi)\bigr)-\frac{1}{v^2}h^{\mu\nu}\\
&-&\frac{1}{v^2}\bigl(2n^{\mu}h^{\nu\sigma}n_{\sigma}
+2n^{\nu}h^{\mu\sigma}n_{\sigma}-3\varphi{n}^{\mu}n^{\nu}\bigr)+\cdots,\nonumber
\end{eqnarray}
where we have used the definitions
\begin{eqnarray}
\label{sec32-g-h-phi}
h=\eta^{\mu\nu}h_{\mu\nu},~~\varphi=-{n}^{\mu}n^{\nu}h_{\mu\nu}.
\end{eqnarray}
We see that $g_{\mu\nu}$ and $g^{\mu\nu}$ can be expressed by $h_{\mu\nu}$, and $n_{\beta}^{\alpha}$ in Eq.~(\ref{sec32-n}) can also be expressed by $h_{\mu\nu}$ using Eq.~(\ref{sec32-h-u}). Using the above expressions, we can obtain the expressions for the higher derivative part of the Lagrangian~(\ref{sec3-ym-lag-re})
\begin{eqnarray}
\label{sec32-lag-4}
4\kappa^2v^2\mathscr{L}^{(4)}&=&-\frac{1}{2}(\eta^{\mu\nu}+n^{\mu}n^{\nu})
\partial_{\mu}\partial_{\rho}h_{\alpha\beta}\partial_{\nu}\partial^{\rho}h^{\alpha\beta}
+\frac{1}{2}(2\eta^{\rho\sigma}+n^{\rho}n^{\sigma})
\partial_{\rho}\partial_{\alpha}h_{\nu\beta}\partial_{\sigma}\partial^{\nu}h^{\alpha\beta}\nonumber\\
&-&\frac{1}{2}
(\partial_{\alpha}\partial_{\beta}h^{\alpha\beta})^2
-\frac{1}{2}\partial_{\mu}\partial_{\nu}u_{\sigma}\partial^{\mu}\partial^{\nu}u^{\sigma}
+\frac{1}{2}\partial_{\mu}\partial_{\sigma}u_{\nu}\partial^{\mu}\partial^{\nu}u^{\sigma}
+\frac{2}{3}n^{\rho}\partial_{\rho}\partial_{\sigma}(h+\varphi)\partial^{\sigma}\partial_{\alpha}u^{\alpha}\nonumber\\
&+&n^{\rho}\partial_{\rho}\partial_{\sigma}h_{\mu\nu}\partial^{\mu}\partial^{\nu}u^{\sigma}
-n^{\rho}\partial_{\rho}\partial_{\mu}h_{\sigma\nu}\partial^{\mu}\partial^{\nu}u^{\sigma}
-\frac{1}{3}\partial_{\alpha}\partial_{\beta}(h+\varphi)\partial_{\rho}\partial^{\rho}h^{\alpha\beta}\\
&+&\frac{1}{9}(\eta^{\mu\nu}+n^{\mu}n^{\nu})\partial_{\mu}\partial_{\rho}(2h-\varphi)\partial_{\nu}\partial^{\rho}(h+\varphi)
-\frac{1}{3}\partial_{\mu}\partial_{\nu}\varphi\partial^{\mu}\partial^{\nu}(h+\varphi).\nonumber
\end{eqnarray}
The quadratic derivative part in~(\ref{sec3-ym-lag-re-2}) is
\begin{eqnarray}
\label{sec32-lag-2}
-4\kappa^2v\mathscr{L}^{(2)}&=&-\frac{1}{2}(3\eta^{\mu\nu}+4n^{\mu}n^{\nu})\partial_{\mu}h_{\alpha\beta}\partial_{\nu}h^{\alpha\beta}
+3\partial^{\mu}h_{\mu\alpha}\partial_{\nu}h^{\nu\alpha}\\
&+&4n^{\rho}\partial_{\rho}h\partial_{\alpha}u^{\alpha}
-6n^{\rho}\partial_{\rho}h_{\alpha\beta}\partial_{\alpha}u^{\beta}
-2\partial_{\alpha}u_{\beta}\partial^{\alpha}u^{\beta}
+2\partial_{\alpha}(\varphi-h)\partial_{\beta}h^{\alpha\beta}\nonumber\\
&-&\partial_{\mu}\varphi\partial^{\mu}(\varphi+2h)-\frac{1}{6}\partial_{\mu}h\partial^{\mu}(h-2\varphi)
+(\eta^{\mu\nu}+n^{\mu}n^{\nu})\partial_{\mu}h\partial_{\nu}h.\nonumber
\end{eqnarray}
The Fierz-Pauli type massive term in Eq.~(\ref{sec3-ym-lag-re-0}) is expressed as
\begin{eqnarray}
\label{sec32-lag-0}
-4\kappa^2\mathscr{L}^{(0)}=6v^2-h_{\alpha\beta}h^{\alpha\beta}+\frac{1}{3}h^2-\frac{2}{3}\varphi{h}+4\varphi^2-8v\varphi.
\end{eqnarray}
And we also have the term in Eq.~(\ref{sec3-ym-lag-re-3}) including the antisymmetric tensor
\begin{eqnarray}
\label{sec32-lag-3}
-4\kappa^2v^{\frac{3}{2}}\mathscr{L}^{(3)}&=&\varepsilon^{\tau\alpha\beta\theta}
n_{\tau}\partial_{\mu}\partial_{\theta}u_{\beta}\partial^{\mu}u_{\alpha}
+\varepsilon^{\mu\rho\sigma\theta}n_{\mu}n^{\alpha}\partial_{\alpha}\partial_{\theta}h_{\rho\beta}
n^{\tau}\partial_{\tau}h^{\beta}_{\sigma}\nonumber\\
&+&2\varepsilon^{\alpha\mu\rho\sigma}n_{\alpha}\partial_{\beta}\partial_{\rho}u_{\mu}n^{\tau}\partial_{\tau}h^{\beta}_{\sigma}.
\end{eqnarray}
The quadratic term of Eq.~(\ref{sec3-ym-lag-re-1}) is zero. In the above expressions, $u_{\mu}$ is determined by Eq.~(\ref{sec32-h-u}). So the above equations give the kinetic part of the tensor $h_{\mu\nu}$. In order to obtain the kinetic operator, we can define the operators
 \begin{eqnarray}
\label{sec32-lag-op-def-1}
\mathcal{P}^{(k,1)}_{\mu\nu,\alpha\beta}&=&\eta_{\mu\nu}\eta_{\alpha\beta},~~
\mathcal{P}^{(k,2)}_{\mu\nu,\alpha\beta}=\eta_{\mu\alpha}\eta_{\nu\beta}+\eta_{\mu\beta}\eta_{\nu\alpha},\nonumber\\
\mathcal{P}^{(k,3)}_{\mu\nu,\alpha\beta}
&=&\eta_{\mu\nu}\partial_{\alpha}\partial_{\beta}+\eta_{\alpha\beta}\partial_{\mu}\partial_{\nu},~~
\mathcal{P}^{(k,5)}_{\mu\nu,\alpha\beta}=\partial_{\mu}\partial_{\nu}\partial_{\alpha}\partial_{\beta},\\
\mathcal{P}^{(k,4)}_{\mu\nu,\alpha\beta}
&=&(\eta_{\mu\alpha}\partial_{\nu}\partial_{\beta}+\eta_{\nu\alpha}\partial_{\mu}\partial_{\beta})+(\alpha\leftrightarrow\beta).\nonumber
\end{eqnarray}
These operators belong to the class of the projection operators in appendix~\ref{propagator}. However, here we also have three operators constructed from $n_{\mu}$
 \begin{eqnarray}
\label{sec32-lag-op-def-2n}
\mathcal{P}^{(n,1)}_{\mu\nu,\alpha\beta}
&=&\eta_{\mu\nu}n_{\alpha}n_{\beta}+\eta_{\alpha\beta}n_{\mu}n_{\nu},~~
\mathcal{P}^{(n,3)}_{\mu\nu,\alpha\beta}=n_{\mu}n_{\nu}n_{\alpha}n_{\beta},\\
\mathcal{P}^{(n,2)}_{\mu\nu,\alpha\beta}&=&
(\eta_{\mu\alpha}n_{\nu}n_{\beta}+\eta_{\nu\alpha}n_{\mu}n_{\beta})+(\alpha\leftrightarrow\beta),\nonumber
\end{eqnarray}
and six composite operators constructed from $n_{\mu}$ and the partial derivative
 \begin{eqnarray}
\label{sec32-lag-op-def-2nk}
\mathcal{P}^{(nk,1)}_{\mu\nu,\alpha\beta}
&=&n_{\mu}n_{\nu}\partial_{\alpha}\partial_{\beta}+n_{\alpha}n_{\beta}\partial_{\mu}\partial_{\nu},~
\mathcal{P}^{(nk,2)}_{\mu\nu,\alpha\beta}
=(n_{\mu}n_{\alpha}\partial_{\nu}\partial_{\beta}+n_{\nu}n_{\alpha}\partial_{\mu}\partial_{\beta})
+(\alpha\leftrightarrow\beta),\nonumber\\
\mathcal{P}^{(nk,3)}_{\mu\nu,\alpha\beta}
&=&\eta_{\mu\nu}(n_{\alpha}\partial_{\beta}+n_{\beta}\partial_{\alpha})
+\eta_{\alpha\beta}(n_{\mu}\partial_{\nu}+n_{\nu}\partial_{\mu}),\nonumber\\
\mathcal{P}^{(nk,4)}_{\mu\nu,\alpha\beta}
&=&n_{\mu}n_{\nu}(n_{\alpha}\partial_{\beta}+n_{\beta}\partial_{\alpha})
+n_{\alpha}n_{\beta}(n_{\mu}\partial_{\nu}+n_{\nu}\partial_{\mu}),\\
\mathcal{P}^{(nk,5)}_{\mu\nu,\alpha\beta}
&=&(n_{\mu}\partial_{\nu}+n_{\nu}\partial_{\mu})\partial_{\alpha}\partial_{\beta}
+(n_{\alpha}\partial_{\beta}+n_{\beta}\partial_{\alpha})\partial_{\mu}\partial_{\nu},\nonumber\\
\mathcal{P}^{(nk,6)}_{\mu\nu,\alpha\beta}
&=&\bigl(\eta_{\mu\alpha}(n_{\nu}\partial_{\beta}+n_{\beta}\partial_{\nu})
+\eta_{\nu\alpha}(n_{\mu}\partial_{\beta}+n_{\beta}\partial_{\mu})\bigr)+(\alpha\leftrightarrow\beta).\nonumber
\end{eqnarray}
And Eq.~(\ref{sec32-lag-3}) also provides four operators relevant to the antisymmetric tensor
\begin{eqnarray}
\label{sec32-lag-op-def-3}
\mathcal{P}^{(\varepsilon,1)}_{\mu\nu,\alpha\beta}&=&
(\eta_{\mu\alpha}\varepsilon_{\nu\beta\rho\sigma}n^{\sigma}\partial^{\rho}
+\eta_{\nu\alpha}\varepsilon_{\mu\beta\rho\sigma}n^{\sigma}\partial^{\rho})
+(\alpha\leftrightarrow\beta),\nonumber\\
\mathcal{P}^{(\varepsilon,2)}_{\mu\nu,\alpha\beta}&=&
(n_{\mu}n_{\alpha}\varepsilon_{\nu\beta\rho\sigma}n^{\sigma}\partial^{\rho}
+n_{\nu}n_{\alpha}\varepsilon_{\mu\beta\rho\sigma}n^{\sigma}\partial^{\rho})
+(\alpha\leftrightarrow\beta),\\
\mathcal{P}^{(\varepsilon,3)}_{\mu\nu,\alpha\beta}&=&
(\varepsilon_{\nu\beta\rho\sigma}n^{\sigma}\partial^{\rho}\partial_{\mu}\partial_{\alpha}
+\varepsilon_{\mu\beta\rho\sigma}n^{\sigma}\partial^{\rho}\partial_{\nu}\partial_{\alpha})
+(\alpha\leftrightarrow\beta),\nonumber\\
\mathcal{P}^{(\varepsilon,4)}_{\mu\nu,\alpha\beta}&=&
\bigl(\varepsilon_{\nu\beta\rho\sigma}n^{\sigma}\partial^{\rho}(n_{\mu}\partial_{\alpha}+n_{\alpha}\partial_{\mu})
+\varepsilon_{\mu\beta\rho\sigma}n^{\sigma}\partial^{\rho}(n_{\nu}\partial_{\alpha}+n_{\alpha}\partial_{\nu})\bigr)
+(\alpha\leftrightarrow\beta).\nonumber
\end{eqnarray}
Using the above operators, the kinetic operator of $h_{\mu\nu}$ can be written as
\begin{eqnarray}
\label{sec32-lag-op-k}
\mathcal{D}_{\mu\nu,\alpha\beta}&=&\sum_{i=1}^{5}a_{i}\mathcal{P}^{(k,i)}_{\mu\nu,\alpha\beta}
+\sum_{i=1}^{3}b_{i}\mathcal{P}^{(n,i)}_{\mu\nu,\alpha\beta}
+\sum_{i=1}^{6}f_{i}\mathcal{P}^{(nk,i)}_{\mu\nu,\alpha\beta}
+\sum_{i=1}^{4}x_{i}\mathcal{P}^{(\varepsilon,i)}_{\mu\nu,\alpha\beta}.
\end{eqnarray}
The coefficients of this operator can be obtained from Eqs.~(\ref{sec32-lag-4})-(\ref{sec32-lag-3}). The coefficients $a_{i}$ are
\begin{eqnarray}
\label{sec32-lag-op-k-co-a}
a_{1}&=&-\frac{2}{9v^2}\bigl(\square+(n\cdot\partial)^2\bigr)\square
-\frac{1}{6v}\bigl(5\square+6(n\cdot\partial)^2\bigr)+\frac{1}{3},\nonumber\\
a_{2}&=&\frac{1}{4v^2}\bigl(\square+(n\cdot\partial)^2\bigr)\square
+\frac{1}{4v}\bigl(3\square+4(n\cdot\partial)^2\bigr)-\frac{1}{2},\\
a_{3}&=&\frac{1}{6v^2}\square+\frac{1}{v},~
a_{4}=-\frac{1}{8v^2}\bigl(2\square+(n\cdot\partial)^2\bigr)-\frac{3}{4v},~a_{5}=\frac{1}{2v^2},\nonumber
\end{eqnarray}
where $\square=\eta^{\mu\nu}\partial_{\mu}\partial_{\nu}$ is the Laplace operator, and $n\cdot\partial=\eta^{\mu\nu}n_{\mu}\partial_{\nu}$. The coefficients $b_{i}$ are
\begin{eqnarray}
\label{sec32-lag-op-k-co-b}
b_{1}&=&-\frac{1}{18v^2}\bigl(2\square-(n\cdot\partial)^2\bigr)\square
-\frac{5}{6v}\square+\frac{1}{3},\\
b_{2}&=&\frac{1}{8v^2}(\square+4v)\square,~
b_{3}=\frac{1}{9v^2}\bigl(4\square+(n\cdot\partial)^2\bigr)\square
+\frac{1}{v}\square+4.\nonumber
\end{eqnarray}
The coefficients $f_{i}$ are
\begin{eqnarray}
\label{sec32-lag-op-k-co-f}
f_{1}&=&-\frac{1}{6v^2}\square+\frac{1}{v},~f_{2}=-\frac{1}{8v^2}\square,~
f_{3}=\frac{1}{6v^2}(\square+6v)(n\cdot\partial),\\
f_{4}&=&-\frac{1}{6v^2}(n\cdot\partial)\square,~f_{5}=\frac{1}{4v^2}(n\cdot\partial),
~f_{6}=-\frac{1}{8v^2}(\square+6v)(n\cdot\partial),\nonumber
\end{eqnarray}
and the coefficients $x_{i}$ are
\begin{eqnarray}
\label{sec32-lag-op-k-co-x}
x_{1}&=&-\frac{1}{4v^{\frac{3}{2}}}(n\cdot\partial)^2,~x_{2}=-\frac{1}{4v^{\frac{3}{2}}}\square,
~x_{3}=0,~x_{4}=\frac{1}{4v^{\frac{3}{2}}}(n\cdot\partial).
\end{eqnarray}
In Eqs.~(\ref{sec32-lag-4})-(\ref{sec32-lag-3}), we have given the expressions for the quadratic part of the Lagrangian. The interactive part can also be derived. If we use $V_1$ and $V_3$ stand for the number of the vertex of one and three derivatives respectively. Then similar to the discussion in the 3D case in subsection~\ref{sec:2.2}, we can obtain the superficial degree of divergence
\begin{eqnarray}
\label{sec32-ym-dod}
\omega=4-3V_{1}-2V_{2}-V_{3}.
\end{eqnarray}
We see that $\omega$ is suppressed by the numbers of vertex, so the 4D case is still renormalizable by power counting. However, compared to the 3D case in Eq.~(\ref{sec2-ym-lag-3d-dod-l-1}), $\omega$ in Eq.~(\ref{sec32-ym-dod}) is not further suppressed by the number of loops. Moreover, the renormalizability by power counting does not ensure that the theory is renormalizable. An alternative approach to analyze the effective potential is using the functional renormalization group~\cite{Wetterich:1992yh,Morris:1993qb,Berges:2000ew}, which does not require the theory is renormalizable in the conventional meaning. The functional renormalization group has been used to analyzing the gravity theory, in which the asymptotic safety property is expected~\cite{Reuter:1996cp,Lauscher:2001ya,Reuter:2012id}.

In order to compute the one-loop effective action, we need to compute the determinant
\begin{eqnarray}
\label{sec32-ym-det-ten}
\bigl[\mathrm{pdet}(\mathcal{D}_{\mu\nu,\alpha\beta})\bigr]^{-\frac{1}{2}}.
\end{eqnarray}
Here $\mathrm{pdet}$ means that we only consider the nonzero eigenvalues, because $\mathcal{D}_{\mu\nu,\alpha\beta}$ in Eq.~(\ref{sec32-lag-op-k}) has one zero eigenvalue. This zero eigenvalue contributes an overall volume factor, which is hence can be removed~\cite{Coleman:1985,Dunne:2007rt}. After computing the determinant of the tensor sector by considering the nonzero eigenvalues, we obtain the determinant
\begin{eqnarray}
\label{sec32-ym-det}
\mathrm{det}\biggl(-\frac{1}{96}W_{1,1}\bigl(1+\frac{W_{1,2}}{W_{1,1}}\bigr)
\cdot{W}_{2,1}\bigl(1+\frac{W_{2,2}}{W_{2,1}}\bigr)\cdot{W}_{3,1}\bigl(1+\frac{W_{3,2}}{W_{3,1}}\bigr)\biggr).
\end{eqnarray}
In the above, $W_{1,1}$ and $W_{1,2}$ are
\begin{eqnarray}
\label{sec32-ym-det-w-11}
W_{1,1}&=&\bigl(\frac{1}{v}\square+1\bigr)\bigl(\frac{1}{v}\square+2\bigr)^2,\\
\label{sec32-ym-det-w-12}
W_{1,2}&=&\frac{1}{2v^3}\bigl(\square^2+11v\square+14v^2\bigr)(n\cdot\partial)^2-\frac{1}{2v^3}\square(n\cdot\partial)^4.
\end{eqnarray}
$W_{2,1}$ and $W_{2,2}$ are
\begin{eqnarray}
\label{sec32-ym-det-w-21}
W_{2,1}&=&\bigl(\frac{1}{v}\square+b_{+}\bigr)^2\bigl(\frac{1}{v}\square-b_{-}\bigr)^2,\\
\label{sec32-ym-det-w-22}
W_{2,2}&=&\frac{2}{v^4}\bigl(\square^3+7v\square^2+10v^2\square-8v^3\bigr)(n\cdot\partial)^2\\
&+&\frac{1}{v^4}\bigl(\square^2+12v\square+16v^2\bigr)(n\cdot\partial)^4
+\frac{4}{v^3}\square(n\cdot\partial)^6,\nonumber
\end{eqnarray}
where $b_{+}$ and $b_{-}$ have been given in Eq.~(\ref{sec2-ym-lag-re-prop-co-1-fac-root}). $W_{3,1}$ and $W_{3,2}$ are
\begin{eqnarray}
\label{sec32-ym-det-w-31}
W_{3,1}&=&\frac{1}{v^4}\bigl(\square^2+8v\square-2v^2\bigr)\bigl(\square^2-2v^2\bigr)+12,\\
\label{sec32-ym-det-w-32}
W_{3,2}&=&\frac{4}{v^3}\bigl(3\square^2+4v\square-8v^2\bigr)(n\cdot\partial)^2\\
&-&\frac{2}{v^4}\bigl(\square^2+2v\square-12v^2\bigr)(n\cdot\partial)^4
-\frac{8}{v^3}(n\cdot\partial)^6+\frac{1}{v^4}(n\cdot\partial)^8.\nonumber
\end{eqnarray}
$W_{3,1}$ can also be factorized as
\begin{eqnarray}
\label{sec32-ym-det-w-31-fac}
W_{3,1}=\bigl(\frac{1}{v}\square+r_{1}\bigr)\bigl(\frac{1}{v}\square+r_{2}\bigr)
\bigl(\frac{1}{v}\square+z\bigr)\bigl(\frac{1}{v}\square+\bar{z}\bigr),
\end{eqnarray}
where
\begin{eqnarray}
\label{sec32-ym-det-w-31-fac-r}
r_{1}\approx {8.2210},~~r_{2}\approx {1.7144},
\end{eqnarray}
are numerical values of real roots, and
\begin{eqnarray}
\label{sec32-ym-det-w-31-fac-z}
z\approx {-0.96772+0.44578i},~~\bar{z}\approx {-0.96772-0.44578i},
\end{eqnarray}
are a pair of conjugate complex roots. From the determinant in Eq.~(\ref{sec32-ym-det}), the effective potential can be expressed as
\begin{eqnarray}
\label{sec32-ym-lag-4d-eff-tr}
V_{\mathrm{eff}}&=&\frac{3}{2\kappa^2}v^2
-\frac{i}{2}\mathrm{tr}\bigl(\mathrm{log}W_{1,1}+\mathrm{log}W_{2,1}+\mathrm{log}W_{3,1}\bigr)\\
&-&\frac{i}{2}\mathrm{tr}\biggl(\mathrm{log}\bigl(1+\frac{W_{1,2}}{W_{1,1}}\bigr)
+\mathrm{log}\bigl(1+\frac{W_{2,2}}{W_{2,1}}\bigr)+\mathrm{log}\bigl(1+\frac{W_{3,2}}{W_{3,1}}\bigr)\biggr)+\cdots.\nonumber
\end{eqnarray}
The operators in the second line of Eq.~(\ref{sec32-ym-lag-4d-eff-tr}) depend on the vector $n_{\mu}$. As discussed below Eq.~(\ref{sec32-u-nn}), $n_{\mu}$ can be eliminated by Eq.~(\ref{sec32-u-nn}) after taking the trace operation, so $V_{\mathrm{eff}}$ shall be independent of $n_{\mu}$. These operators can be computed as perturbative series, but it is difficult to obtain compact results. In this subsection, we shall only consider the contribution from the first line of Eq.~(\ref{sec32-ym-lag-4d-eff-tr}). In momentum space, the first line of Eq.~(\ref{sec32-ym-lag-4d-eff-tr}) can be computed as
\begin{eqnarray}
\label{sec32-ym-lag-4d-eff}
V_{\mathrm{eff}}&=&\frac{3}{2\kappa^2}v^2+\frac{1}{2}V^{(1)}_{\mathrm{eff}}
+\frac{1}{2}V^{(2)}_{\mathrm{eff}}+\frac{1}{2}V^{(3)}_{\mathrm{eff}}+\cdots.
\end{eqnarray}
In the above, the contributions from the $W_{1,1}$~and $W_{2,1}$ terms are
\begin{eqnarray}
\label{sec32-ym-lag-4d-eff-1}
V^{(1)}_{\mathrm{eff}}&=&\int\frac{d^4k_{\mathrm{E}}}{(2\pi)^4}\biggl[\mathrm{log}\bigl(\frac{1}{v}k^2_{\mathrm{E}}+1\bigr)+
2\mathrm{log}\bigl(\frac{1}{v}k^2_{\mathrm{E}}+2\bigr)\biggr],\\
\label{sec32-ym-lag-4d-eff-2}
V^{(2)}_{\mathrm{eff}}&=&\int\frac{d^4k_{\mathrm{E}}}{(2\pi)^4}\biggl[2\mathrm{log}\bigl(\frac{1}{v}k^2_{\mathrm{E}}+b_{+}\bigr)
+2\mathrm{log}\bigl\lvert\frac{1}{v}k^2_{\mathrm{E}}-b_{-}\bigr\rvert\biggr].
\end{eqnarray}
We see that the second term of Eq.~(\ref{sec32-ym-lag-4d-eff-1}) and the two terms of Eq.~(\ref{sec32-ym-lag-4d-eff-2}) coincide with the contributions~(\ref{sec2-ym-lag-3d-eff}) in 3D. The contribution from the $W_{3,1}$ term is
\begin{eqnarray}
\label{sec32-ym-lag-4d-eff-3}
V^{(3)}_{\mathrm{eff}}&=&\int\frac{d^4k_{\mathrm{E}}}{(2\pi)^4}\biggl[\mathrm{log}\bigl(\frac{1}{v}k^2_{\mathrm{E}}+r_{1}\bigr)+
\mathrm{log}\bigl(\frac{1}{v}k^2_{\mathrm{E}}+r_{2}\bigr)\biggr],\\
&+&\int\frac{d^4k_{\mathrm{E}}}{(2\pi)^4}\biggl[\mathrm{log}\bigl(\frac{1}{v}k^2_{\mathrm{E}}+z\bigr)
+\mathrm{log}\bigl(\frac{1}{v}k^2_{\mathrm{E}}+\bar{z}\bigr)\biggr].\nonumber
\end{eqnarray}
Using the dimensional regularization in appendix~\ref{dimreg}, we obtain the result
\begin{eqnarray}
\label{sec32-ym-lag-4d-eff-dr}
V_{\mathrm{eff}}&=&\frac{3}{2\kappa^2}v^2-\frac{c_1}{128\pi^2}v^2-\frac{c_2}{64\pi^2}v^2\mathrm{log}\frac{\mu^2}{v}
-\frac{c_2}{64\pi^2}v^2\bigl(\frac{1}{\epsilon}+\mathrm{log}4\pi-\gamma_{\mathrm{E}}\bigr),\\
c_1&=&s(1)+2s(2)+2s(b_{+})+2s(b_{-})+s(r_{1})+s(r_{2})+s(z)+s(\bar{z}),\nonumber\\
c_2&=&1+2\cdot{2^2}+2b_{+}^{2}+2b_{-}^{2}+r_{1}^{2}+r_{2}^{2}+{z}^2+\bar{z}^2=133,\nonumber
\end{eqnarray}
where $s(x)$ is the function
\begin{eqnarray}
\label{sec32-ym-lag-4d-eff-dr-s}
s(x)=x^2(3-2\mathrm{log}x),
\end{eqnarray}
and $\mu$ is the renormalization scale. The fourth term of Eq.~(\ref{sec32-ym-lag-4d-eff-dr}) can be subtracted in the minimal subtraction theme. In Eq.~(\ref{sec2-ym-lag-3d-eff-dg}) for the 3D case, we know that the tachyonic term $b_{-}$ vanishes. However, the tachyonic term $b_{-}$ in Eq.~(\ref{sec32-ym-lag-4d-eff-dr}) behaves like the other massive modes. We can define the running coupling constant $\bar{\kappa}$
\begin{eqnarray}
\label{sec32-ym-lag-4d-eff-dr-k}
\frac{3}{\bar{\kappa}^2}=\frac{\partial^2V_{\mathrm{eff}}}{\partial{v}^2}\big\vert_{v=\bar{\mu}^2}
=\frac{3}{{\kappa}^2}-\frac{c_1}{64\pi^2}+\frac{c_2}{64\pi^2}\bigl(2\mathrm{log}\frac{\bar{\mu}^2}{\mu^2}+3\bigr).
\end{eqnarray}
From this definition, we obtain the renormalization group equation for $\bar{\kappa}$
\begin{eqnarray}
\label{sec32-ym-lag-4d-eff-dr-k-rg}
\bar{\mu}\frac{\partial\bar{\kappa}}{\partial\bar{\mu}}
=-\frac{c_2}{96\pi^2}\bar{\kappa}^3=-\bigl(3\frac{11}{24}+\frac{1}{96}\bigr)\frac{\bar{\kappa}^3}{\pi^2}.
\end{eqnarray}
This equation is similar to the definition of the beta function in the pure $SU(2)$ Yang-Mills theory. However, the value of this beta function is about 3 times of that in the $SU(2)$ Yang-Mills theory. This difference could be attributed to the contributions from the quadratic constraint term as we have discussed below Eq.~(\ref{sec3-ym-pf-norm-trans-mdef-e}). Note that the above results are obtained assuming that we only consider the first line of Eq.~(\ref{sec32-ym-lag-4d-eff-tr}). It would be interesting to see how the beta function can be corrected by the second line of Eq.~(\ref{sec32-ym-lag-4d-eff-tr}). In terms of $\bar{\kappa}$, the effective potential can be rewritten as
\begin{eqnarray}
\label{sec32-ym-lag-4d-eff-dr-rg}
V_{\mathrm{eff}}=\frac{3}{2\bar{\kappa}^2}v^2+\frac{c_2}{64\pi^2}v^2\bigl(\mathrm{log}\frac{v}{\bar{\mu}^2}-\frac{3}{2}\bigr),
\end{eqnarray}
which has the minimum value at
\begin{eqnarray}
\label{sec32-ym-lag-4d-eff-dr-v}
\langle{v}\rangle=\bar{\mu}^2\mathrm{exp}\bigl(-\frac{96\pi^2}{133\bar{\kappa}^2}+1\bigr).
\end{eqnarray}
This is the typical dimensional transmutation phenomenon for classically scale invariant theories.

In the end of subsection~\ref{sec:2.2}, we have discussed the effectiveness of the one-loop computation in 3D. In 3D, we found that all interactive vertices are suppressed by the dimension-2 condensate $v$ with no exception. In 4D, the situation is a little more complicated. In 4D, we make the redefinitions $\tilde{h}_{\mu\nu}=\frac{1}{\kappa\sqrt{v}}{h}_{\mu\nu}$, then $\tilde{h}_{\mu\nu}$ has the canonical dimension. After this redefinition, the interactive vertices produced by Eqs.~(\ref{sec3-ym-lag-re-4}), (\ref{sec3-ym-lag-re-2}) and (\ref{sec3-ym-lag-re-3}) are all suppressed by $v$. The vertex of the lowest dimension is $\frac{\kappa}{\sqrt{v}}\tilde{h}\partial\tilde{h}\partial\tilde{h}$. These are similar to the cases in 3D. However, because the torsion is non-zero, a different point arises in the 4D case. The vertices, which only depend on $\kappa$ but are not suppressed by $v$, could be produced by the linear torsion term~(\ref{sec3-ym-lag-re-1}). Nevertheless, using the torsion in Eq.~(\ref{sec3-gf-res-g-sol-1}) and the metric in Eq.~(\ref{sec32-g-para-a}), this term is exactly zero. Therefore, all interactive vertices are also suppressed by $v$ in 4D. The effectiveness of one-loop computation is actually controlled by the effective coupling $\frac{\kappa}{\sqrt{v}}$.

In addition, we compare the above results with the magnetic instability discovered in~\cite{Nielsen:1978rm}. In~\cite{Nielsen:1978rm}, it was discovered that the constant Savvidy background~\cite{Savvidy:1977as} is unstable due to the appearance of tachyonic modes\footnote{In~\cite{Cho:2012pq}, it was proposed that this instability could be eliminated by considering the gauge invariant background and the parity of the modes.}. In a series of papers~\cite{Nielsen:1978nk,Ambjorn:1979xi}, the authors proposed that the instability can be removed by using space-time dependent backgrounds, which have the periodic structure and share similarities with a type II superconductor. In terms of the metric-like variables, we also find the tachyonic mode in Eq.~(\ref{sec32-ym-det-w-21}). However, the tachyonic mode appears in pair, and the term associated to the tachyon mode in the functional determinant appears to be completely square, which ensures that the operator is non-negative. So the one-loop effective potential is still stable. A question is that if space-time dependent backgrounds are also stable and can even give lower vacuum energy. The answer is probably negative. The reason is that the reformulated theory is a higher derivative theory. Higher derivative theories generally suffer from the Ostrogradski instability especially when the field configurations are space-time dependent~\cite{Woodard:2006nt}. In the situation here, the space-time dependent background may break the square structure of the tachyon mode in the functional determinant, and hence may yield imaginary contributions to the effective potential. However, the analysis of this subsection do not mean that the analysis of the Savvidy instability is invalid. The Savvidy background can be expressed by the gauge field $A^{a}_{\mu}$. The dimension-2 condensate is in terms of the frame field $e^{a}_{\mu}$, which is basically a non-local function of $A^{a}_{\mu}$ according to the decompositions in subsections~\ref{sec:2.1} and~\ref{sec:3.1}. In~\cite{Gubarev:2000eu}, the dimension-2 condensate is proposed to be expressed by a non-local operator. In~\cite{Nielsen:1978rm}, the authors also proposed that the stable vacuum could be reached after a cascade decay process, but the vacuum is characterized by non-local operators.

\subsection{Gauge Invariant Gluon Polarization}\label{sec:3.3}

In 4D, the gluon angular momentum operator are similar to that in Eqs.~(\ref{sec2.3-gp-def-gs}) and (\ref{sec2.3-gp-def-go}) in 3D. Using the variables in the subsection~\ref{sec:3.1}, we have
 \begin{eqnarray}
\label{sec33-gp-def-gs-t}
\kappa^2M_{gs}^{\mu\alpha\beta}=-\eta^{\mu\tau}
(\eta^{\alpha\theta}\eta^{\beta\gamma}-\eta^{\beta\theta}\eta^{\alpha\gamma})(\frac{1}{2}R^{\sigma}_{\hspace{1mm}\rho\tau\theta}
g^{\rho\lambda}\mathcal{E}_{\gamma\lambda\sigma}
+2\Sigma^{\rho}_{\tau\theta}g_{\rho\gamma}+\mathcal{E}_{\gamma\tau\theta}).
\end{eqnarray}
In the low momentum limit, $M_{gs}^{\mu\alpha\beta}$ shall be dominated by
 \begin{eqnarray}
\label{sec33-gp-def-gs-t-l}
\kappa^2M_{gs}^{\mu\alpha\beta}\approx -2\eta^{\mu\tau}\eta^{\alpha\rho}\eta^{\beta\sigma}\mathcal{E}_{\tau\rho\sigma}.
\end{eqnarray}
The orbital angular momentum $M_{go}^{\mu\alpha\beta}$ can also be expressed by space-time tensor fields. Many results are similar to the expression in subsection~\ref{sec:2.3}. Because the torsion~(\ref{sec3-gf-res-g-sol-2}) is not zero, $M_{go}^{\mu\alpha\beta}$ also includes the contribution from the torsion.

For the Maxwell field, it has been shown that the Laguerre-Gaussian laser modes have a well-defined angular momentum~\cite{Allen:1992zz,Barnett:94B}. The Laguerre-Gaussian laser modes are described by classical solutions of vacuum Maxwell equations. In the above, we have given the gauge invariant expressions of the $SU(2)$ gluon angular momentum. For another different expression, see~\cite{Guo:2013jia}. It is interesting to consider whether there exist classical solutions of the Lagrangian~(\ref{sec3-ym-lag-re}) which have well-defined angular momentum. The above expressions of $M_{gs}^{\mu\alpha\beta}$ and $M_{go}^{\mu\alpha\beta}$ provide appropriate variables to perform the computations.

In the end of subsection~\ref{sec:2.3}, we have discussed how to solve $e^{a}_{\mu}$ as a function of $A^{a}_{\mu}$. The results are that a solution based on the Taylor expansion can be found, and the solution is not unique. The four dimensional case is more complicated. In order to solve $e^{a}_{\mu}$ in terms of $A^{a}_{\mu}$, we need to solve two constraints~(\ref{sec3-ym-pf-tor-j}) and~(\ref{sec3-ym-pf-tor-n}). A similar solution based on the Taylor series can be obtained, although the solution have more complicated expressions than the case in 3D.

\section{Conclusions}\label{sec:4}

We employed new variables in terms of space-time tensor fields to recast $SU(2)$ Yang-Mills theories into gravity-like formulations, which are proposed as effective descriptions of the infrared region. The gravity-like formulations include the square of Riemann curvature, which are higher derivative theories. These higher derivative theories are shown to be renormalizable by power counting. The gauge invariant metric-like variables are then used to analyze the dimension two condensate.

In 3 dimensions, the one-loop effective potential is finite, and a nonzero dimension two condensate is obtained. In 4 dimensions, the one-loop effective potential provides the dimensional transmutation phenomenon, and the beta function of running coupling is also obtained, which is negative but with numerical difference with the conventional Yang-Mills theory. Although we find tachyonic modes both in 3 and 4 dimensions, these tachyonic modes do not induce instability of the effective potential because of the special structure of kinetic terms.

In subsections~\ref{sec:2.2} and~\ref{sec:3.2}, we discuss a new phenomenon in the reformulated theories. In Yang-Mills theories, the interaction vertices are measured by a unique coupling $\kappa$. In contrast, the interactions in the reformulated theories are controlled by effective couplings, which are functions of $\kappa$ and the dimension two condensate, and the dimension two condensate has the effect of suppressing the interaction vertices. In 3 dimensions, this effect has the interpretation that the higher loop expansions are actually strong coupling expansions.

In 4 dimensions, we also discuss the additional unit 4-vector in subsection~\ref{sec:3.2a}. The obtained scalar-vector sector of the Lagrangian has the property of duality between the dimension two condensate and the unit 4-vector, which provides a novel dual property of the $SU(2)$ Yang-Mills theory in 4 dimensions.

An interesting question is whether the $SU(3)$ Yang-Mills theories can also be reformulated into gravity-like formulations. In this regard, the methods used in~\cite{Campoleoni:2012hp,Fujisawa:2012dk,Fredenhagen:2014oua,Guo:2014pta} could be very relevant. For example, the $SU(3)$ Yang-Mills theory in 3 dimension can be straightforwardly reformulated by introducing the higher-spin field variables in~\cite{Guo:2014pta}.

\acknowledgments
This work was supported in part by Fondecyt~(Chile) grant 1100287 and by Project Basal under Contract No.~FB0821. We thank the tensor algebra package xAct~\cite{Martin-Garcia:2008} for help in the computations.

\appendix

\section{Conventions}\label{convention}

Because we have two metrics $\eta_{\mu\nu}$ and $g_{\mu\nu}$, the convention is that we raise up or down the tensor indices by $\eta_{\mu\nu}$ and its inverse $\eta^{\mu\nu}$. However, we claim several exceptions in this paper. $g_{\mu\nu}$ is an exception. $g^{\mu\nu}$ in section~\ref{sec:2} is its inverse or is its generalized inverse in section~\ref{sec:3}, but is not obtained by raising up indices by $\eta^{\mu\nu}$ from $g_{\mu\nu}$. $\gamma^{\mu\nu}$ and $\mathcal{E}^{\alpha\beta\gamma}$ in section~\ref{sec:3} are also exceptions. In most of cases, we give the summation manifestly to avoid confusions.

\section{Projection Operator of Propagator}\label{propagator}

In $D$ dimensional space-time, we define the following projection operators~\cite{Stelle:1976gc}
\begin{eqnarray}
\label{prop-1}
P^{(2)}_{\mu\nu,\alpha\beta}&=&\frac{1}{2}
\left(\theta_{\mu\alpha}\theta_{\nu\beta}+\theta_{\mu\beta}\theta_{\nu\alpha}-\frac{2}{D-1}\theta_{\mu\nu}\theta_{\alpha\beta}\right),\\
\label{prop-2}
P^{(1)}_{\mu\nu,\alpha\beta}&=&\frac{1}{2}
\left(\theta_{\mu\alpha}\omega_{\nu\beta}+\theta_{\mu\beta}\omega_{\nu\alpha}
+\theta_{\nu\alpha}\omega_{\mu\beta}+\theta_{\nu\beta}\omega_{\mu\alpha}\right),\\
\label{prop-3}
P^{(0,s)}_{\mu\nu,\alpha\beta}&=&\frac{1}{D-1}\theta_{\mu\nu}\theta_{\alpha\beta},
~~P^{(0,\omega)}_{\mu\nu,\alpha\beta}=\omega_{\mu\nu}\omega_{\alpha\beta},
\end{eqnarray}
and two additional operators
\begin{eqnarray}
\label{prop-4}
P^{(0,s\omega)}_{\mu\nu,\alpha\beta}=\frac{1}{\sqrt{D-1}}\theta_{\mu\nu}\omega_{\alpha\beta},
~~P^{(0,\omega{s})}_{\mu\nu,\alpha\beta}=\frac{1}{\sqrt{D-1}}\omega_{\mu\nu}\theta_{\alpha\beta},
\end{eqnarray}
where in the coordinate space
 \begin{eqnarray}
\label{prop-1-def-q}
\theta_{\mu\nu}=\eta_{\mu\nu}-\frac{\partial_{\mu}\partial_{\nu}}{\eta^{\rho\sigma}\partial_{\rho}\partial_{\sigma}},
~~\omega_{\mu\nu}=\frac{\partial_{\mu}\partial_{\nu}}{\eta^{\rho\sigma}\partial_{\rho}\partial_{\sigma}},
\end{eqnarray}
or in the momentum space
 \begin{eqnarray}
\label{prop-1-def-p}
\theta_{\mu\nu}=\eta_{\mu\nu}-\frac{k_{\mu}k_{\nu}}{\eta^{\rho\sigma}k_{\rho}k_{\sigma}},
~~\omega_{\mu\nu}=\frac{k_{\mu}k_{\nu}}{\eta^{\rho\sigma}k_{\rho}k_{\sigma}}.
\end{eqnarray}

\section{Calculation by Dimensional Regularization }\label{dimreg}

We calculate the integral in $n$ dimensions
\begin{eqnarray}
\label{dr-int-po}
\int\frac{d^n k_{\mathrm{E}}}{(2\pi)^3}\mathrm{log}\left(a{k}^2_{\mathrm{E}}+b\right)
=\frac{\pi^{\frac{n}{2}}}{(2\pi)^{n}\Gamma(\frac{n}{2})}a^{-\frac{n}{2}}\int_{0}^{\infty}
du~{u}^{\frac{n-2}{2}}\mathrm{log}(u+b),
\end{eqnarray}
where we suppose $a>0$ and $b>0$. If we define~\cite{Peskin:1995ev}
\begin{eqnarray}
\label{dr-int-po-f}
f(\alpha)=-\int_{0}^{\infty}
du~{u}^{\frac{n-2}{2}}\frac{1}{(u+b)^{\alpha}},
\end{eqnarray}
then we have
\begin{eqnarray}
\label{dr-int-po-f-pr}
f^{\prime}(0)=\int_{0}^{\infty}
du~{u}^{\frac{n-2}{2}}\mathrm{log}(u+b),~~f^{\prime}(\alpha)=\frac{df}{d\alpha}.
\end{eqnarray}
The integral $f(\alpha)$ can be computed to be
\begin{eqnarray}
\label{dr-int-po-f-re}
f(\alpha)&=&-\frac{2}{n}b^{\frac{n}{2}-\alpha}\biggl(\frac{u}{u+b}\biggr)^{\frac{n}{2}}
{_{2}F_{1}}\bigl(\frac{n}{2},1+\frac{n}{2}-\alpha;1+\frac{n}{2},\frac{u}{u+b}\bigr)\bigl\vert^{\infty}_{0}\nonumber\\
&=&-b^{\frac{n}{2}-\alpha}\frac{\Gamma(\frac{n}{2})\Gamma(\alpha-\frac{n}{2})}{\Gamma(\alpha)},
\end{eqnarray}
which yields
\begin{eqnarray}
\label{dr-int-po-f-pr-re}
f^{\prime}(0)=-b^{\frac{n}{2}}\Gamma(\frac{n}{2})\Gamma(-\frac{n}{2}).
\end{eqnarray}
Another type of integral is
\begin{eqnarray}
\label{dr-int-ne}
\int\frac{d^n k_{\mathrm{E}}}{(2\pi)^3}\mathrm{log}\lvert{a}{k}^2_{\mathrm{E}}-b\rvert
=\frac{\pi^{\frac{n}{2}}}{(2\pi)^{n}\Gamma(\frac{n}{2})}a^{-\frac{n}{2}}\int_{0}^{\infty}
du~{u}^{\frac{n-2}{2}}\mathrm{log}\lvert{u-b}\rvert.
\end{eqnarray}
The abstract sign can be removed by
\begin{eqnarray}
\label{dr-int-ne-dec}
\int_{0}^{\infty}
du~{u}^{\frac{n-2}{2}}\mathrm{log}\lvert{u-b}\rvert&=&\int_{0}^{b}
du~{u}^{\frac{n-2}{2}}\mathrm{log}(b-u)
+\int_{b}^{\infty}du~{u}^{\frac{n-2}{2}}\mathrm{log}(u-b).
\end{eqnarray}
If we define
\begin{eqnarray}
\label{dr-int-ne-f-1}
f_{1}(\alpha)&=&-\int_{0}^{b}
du~{u}^{\frac{n-2}{2}}\frac{1}{(b-u)^{\alpha}},\\
\label{dr-int-ne-f-2}
f_{2}(\alpha)&=&-\int_{b}^{\infty}
du~{u}^{\frac{n-2}{2}}\frac{1}{(u-b)^{\alpha}}=-\int_{0}^{\infty}
du~{(u+b)}^{\frac{n-2}{2}}\frac{1}{u^{\alpha}},
\end{eqnarray}
then we have
\begin{eqnarray}
\label{dr-int-ne-f-1-re}
f_{1}(\alpha)&=&-\frac{2}{n}u^{\frac{n}{2}}b^{-\alpha}
{_{2}F_{1}}\bigl(\frac{n}{2},\alpha;1+\frac{n}{2},\frac{u}{b}\bigr)\bigl\vert^{\infty}_{0}
=-b^{\frac{n}{2}-\alpha}\frac{\Gamma(\frac{n}{2})\Gamma(1-\alpha)}{\Gamma(1+\frac{n}{2}-\alpha)},\\
\label{dr-int-ne-f-2-re}
f_{2}(\alpha)&=&\frac{1}{\alpha-1}b^{\frac{n}{2}-\alpha}\biggl(\frac{u}{u+b}\biggr)^{1-\alpha}
{_{2}F_{1}}\bigl(1+\frac{n}{2}-\alpha,1-\alpha;2-\alpha,\frac{u}{u+b}\bigr)\bigl\vert^{\infty}_{0}\nonumber\\
&=&-b^{\frac{n}{2}-\alpha}\frac{\Gamma(1-\alpha)\Gamma(\alpha-\frac{n}{2})}{\Gamma(1-\frac{n}{2})}.
\end{eqnarray}
From the above, we further obtain
\begin{eqnarray}
\label{dr-int-ne-f-1-re-de}
f^{\prime}_{1}(0)&=&\frac{2b^{\frac{n}{2}}}{n}\bigl[\mathrm{log}(b)
-\gamma_{\mathrm{E}}-\psi^{(0)}(1+\frac{n}{2})\bigr],\\
\label{dr-int-ne-f-2-re-de}
f^{\prime}_{2}(0)&=&-\frac{2b^{\frac{n}{2}}}{n}\bigl[\mathrm{log}(b)
-\gamma_{\mathrm{E}}-\psi^{(0)}(-\frac{n}{2})\bigr],
\end{eqnarray}
where
\begin{eqnarray}
\label{dr-int-ne-f-1-pg}
\psi^{(0)}(z)=\frac{1}{\Gamma(z)}\frac{d\Gamma(z)}{dz}
\end{eqnarray}
is the poly gamma function and $\gamma_{\mathrm{E}}$ is the Euler gamma constant.

\section{Property of Projection Tensor}\label{append-0}

As we have mentioned in section~\ref{sec:3}, the tensors $(\delta^{\alpha}_{\beta}-n^{\alpha}_{\beta})$ and $n^{\alpha}_{\beta}$ behave like projection tensors. Using the definition~(\ref{sec3-gf-res-e-inv-n}), we can prove that they have the properties
\begin{eqnarray}
\label{ap0-pro-n}
n^{\alpha}_{\sigma}n^{\sigma}_{\beta}=n^{\alpha}_{\beta},
~~(\delta^{\alpha}_{\sigma}-n^{\alpha}_{\sigma})(\delta^{\sigma}_{\beta}-n^{\sigma}_{\beta})=(\delta^{\alpha}_{\beta}-n^{\alpha}_{\beta}),
~~n^{\alpha}_{\sigma}(\delta^{\sigma}_{\beta}-n^{\sigma}_{\beta})=0.
\end{eqnarray}
Acting on the metric $g_{\alpha\beta}$, they yield
\begin{eqnarray}
\label{ap0-pro-n-g}
n^{\sigma}_{\alpha}g_{\sigma\beta}=g_{\alpha\beta},~~(\delta^{\sigma}_{\alpha}-n^{\sigma}_{\alpha})g_{\sigma\beta}=0.
\end{eqnarray}
Similarly, acting on the generalized inverse metric $g^{\alpha\beta}$, they yield
\begin{eqnarray}
\label{ap0-pro-n-g-in}
n^{\alpha}_{\sigma}g^{\sigma\beta}=g^{\alpha\beta},~~(\delta^{\alpha}_{\sigma}-n^{\alpha}_{\sigma})g^{\sigma\beta}=0.
\end{eqnarray}
$(\delta^{\alpha}_{\beta}-n^{\alpha}_{\beta})$ and $n^{\alpha}_{\beta}$ can also be used to decompose tensors. For example, for the tensor $\Sigma^{\rho}_{\alpha\beta}$, we have the identity
\begin{eqnarray}
\label{ap0-pro-n-sig}
\Sigma^{\rho}_{\alpha\beta}
&=&n^{\sigma}_{\alpha}n^{\tau}_{\beta}n^{\rho}_{\theta}\Sigma^{\theta}_{\sigma\tau}
+n^{\sigma}_{\alpha}n^{\tau}_{\beta}(\delta^{\rho}_{\theta}-n^{\rho}_{\theta})\Sigma^{\theta}_{\sigma\tau}
+n^{\sigma}_{\alpha}(\delta^{\tau}_{\beta}-n^{\tau}_{\beta})n^{\rho}_{\theta}\Sigma^{\theta}_{\sigma\tau}\\
&+&n^{\sigma}_{\alpha}(\delta^{\tau}_{\beta}-n^{\tau}_{\beta})(\delta^{\rho}_{\theta}-n^{\rho}_{\theta})\Sigma^{\theta}_{\sigma\tau}
+(\delta^{\sigma}_{\alpha}-n^{\sigma}_{\alpha})n^{\tau}_{\beta}n^{\rho}_{\theta}\Sigma^{\theta}_{\sigma\tau}
+(\delta^{\sigma}_{\alpha}-n^{\sigma}_{\alpha})n^{\tau}_{\beta}(\delta^{\rho}_{\theta}-n^{\rho}_{\theta})\Sigma^{\theta}_{\sigma\tau}
\nonumber\\
&+&(\delta^{\sigma}_{\alpha}-n^{\sigma}_{\alpha})(\delta^{\tau}_{\beta}-n^{\tau}_{\beta})n^{\rho}_{\theta}\Sigma^{\theta}_{\sigma\tau}
+(\delta^{\sigma}_{\alpha}-n^{\sigma}_{\alpha})(\delta^{\tau}_{\beta}-n^{\tau}_{\beta})
(\delta^{\rho}_{\theta}-n^{\rho}_{\theta})\Sigma^{\theta}_{\sigma\tau},\nonumber
\end{eqnarray}
that is, $\Sigma^{\rho}_{\alpha\beta}$ has 8 different parts under the projection decomposition. The above discussions are useful to find solutions of connection with metric compatibility.

\section{Solution of Connection in 4 Dimensions}\label{append-1}

In this appendix, we give the procedure to solve Eq.~(\ref{sec3-gf-res-g}). We attempt to find a connection which satisfies
\begin{eqnarray}
\label{ap1-gf-res-g}
\partial_{\mu}g_{\alpha\beta}=\Gamma^{\rho}_{\mu\alpha}g_{\rho\beta}+\Gamma^{\rho}_{\mu\beta}g_{\rho\alpha}
\end{eqnarray}
From Eq.~(\ref{ap1-gf-res-g}), we have
\begin{eqnarray}
\label{ap1-gf-res-g-re}
\Gamma_{\sigma,\alpha\beta}=\Gamma^{\rho}_{(\alpha\beta)}g_{\rho\sigma}+\Sigma^{\rho}_{\alpha\sigma}g_{\rho\beta}
+\Sigma^{\rho}_{\beta\sigma}g_{\rho\alpha},
\end{eqnarray}
where $\Gamma^{\rho}_{(\alpha\beta)}=\frac{1}{2}(\Gamma^{\rho}_{\alpha\beta}+\Gamma^{\rho}_{\beta\alpha})$ is the symmetric part of the connection and $\Sigma^{\rho}_{\alpha\beta}=\frac{1}{2}(\Gamma^{\rho}_{\alpha\beta}-\Gamma^{\rho}_{\beta\alpha})$ is its antisymmetric part. $\Gamma_{\sigma,\alpha\beta}$ has been defined in Eq.~(\ref{sec3-gf-res-g-sol-2}). From Eq.~(\ref{ap1-gf-res-g-re}), we have
\begin{eqnarray}
\label{ap1-gf-res-g-re-cs}
\Gamma^{\rho}_{(\alpha\beta)}n_{\rho}^{\sigma}=g^{\sigma\rho}\Gamma_{\rho,\alpha\beta}
-g^{\sigma\tau}(\Sigma^{\rho}_{\alpha\tau}g_{\rho\beta}+\Sigma^{\rho}_{\beta\tau}g_{\rho\alpha}).
\end{eqnarray}
Because we have the decomposition
\begin{eqnarray}
\label{ap1-gf-con}
\Gamma^{\sigma}_{\alpha\beta}&=&\Gamma^{\sigma}_{(\alpha\beta)}+\Sigma^{\sigma}_{\alpha\beta}
=\Gamma^{\rho}_{(\alpha\beta)}n_{\rho}^{\sigma}+\Gamma^{\rho}_{(\alpha\beta)}(\delta_{\rho}^{\sigma}-n_{\rho}^{\sigma})
+\Sigma^{\sigma}_{\alpha\beta}\nonumber\\
&=&g^{\sigma\rho}\Gamma_{\rho,\alpha\beta}
-g^{\sigma\tau}(\Sigma^{\rho}_{\alpha\tau}g_{\rho\beta}+\Sigma^{\rho}_{\beta\tau}g_{\rho\alpha})
+\Gamma^{\rho}_{(\alpha\beta)}(\delta_{\rho}^{\sigma}-n_{\rho}^{\sigma})
+\Sigma^{\sigma}_{\alpha\beta}.
\end{eqnarray}
We have used Eq.~(\ref{ap1-gf-res-g-re-cs}) to replace the part $\Gamma^{\rho}_{(\alpha\beta)}n_{\rho}^{\sigma}$ of $\Gamma^{\sigma}_{\alpha\beta}$. Substituting Eq.~(\ref{ap1-gf-con}) into Eq.~(\ref{ap1-gf-res-g-re}), we find that Eq.~(\ref{ap1-gf-res-g-re}) is satisfied if
\begin{eqnarray}
\label{ap1-gf-tor}
(\delta^{\rho}_{\sigma}-n^{\rho}_{\sigma})\Gamma_{\rho,\alpha\beta}=
(\delta^{\tau}_{\sigma}-n^{\tau}_{\sigma})(\Sigma^{\rho}_{\alpha\tau}g_{\rho\beta}+\Sigma^{\rho}_{\beta\tau}g_{\rho\alpha}).
\end{eqnarray}
We can check that Eqs.~(\ref{ap1-gf-tor}) and~(\ref{ap1-gf-res-g-re-cs}) make Eq.~(\ref{ap1-gf-res-g}) satisfied. Multiplying the two sides of Eq.~(\ref{ap1-gf-tor}) by the projection tensor $(\delta^{\alpha}_{\theta}-n^{\alpha}_{\theta})$, we have
\begin{eqnarray}
\label{ap1-gf-tor-m}
(\delta^{\alpha}_{\theta}-n^{\alpha}_{\theta})(\delta^{\rho}_{\sigma}-n^{\rho}_{\sigma})\Gamma_{\rho,\alpha\beta}=
(\delta^{\alpha}_{\theta}-n^{\alpha}_{\theta})(\delta^{\tau}_{\sigma}-n^{\tau}_{\sigma})\Sigma^{\rho}_{\alpha\tau}g_{\rho\beta},
\end{eqnarray}
which means
\begin{eqnarray}
\label{ap1-gf-tor-p1}
(\delta^{\alpha}_{\theta}-n^{\alpha}_{\theta})(\delta^{\tau}_{\sigma}-n^{\tau}_{\sigma})\Sigma^{\rho}_{\alpha\tau}n_{\rho}^{\lambda}
=(\delta^{\alpha}_{\theta}-n^{\alpha}_{\theta})(\delta^{\rho}_{\sigma}-n^{\rho}_{\sigma})g^{\lambda\beta}\Gamma_{\rho,\alpha\beta},
\end{eqnarray}
From Eq.~(\ref{ap1-gf-tor}), we also have
\begin{eqnarray}
\label{ap1-gf-tor-p2-st}
(\delta^{\rho}_{\sigma}-n^{\rho}_{\sigma})n^{\theta}_{\alpha}n^{\tau}_{\beta}\Gamma_{\rho,\theta\tau}=
(\delta^{\tau}_{\sigma}-n^{\tau}_{\sigma})(n^{\theta}_{\alpha}\Sigma^{\rho}_{\theta\tau}g_{\rho\beta}
+n^{\theta}_{\beta}\Sigma^{\rho}_{\theta\tau}g_{\rho\alpha}).
\end{eqnarray}
We have the decomposition
\begin{eqnarray}
\label{ap1-gf-tor-p2-st1}
(\delta^{\tau}_{\sigma}-n^{\tau}_{\sigma})n^{\theta}_{\alpha}\Sigma^{\rho}_{\theta\tau}g_{\rho\beta}&=&
\frac{1}{2}(\delta^{\tau}_{\sigma}-n^{\tau}_{\sigma})(n^{\theta}_{\alpha}\Sigma^{\rho}_{\theta\tau}g_{\rho\beta}
+n^{\theta}_{\beta}\Sigma^{\rho}_{\theta\tau}g_{\rho\alpha})\\
&+&\frac{1}{2}(\delta^{\tau}_{\sigma}-n^{\tau}_{\sigma})
(n^{\theta}_{\alpha}\Sigma^{\rho}_{\theta\tau}g_{\rho\beta}
-n^{\theta}_{\beta}\Sigma^{\rho}_{\theta\tau}g_{\rho\alpha}).\nonumber
\end{eqnarray}
Using Eq.~(\ref{ap1-gf-tor-p2-st}) to replace the symmetrical part of Eq.~(\ref{ap1-gf-tor-p2-st1}), then from Eq.~(\ref{ap1-gf-tor-p2-st1}), we can obtain
\begin{eqnarray}
\label{ap1-gf-tor-p2}
(\delta^{\tau}_{\sigma}-n^{\tau}_{\sigma})n^{\theta}_{\alpha}\Sigma^{\rho}_{\theta\tau}n_{\rho}^{\beta}&=&
\frac{1}{2}(\delta^{\rho}_{\sigma}-n^{\rho}_{\sigma})n^{\theta}_{\alpha}g^{\beta\tau}\Gamma_{\rho,\theta\tau}\\
&+&\frac{1}{2}(\delta^{\tau}_{\sigma}-n^{\tau}_{\sigma})(n^{\theta}_{\alpha}\Sigma^{\rho}_{\theta\tau}n_{\rho}^{\beta}
-g^{\beta\theta}\Sigma^{\rho}_{\theta\tau}g_{\rho\alpha}).\nonumber
\end{eqnarray}
Eqs.~(\ref{ap1-gf-tor-p1}) and (\ref{ap1-gf-tor-p2}) together yield
\begin{eqnarray}
\label{ap1-gf-tor-dec-b}
(\delta^{\tau}_{\mu}-n^{\tau}_{\mu})\Sigma^{\rho}_{\alpha\tau}n^{\sigma}_{\rho}&=&
(\delta^{\tau}_{\mu}-n^{\tau}_{\mu})(\delta^{\theta}_{\alpha}-n^{\theta}_{\alpha})\Sigma^{\rho}_{\theta\tau}n^{\sigma}_{\rho}
+(\delta^{\tau}_{\mu}-n^{\tau}_{\mu})n^{\theta}_{\alpha}\Sigma^{\rho}_{\theta\tau}n^{\sigma}_{\rho}\nonumber\\
&=&(\delta^{\theta}_{\alpha}-n^{\theta}_{\alpha})(\delta^{\rho}_{\mu}-n^{\rho}_{\mu})g^{\sigma\tau}\Gamma_{\rho,\theta\tau}
+\frac{1}{2}(\delta^{\rho}_{\mu}-n^{\rho}_{\mu})n^{\theta}_{\alpha}g^{\sigma\tau}\Gamma_{\rho,\theta\tau}\nonumber\\
&+&\frac{1}{2}(\delta^{\tau}_{\mu}-n^{\tau}_{\mu})(n^{\theta}_{\alpha}\Sigma^{\rho}_{\theta\tau}n_{\rho}^{\sigma}
-g^{\sigma\theta}\Sigma^{\rho}_{\theta\tau}g_{\rho\alpha}),
\end{eqnarray}
which further yields
\begin{eqnarray}
\label{ap1-gf-tor-dec-a}
\Sigma^{\sigma}_{\alpha\mu}
&=&
(\delta^{\tau}_{\mu}-n^{\tau}_{\mu})\Sigma^{\rho}_{\alpha\tau}n^{\sigma}_{\rho}
+(\delta^{\tau}_{\mu}-n^{\tau}_{\mu})\Sigma^{\rho}_{\alpha\tau}(\delta^{\sigma}_{\rho}-n^{\sigma}_{\rho})
+n^{\tau}_{\mu}\Sigma^{\sigma}_{\alpha\tau}\nonumber\\
&=&(\delta^{\theta}_{\alpha}-n^{\theta}_{\alpha})(\delta^{\rho}_{\mu}-n^{\rho}_{\mu})g^{\sigma\tau}\Gamma_{\rho,\theta\tau}
+\frac{1}{2}(\delta^{\rho}_{\mu}-n^{\rho}_{\mu})n^{\theta}_{\alpha}g^{\sigma\tau}\Gamma_{\rho,\theta\tau}\\
&+&\frac{1}{2}(\delta^{\tau}_{\mu}-n^{\tau}_{\mu})(n^{\theta}_{\alpha}\Sigma^{\rho}_{\theta\tau}n_{\rho}^{\sigma}
-g^{\sigma\theta}\Sigma^{\rho}_{\theta\tau}g_{\rho\alpha})
+(\delta^{\tau}_{\mu}-n^{\tau}_{\mu})\Sigma^{\rho}_{\alpha\tau}(\delta^{\sigma}_{\rho}-n^{\sigma}_{\rho})
+n^{\tau}_{\mu}\Sigma^{\sigma}_{\alpha\tau}.\nonumber
\end{eqnarray}
Using the expression~(\ref{ap1-gf-tor-dec-a}), and the identity
\begin{eqnarray}
\label{ap1-gf-tor-id}
(\delta^{\rho}_{\sigma}-n^{\rho}_{\sigma})(\delta^{\theta}_{\alpha}-n^{\theta}_{\alpha})
(n^{\tau}_{\beta}-n^{\tau}_{\beta})\Gamma_{\rho,\theta\tau}=0,
\end{eqnarray}
we can prove that Eq.~(\ref{ap1-gf-tor}) is satisfied. $\Sigma^{\sigma}_{\alpha\mu}$ is antisymmetric about $\alpha$ and $\mu$, but the expression of Eq.~(\ref{ap1-gf-tor-dec-a}) is not. However, multiplying the two sides of Eq.~(\ref{ap1-gf-tor-dec-a}) by $(\delta^{\tau}_{\mu}-n^{\tau}_{\mu})$ and $n^{\theta}_{\alpha}$, we have
\begin{eqnarray}
\label{ap1-gf-tor-dec-a-1}
(\delta^{\tau}_{\mu}-n^{\tau}_{\mu})n^{\theta}_{\alpha}\Sigma^{\sigma}_{\theta\tau}
&=&\frac{1}{2}(\delta^{\rho}_{\mu}-n^{\rho}_{\mu})n^{\theta}_{\alpha}g^{\sigma\tau}\Gamma_{\rho,\theta\tau}
+(\delta^{\tau}_{\mu}-n^{\tau}_{\mu})n^{\theta}_{\alpha}\Sigma^{\rho}_{\theta\tau}(\delta^{\sigma}_{\rho}-n^{\sigma}_{\rho})\\
&+&\frac{1}{2}(\delta^{\tau}_{\mu}-n^{\tau}_{\mu})(n^{\theta}_{\alpha}\Sigma^{\rho}_{\theta\tau}n_{\rho}^{\sigma}
-g^{\sigma\theta}\Sigma^{\rho}_{\theta\tau}g_{\rho\alpha}).\nonumber
\end{eqnarray}
Similarly, we have
\begin{eqnarray}
\label{ap1-gf-tor-dec-a-2}
(\delta^{\tau}_{\mu}-n^{\tau}_{\mu})(\delta^{\theta}_{\alpha}-n^{\theta}_{\alpha})\Sigma^{\sigma}_{\theta\tau}
&=&(\delta^{\theta}_{\alpha}-n^{\theta}_{\alpha})(\delta^{\rho}_{\mu}-n^{\rho}_{\mu})g^{\sigma\tau}\Gamma_{\rho,\theta\tau}\\
&+&(\delta^{\tau}_{\mu}-n^{\tau}_{\mu})(\delta^{\theta}_{\alpha}-n^{\theta}_{\alpha})
\Sigma^{\rho}_{\theta\tau}(\delta^{\sigma}_{\rho}-n^{\sigma}_{\rho}).\nonumber
\end{eqnarray}
We have decomposition
\begin{eqnarray}
\label{ap1-gf-tor-dec-c}
\Sigma^{\sigma}_{\alpha\mu}=n^{\theta}_{\alpha}(\delta^{\tau}_{\mu}-n^{\tau}_{\mu})\Sigma^{\sigma}_{\theta\tau}+
(\delta^{\theta}_{\alpha}-n^{\theta}_{\alpha})(\delta^{\tau}_{\mu}-n^{\tau}_{\mu})\Sigma^{\sigma}_{\theta\tau}
+n^{\theta}_{\alpha}n^{\tau}_{\mu}\Sigma^{\sigma}_{\theta\tau}
+(\delta^{\theta}_{\alpha}-n^{\theta}_{\alpha})n^{\tau}_{\mu}\Sigma^{\sigma}_{\theta\tau},
\end{eqnarray}
whose right side is transparently antisymmetric about $\alpha$ and $\mu$. Substituting Eqs.~(\ref{ap1-gf-tor-dec-a-1}) and (\ref{ap1-gf-tor-dec-a-2}) into the above decomposition, we have
\begin{eqnarray}
\label{ap1-gf-tor-dec-c-re}
\Sigma^{\sigma}_{\alpha\mu}&=&\frac{1}{2}(\delta^{\tau}_{\mu}-n^{\tau}_{\mu})g^{\sigma\rho}\Gamma_{\tau,\rho\alpha}
-\frac{1}{2}(\delta^{\tau}_{\alpha}-n^{\tau}_{\alpha})g^{\sigma\rho}\Gamma_{\tau,\rho\mu}\\
&+&\frac{1}{2}(\delta^{\tau}_{\mu}-n^{\tau}_{\mu})
(n^{\theta}_{\alpha}\Sigma^{\rho}_{\theta\tau}n^{\sigma}_{\rho}-g^{\sigma\theta}\Sigma^{\rho}_{\theta\tau}g_{\rho\alpha})
-\frac{1}{2}(\delta^{\tau}_{\alpha}-n^{\tau}_{\alpha})
(n^{\theta}_{\mu}\Sigma^{\rho}_{\theta\tau}n^{\sigma}_{\rho}-g^{\sigma\theta}\Sigma^{\rho}_{\theta\tau}g_{\rho\mu})\nonumber\\
&+&(\delta^{\tau}_{\mu}-n^{\tau}_{\mu})n^{\theta}_{\alpha}\Sigma^{\rho}_{\theta\tau}(\delta^{\sigma}_{\rho}-n^{\sigma}_{\rho})
-(\delta^{\tau}_{\alpha}-n^{\tau}_{\alpha})n^{\theta}_{\mu}\Sigma^{\rho}_{\theta\tau}
(\delta^{\sigma}_{\rho}-n^{\sigma}_{\rho})\nonumber\\
&+&(\delta^{\tau}_{\mu}-n^{\tau}_{\mu})(\delta^{\theta}_{\alpha}-n^{\theta}_{\alpha})
\Sigma^{\rho}_{\theta\tau}(\delta^{\sigma}_{\rho}-n^{\sigma}_{\rho})
+n^{\tau}_{\mu}n^{\theta}_{\alpha}\Sigma^{\sigma}_{\theta\tau}.\nonumber
\end{eqnarray}
The right side of this expression is antisymmetric about $\alpha$ and $\mu$, and we can check that $\Sigma^{\sigma}_{\alpha\mu}$ given by Eq.~(\ref{ap1-gf-tor-dec-c-re}) satisfies Eq.~(\ref{ap1-gf-tor}).

In the above, we have obtained the connection which satisfies the metric compatibility condition~(\ref{ap1-gf-res-g}). Now we further require that the connection satisfies
\begin{eqnarray}
\label{ap1-gf-res-g-inv}
\partial_{\mu}g^{\alpha\beta}=-\Gamma^{\beta}_{\mu\rho}g^{\rho\alpha}-\Gamma^{\alpha}_{\mu\rho}g^{\rho\beta}.
\end{eqnarray}
If $g_{\alpha\beta}$ is invertible, then we can derive Eq.~(\ref{ap1-gf-res-g-inv}) from Eq.~(\ref{ap1-gf-res-g}). However, $g_{\alpha\beta}$ is not invertible here. We also need to solve Eq.~(\ref{ap1-gf-res-g-inv}). Using the projection tensor, from Eq.~(\ref{ap1-gf-res-g-inv}), we have
\begin{eqnarray}
\label{ap1-gf-ig-st}
(\delta^{\alpha}_{\sigma}-n^{\alpha}_{\sigma})\Gamma^{\sigma}_{\mu\rho}g^{\rho\beta}=
-(\delta^{\alpha}_{\sigma}-n^{\alpha}_{\sigma})\partial_{\mu}g^{\sigma\beta},
\end{eqnarray}
which further yields
\begin{eqnarray}
\label{ap1-gf-ig-st-1}
(\delta^{\alpha}_{\sigma}-n^{\alpha}_{\sigma})\Gamma^{\sigma}_{\mu\rho}n^{\rho}_{\beta}=
-n^{\sigma}_{\beta}\partial_{\mu}n^{\alpha}_{\sigma},
\end{eqnarray}
which is equivalent to
\begin{eqnarray}
\label{ap1-gf-ig-st-1-a}
(\delta^{\alpha}_{\sigma}-n^{\alpha}_{\sigma})\Gamma^{\sigma}_{(\mu\rho)}n^{\rho}_{\beta}=
-(\delta^{\alpha}_{\sigma}-n^{\alpha}_{\sigma})\Sigma^{\sigma}_{\mu\rho}n^{\rho}_{\beta}
-n^{\sigma}_{\beta}\partial_{\mu}n^{\alpha}_{\sigma}.
\end{eqnarray}
we can derive Eq.~(\ref{ap1-gf-res-g-inv}) from Eqs.~(\ref{ap1-gf-res-g}) and (\ref{ap1-gf-ig-st-1}). So Eq.~(\ref{ap1-gf-res-g-inv}) provides a new constraint~(\ref{ap1-gf-ig-st-1}) for the connection~(\ref{ap1-gf-con}). From Eqs.~(\ref{ap1-gf-ig-st-1}) and~(\ref{ap1-gf-ig-st-1-a}), we can obtain
\begin{eqnarray}
\label{ap1-gf-ig-st-2a}
n^{\tau}_{\alpha}n^{\theta}_{\beta}\Gamma^{\rho}_{(\tau\theta)}(\delta^{\sigma}_{\rho}-n^{\sigma}_{\rho})&=&
-\frac{1}{2}n^{\tau}_{\alpha}n^{\theta}_{\beta}(\partial_{\tau}n^{\sigma}_{\theta}+\partial_{\theta}n^{\sigma}_{\tau}),\\
\label{ap1-gf-ig-st-2b}
n^{\tau}_{\alpha}n^{\theta}_{\beta}\Sigma^{\rho}_{\tau\theta}(\delta^{\sigma}_{\rho}-n^{\sigma}_{\rho})&=&
-\frac{1}{2}n^{\tau}_{\alpha}n^{\theta}_{\beta}(\partial_{\tau}n^{\sigma}_{\theta}-\partial_{\theta}n^{\sigma}_{\tau}),\\
\label{ap1-gf-ig-st-2c}
(\delta^{\tau}_{\alpha}-n^{\tau}_{\alpha})n^{\theta}_{\beta}\Gamma^{\rho}_{(\tau\theta)}(\delta^{\sigma}_{\rho}-n^{\sigma}_{\rho})
&=&
-(\delta^{\tau}_{\alpha}-n^{\tau}_{\alpha})n^{\theta}_{\beta}\Sigma^{\rho}_{\tau\theta}(\delta^{\sigma}_{\rho}-n^{\sigma}_{\rho})
-(\delta^{\tau}_{\alpha}-n^{\tau}_{\alpha})n^{\theta}_{\beta}\partial_{\tau}n^{\sigma}_{\theta}.
\end{eqnarray}
Inversely, Eqs.~(\ref{ap1-gf-ig-st-2a}), (\ref{ap1-gf-ig-st-2b}) and (\ref{ap1-gf-ig-st-2c}) make Eq.~(\ref{ap1-gf-ig-st-1}) satisfied. We have
\begin{eqnarray}
\label{ap1-gf-con-st-1-2c}
\Gamma^{\rho}_{(\alpha\mu)}(\delta^{\sigma}_{\rho}-n^{\sigma}_{\rho})&=&
n^{\tau}_{\mu}\Gamma^{\rho}_{(\alpha\tau)}(\delta^{\sigma}_{\rho}-n^{\sigma}_{\rho})
+n^{\theta}_{\alpha}(\delta^{\tau}_{\mu}-n^{\tau}_{\mu})\Gamma^{\rho}_{(\theta\tau)}
(\delta^{\sigma}_{\rho}-n^{\sigma}_{\rho})\nonumber\\
&+&(\delta^{\theta}_{\alpha}-n^{\theta}_{\alpha})(\delta^{\tau}_{\mu}-n^{\tau}_{\mu})
\Gamma^{\rho}_{\theta\tau}(\delta^{\sigma}_{\rho}-n^{\sigma}_{\rho})\nonumber\\
&=&-n^{\rho}_{\mu}\partial_{\alpha}n^{\beta}_{\rho}-(\delta^{\tau}_{\mu}-n^{\tau}_{\mu})
n^{\rho}_{\alpha}\partial_{\tau}n^{\beta}_{\rho}
-n^{\theta}_{\mu}\Sigma^{\rho}_{\alpha\theta}(\delta^{\sigma}_{\rho}-n^{\sigma}_{\rho})\\
&-&(\delta^{\tau}_{\mu}-n^{\tau}_{\mu})n^{\theta}_{\alpha}\Sigma^{\rho}_{\tau\theta}(\delta^{\sigma}_{\rho}-n^{\sigma}_{\rho})
+(\delta^{\theta}_{\alpha}-n^{\theta}_{\alpha})(\delta^{\tau}_{\mu}-n^{\tau}_{\mu})
\Gamma^{\rho}_{(\theta\tau)}(\delta^{\sigma}_{\rho}-n^{\sigma}_{\rho}),\nonumber
\end{eqnarray}
where we have used Eqs.~(\ref{ap1-gf-ig-st-1-a}) and (\ref{ap1-gf-ig-st-2c}). Using Eq.~(\ref{ap1-gf-ig-st-2b}), $n^{\theta}_{\mu}\Sigma^{\rho}_{\alpha\theta}(\delta^{\sigma}_{\rho}-n^{\sigma}_{\rho})$ and $n^{\tau}_{\alpha}n^{\theta}_{\mu}\Sigma^{\rho}_{\alpha\theta}$ can be decomposed as
\begin{eqnarray}
\label{ap1-gf-con-st-1-2b}
n^{\theta}_{\mu}\Sigma^{\rho}_{\alpha\theta}(\delta^{\sigma}_{\rho}-n^{\sigma}_{\rho})&=&
n^{\tau}_{\alpha}n^{\theta}_{\mu}\Sigma^{\rho}_{\tau\theta}(\delta^{\sigma}_{\rho}-n^{\sigma}_{\rho})
+(\delta^{\tau}_{\alpha}-n^{\tau}_{\alpha})n^{\theta}_{\mu}\Sigma^{\rho}_{\tau\theta}
(\delta^{\sigma}_{\rho}-n^{\sigma}_{\rho})\nonumber\\
&=&-\frac{1}{2}n^{\tau}_{\alpha}n^{\theta}_{\mu}(\partial_{\tau}n^{\sigma}_{\theta}-\partial_{\tau}n^{\sigma}_{\theta})
+(\delta^{\tau}_{\alpha}-n^{\tau}_{\alpha})n^{\theta}_{\mu}\Sigma^{\rho}_{\tau\theta}(\delta^{\sigma}_{\rho}-n^{\sigma}_{\rho}),\\
\label{ap1-gf-tor-st-1-2b}
n^{\tau}_{\alpha}n^{\theta}_{\mu}\Sigma^{\sigma}_{\tau\theta}&=&
n^{\tau}_{\alpha}n^{\theta}_{\mu}\Sigma^{\rho}_{\tau\theta}(\delta^{\sigma}_{\rho}-n^{\sigma}_{\rho})
+n^{\tau}_{\alpha}n^{\theta}_{\mu}\Sigma^{\rho}_{\tau\theta}n^{\sigma}_{\rho}\nonumber\\
&=&-\frac{1}{2}n^{\tau}_{\alpha}n^{\theta}_{\mu}(\partial_{\tau}n^{\sigma}_{\theta}-\partial_{\tau}n^{\sigma}_{\theta})
+n^{\tau}_{\alpha}n^{\theta}_{\mu}\Sigma^{\rho}_{\tau\theta}n^{\sigma}_{\rho}.
\end{eqnarray}
Substituting Eq.~(\ref{ap1-gf-con-st-1-2b}) into (\ref{ap1-gf-con-st-1-2c}), we have
\begin{eqnarray}
\label{ap1-gf-con-st-1-2cb}
\Gamma^{\rho}_{(\alpha\mu)}(\delta^{\sigma}_{\rho}-n^{\sigma}_{\rho})
&=&-(n^{\rho}_{\mu}\partial_{\alpha}n^{\sigma}_{\rho}+n^{\rho}_{\alpha}\partial_{\mu}n^{\sigma}_{\rho})
+\frac{1}{2}n^{\tau}_{\alpha}n^{\theta}_{\mu}(\partial_{\tau}n^{\sigma}_{\theta}+\partial_{\theta}n^{\sigma}_{\tau})\\
&-&(\delta^{\tau}_{\alpha}-n^{\tau}_{\alpha})n^{\theta}_{\mu}\Sigma^{\rho}_{\tau\theta}(\delta^{\sigma}_{\rho}-n^{\sigma}_{\rho})
-(\delta^{\tau}_{\mu}-n^{\tau}_{\mu})n^{\theta}_{\alpha}\Sigma^{\rho}_{\tau\theta}
(\delta^{\sigma}_{\rho}-n^{\sigma}_{\rho})\nonumber\\
&+&(\delta^{\theta}_{\alpha}-n^{\theta}_{\alpha})(\delta^{\tau}_{\mu}-n^{\tau}_{\mu})
\Gamma^{\rho}_{(\theta\tau)}(\delta^{\sigma}_{\rho}-n^{\sigma}_{\rho}).\nonumber
\end{eqnarray}
Substituting Eq.~(\ref{ap1-gf-tor-st-1-2b}) into (\ref{ap1-gf-tor-dec-c-re}), we have
\begin{eqnarray}
\label{ap1-gf-tor-dec-c-re-b}
\Sigma^{\sigma}_{\alpha\mu}&=&\frac{1}{2}(\delta^{\tau}_{\mu}-n^{\tau}_{\mu})g^{\sigma\rho}\Gamma_{\tau,\rho\alpha}
-\frac{1}{2}(\delta^{\tau}_{\alpha}-n^{\tau}_{\alpha})g^{\sigma\rho}\Gamma_{\tau,\rho\mu}
-\frac{1}{2}n^{\tau}_{\alpha}n^{\theta}_{\mu}(\partial_{\tau}n^{\sigma}_{\theta}-\partial_{\theta}n^{\sigma}_{\tau})\\
&+&\frac{1}{2}(\delta^{\tau}_{\mu}-n^{\tau}_{\mu})
(n^{\theta}_{\alpha}\Sigma^{\rho}_{\theta\tau}n^{\sigma}_{\rho}-g^{\sigma\theta}\Sigma^{\rho}_{\theta\tau}g_{\rho\alpha})
-\frac{1}{2}(\delta^{\tau}_{\alpha}-n^{\tau}_{\alpha})
(n^{\theta}_{\mu}\Sigma^{\rho}_{\theta\tau}n^{\sigma}_{\rho}-g^{\sigma\theta}\Sigma^{\rho}_{\theta\tau}g_{\rho\mu})\nonumber\\
&+&(\delta^{\tau}_{\mu}-n^{\tau}_{\mu})n^{\theta}_{\alpha}\Sigma^{\rho}_{\theta\tau}(\delta^{\sigma}_{\rho}-n^{\sigma}_{\rho})
-(\delta^{\tau}_{\alpha}-n^{\tau}_{\alpha})n^{\theta}_{\mu}\Sigma^{\rho}_{\theta\tau}
(\delta^{\sigma}_{\rho}-n^{\sigma}_{\rho})\nonumber\\
&+&(\delta^{\tau}_{\mu}-n^{\tau}_{\mu})(\delta^{\theta}_{\alpha}-n^{\theta}_{\alpha})
\Sigma^{\rho}_{\theta\tau}(\delta^{\sigma}_{\rho}-n^{\sigma}_{\rho})
+n^{\tau}_{\mu}n^{\theta}_{\alpha}\Sigma^{\rho}_{\theta\tau}n^{\sigma}_{\rho}.\nonumber
\end{eqnarray}
Substituting (\ref{ap1-gf-con-st-1-2cb}) and (\ref{ap1-gf-tor-dec-c-re-b}) into Eq.~(\ref{ap1-gf-con}), we then obtain the connection $\Gamma^{\sigma}_{\alpha\beta}$ which satisfies both Eqs.~(\ref{ap1-gf-res-g}) and (\ref{ap1-gf-res-g-inv}). The connection $\Gamma^{\sigma}_{\alpha\beta}$ with the expressions (\ref{ap1-gf-con-st-1-2cb}) and (\ref{ap1-gf-tor-dec-c-re-b}) is the most general solution of Eqs.~(\ref{ap1-gf-res-g}) and (\ref{ap1-gf-res-g-inv}). When $n^{\theta}_{\alpha}=\delta^{\theta}_{\alpha}$, this solution reduces to the conventional metric-compatible connection with torsion.

We give some comments on our above discussions. The basic method which we used to solve Eqs.~(\ref{ap1-gf-res-g}) and (\ref{ap1-gf-res-g-inv}) is: At first we use the projection tensor $n^{\theta}_{\alpha}$ and $(\delta^{\theta}_{\alpha}-n^{\theta}_{\alpha})$ to decompose the connection and the torsion, such as, the torsion is decomposed into 8 different parts in Eq.~(\ref{ap0-pro-n-sig}); Then some parts can be determined from Eqs.~(\ref{ap1-gf-res-g}) and (\ref{ap1-gf-res-g-inv}), and the unconstrained parts are left free. Using this method, we were able to find the most general solution of Eqs.~(\ref{ap1-gf-res-g}) and (\ref{ap1-gf-res-g-inv}). If we set the unconstrained part to be zero, then we obtain the special solutions in Eqs.~(\ref{sec3-gf-res-g-sol-0}) and (\ref{sec3-gf-res-g-sol-1}) in subsection~\ref{sec:3.1}.

There were relevant discussions to find the metric-compatible connection for a degenerate metric. For a different definition of the metric-compatibility condition and the solution of connection, see~\cite{Searight:2003vk}. For a review about the degenerate metric from the mathematical aspect, see~\cite{Shandra:2004}. For a recent discussion of the degenerate metric related to the Newton-Cartan geometry and the Galilean symmetry, see~\cite{Jensen:2014aia}.

\section{Derivation of Generalized Inverse Matrix by Cayley-Hamilton Theorem}\label{append-3}

For a $n\times{n}$ matrix $A$, the Cayley-Hamilton theorem asserts that $A$ satisfy the following matrix identity
\begin{eqnarray}
\label{ap3-ch-th}
A^{n}+e_{1}A^{n-1}+\cdots+e_{n}I=0,
\end{eqnarray}
in which $I$ is the $n\times{n}$ identity matrix. The coefficients $e_{m}$ are given as~\cite{Brown:1994}
\begin{eqnarray}
\label{ap3-ch-th-co}
e_{m}=\frac{(-1)^m}{m!}~\mathrm{det}
\begin{pmatrix}
\mathrm{tr}A& m-1&0&\cdots&\\
\mathrm{tr}A^2& \mathrm{tr}A&m-2&\cdots&\\
\vdots &\vdots & & &\vdots\\
\mathrm{tr}A^{m-1}&\mathrm{tr}A^{m-2} &\cdots &\cdots&1\\
\mathrm{tr}A^{m}&\mathrm{tr}A^{m-1} &\cdots &\cdots&\mathrm{tr}A
\end{pmatrix}.
\end{eqnarray}

In this appendix, we use the Cayley-Hamilton Theorem to derive the generalized inverse of $g_{\mu\nu}$. We first define the matrix
\begin{eqnarray}
\label{ap3-ch-gi-x}
X^{\alpha}_{\beta}=\eta^{\alpha\rho}g_{\rho\beta}.
\end{eqnarray}
We can define the matrix $S^{\alpha}_{\beta}$ as
\begin{eqnarray}
\label{ap3-ch-gi-s}
S^{\alpha}_{\beta}=\frac{1}{e_3}(X^{\alpha}_{\rho}X^{\rho}_{\beta}-X^{\rho}_{\rho}X^{\alpha}_{\beta})
+\frac{e_2}{e_3\cdot{e}_3}(X^{\alpha}_{\rho}X^{\rho}_{\sigma}X^{\sigma}_{\beta}
-X^{\rho}_{\rho}X^{\alpha}_{\sigma}X^{\sigma}_{\beta})+\frac{e_2\cdot{e}_2}{e_3\cdot{e}_3}X^{\alpha}_{\beta},
\end{eqnarray}
where $e_2$ and $e_3$ can be computed from Eq.~(\ref{ap3-ch-th-co}) as
\begin{eqnarray}
\label{ap3-ch-gi-e-2}
e_2&=&\frac{1}{2}(X^{\rho}_{\rho}X^{\sigma}_{\sigma}-X^{\rho}_{\sigma}X^{\sigma}_{\rho}),\\
\label{ap3-ch-gi-e-3}
e_3&=&\frac{1}{6}(X^{\tau}_{\tau}X^{\rho}_{\rho}X^{\sigma}_{\sigma}
-3X^{\rho}_{\rho}X^{\tau}_{\sigma}X^{\sigma}_{\tau}+2X^{\tau}_{\rho}X^{\rho}_{\sigma}X^{\sigma}_{\tau}).
\end{eqnarray}
We can check that the following relations hold
\begin{eqnarray}
\label{ap3-ch-gi-x-s}
X^{\alpha}_{\rho}S^{\rho}_{\sigma}X^{\sigma}_{\beta}=X^{\alpha}_{\beta},~~
S^{\alpha}_{\rho}X^{\rho}_{\sigma}S^{\sigma}_{\beta}=S^{\alpha}_{\beta},
\end{eqnarray}
which mean that $S^{\alpha}_{\beta}$ is the generalized inverse matrix of $X^{\alpha}_{\beta}$. We further define
\begin{eqnarray}
\label{ap3-ch-gi-s-g}
g^{\alpha\beta}=S^{\alpha}_{\rho}\eta^{\rho\beta},
\end{eqnarray}
then $g^{\alpha\beta}$ is the generalized inverse matrix of $g_{\alpha\beta}$.

\bibliographystyle{utphys}
\bibliography{DeGFRef}

\end{document}